\documentclass[prb,reprint,amsmath,amssymb,floatfix,superscriptaddress]{revtex4-1}
\usepackage{graphicx}
\usepackage{subfigure}
\usepackage{relsize}
\usepackage{hyperref}  
\usepackage{enumerate}
\usepackage{bbold}
\usepackage{xcolor}
\usepackage{adjustbox}
\renewcommand{\vec}[1]{\mathbf{#1}}
\usepackage{multirow}
\usepackage{soul}

\newcommand{\vk}{\vec{k}}
\newcommand{\vkprime}{\vec{k}^\prime}
\newcommand{\kprime}{k^\prime}

\newcommand{\ta}{\tilde{a}}

\newcommand{\NN}{\mathrm{NN}}
\newcommand{\NNN}{\mathrm{NNN}}
\newcommand{\RPA}{\mathrm{RPA}}

\begin{document}

\title{Higher angular momentum pairing states in Sr$_2$RuO$_4$ in the presence of longer-range interactions}

\author{Xin Wang}
\affiliation{Department of Physics and Astronomy, McMaster University, Hamilton, Ontario, L8S 4M1, Canada}
\author{Zhiqiang Wang}
\affiliation{James Franck Institute, University of Chicago, Chicago, Illinois, 60637, USA}
\author{Catherine Kallin}
\affiliation{Department of Physics and Astronomy, McMaster University, Hamilton, Ontario, L8S 4M1, Canada}
\date{\today}

\begin{abstract}
The superconducting symmetry of Sr$_2$RuO$_4$ remains a puzzle. 
Time-reversal symmetry breaking $d_{x^2-y^2} + ig_{xy(x^2-y^2)}$ pairing has been proposed for reconciling multiple key experiments. However, its stability remains unclear. 
In this work, we theoretically study the superconducting instabilities in Sr$_2$RuO$_4$, including the effects of spin-orbit coupling (SOC), in the presence of both local and longer-range interactions within a random phase approximation. 
We show that the inclusion of second nearest neighbor repulsions, together with non-local SOC in the $B_{2g}$ channel or orbital-anisotropy of the non-local interactions, can have a significant impact on the stability of both $d_{x^2-y^2}$- and $g$-wave pairing channels. 
We analyze the properties, such as Knight shift and spontaneous edge current, of the realized $d_{x^2-y^2} + ig$, $s^{\prime}+id_{xy}$ and mixed helical pairings in different parameter spaces and find that the $d_{x^2-y^2} + ig$ solution is in better agreement with the experimental data. 
\end{abstract}

\pacs{}

\keywords{}
\maketitle

\section{Introduction}
The nature of the unconventional superconductivity in Sr$_{2}$RuO$_4$ (SRO) remains an outstanding open question after more than 27 years of study, despite this material being simpler than the high-temperature cuprates in many respects.
The samples are clean and superconductivity condenses from a well-defined Fermi liquid normal state so that it is natural to take an itinerant-electron perspective, where superconductivity is an instability of the Fermi surface (FS).
However, despite intense efforts, an order parameter (OP) that is consistent with all the key experimental observations is lacking. 

A multi-component OP is inferred from a variety of experiments, including muon spin rotation ($\mu$SR)~\cite{Luke1998, Grinenko2021}, polar Kerr~\cite{Xia2006}, Josephson relation~\cite{Kidwingira2006} and ultrasound measurements~\cite{Ghosh2021, Benhabib2021, Ghosh2022}.
The multi-components can be degenerate by symmetry, belonging to the two-dimensional irreducible representations (irrep.) of the crystal point symmetry group, or be degenerate accidentally, belonging to two distinct one-dimensional irreps. 

Possible symmetry-related OPs for a crystal with $D_{4h}$ symmetry are spin-triplet $p_x\pm ip_y$ with $E_u$ symmetry and spin-singlet $d_{xz}\pm id_{yz}$ ($E_g$). 
Both are difficult to reconcile with experiments.
The $p_x\pm ip_y$ pairing is inconsistent with the significant drop of the in-plane Knight shift below $T_c$ observed in recent NMR experiments.~\cite{Pustogow2019, Ishida2020}
The $d_{xz}\pm id_{yz}$ has symmetry-protected horizontal line nodes at $k_z=0$ that conflict with thermal conductivity and scanning tunneling microscopy (STM) studies, where vertical line nodes are indicated.~\cite{Hassinger2017, Sharma2020}  
In addition, it would produce a jump in the elastic modulus associated with shear $B_{1g}$ strain which is not observed in experiments.~\cite{Ghosh2021} 
Indeed, no $d_{xz}\pm id_{yz}$ pairing has been found in microscopic calculations for SRO~\cite{Roising2019, Romer2022} except in studies of orbital pairings that include sizable interband pairing~\cite{Suh2020, Clepkens2021a}. 
We briefly discuss interband pairing in the conclusions.

The above difficulties associated with the symmetry-related OPs focused attention on the accidental degeneracy scenario, even though it usually requires fine-tuning.
The need for fine-tuning can be somewhat relaxed by considering inhomogeneous states, where, for example, the second OP is induced by inhomogeneous strains near dislocations.~\cite{Yuan2021, Willa2021}
This scenario is consistent with recent $\mu$SR~\cite{Grinenko2021} and ultrasound attenuation measurements~\cite{Ghosh2022}. 

Recently, a time-reversal symmetry breaking (TRSB) $d_{x^2-y^2} \pm i g_{xy(x^2-y^2)}$ pairing with symmetry-protected vertical line nodes has been proposed to explain multiple key experiments.~\cite{Kivelson2020, Willa2020}
Although $d_{x^2-y^2}$-wave is stable in SRO models in the presence of on-site interactions~\cite{Wang2013, Scaffidi2014, Zhang2018, Gingras2019, Romer2019, Wang2019, Wang2020, Romer2021}, $g$-wave is not favored. 
It has been suggested that the $g$-wave state may be stabilized by longer-range interactions based on studies of single-band Hubbard models.~\cite{Raghu2012, Wolf2018} 
A recent study~\cite{Romer2021} found that neither $d_{x^2-y^2}$- nor $g$-wave pairing is favored in SRO in the presence of orbital-isotropic longer-range Coulomb repulsions.
Instead, an $s^{\prime}+id_{xy}$ solution was suggested with gap minima near (1,1,0), which, like $d_{x^2-y^2}+ig$ order, is also consistent with NMR and ultrasound measurements. ($s^{\prime}$ labels nodal $s$-wave states.)
The calculations in Ref.~\onlinecite{Romer2021} are performed in an intermediate Hubbard-$U$ regime, $U \approx 1.1t$, where $t$ is the primary hopping amplitude. 

It was recently reported in Ref.~\onlinecite{Clepkens2021a} that the $g$-wave pairing could be stabilized in SRO by strong non-local SOC in the $B_{2g}$ channel ($\eta_{B_{2g}}$) within the so-called Hund's coupling mean-field approach. 
In this framework, superconducting pairings are generated by attractive on-site interactions due to strong Hund's coupling. 
However, Refs.~\onlinecite{Roig2022, Romer2022} suggested that, in general, Hund's pairing is less favored than spin-fluctuation pairing in SRO due to its nesting features. 
Therefore, it is of interest to study the effects of $\eta_{B_{2g}}$ in SRO more generally. 

In this work, we study the superconducting instabilities in the presence of both local and longer-range Coulomb repulsions in SRO in a realistic multi-orbital model, with local and non-local SOC, over a range of $U$ and other interaction parameters, including the effects of orbital-anisotropies.
One focus is identifying the effects that stabilize $g$-wave. 
In this paper, the effective interactions are treated within the random phase approximation (RPA).
Our studies include both the weak coupling limit and finite-$U$ RPA.
While RPA includes some higher-order scatterings associated with finite interactions and has been shown to agree with other methods for a one-band model,~\cite{Romer2020b} it is unclear whether RPA provides a more accurate description for SRO beyond weak coupling.

We find that nearest neighbor (NN) Coulomb repulsion, $V^{\NN}$, combined with next-nearest neighbor (NNN) repulsion, $V^{\NNN}$, promotes $g$-wave pairing. 
Depending on the strength of $U$, $g$-wave pairing becomes the leading or the first sub-leading pairing for a substantial range of $V^{\NN}$ and $V^{\NNN}$. 
$\eta_{B_{2g}}$ and orbital-anisotropies of $V^{\NN}$ and $V^{\NNN}$ can further stabilize the $g$-wave phase.
Although $d_{x^2-y^2}$ pairing is not favored in the presence of orbital-independent $V^{\NN}$ and $V^{\NNN}$, it can be stabilized by the effects of $\eta_{B_{2g}}$ and longer-range interaction anisotropies.
As a result, accidentally / near degenerate $d_{x^2-y^2}$ and $g$ pairing can be obtained at the phase boundaries in certain parameter spaces.
We also study the physical properties of the realized $d_{x^2-y^2} \pm i g$ pairing and compare it with another two recently proposed pairing candidates: the $s' \pm id_{xy}$~\cite{Romer2021} and a mixed helical pairing~\cite{Huang2021}.
We find that the $d_{x^2-y^2}+ig$ is somewhat in better agreement with the experiments.

The paper is organized as follows. The microscopic model and method employed are discussed in Sec.~\ref{sec:model} and the results of our RPA calculations are presented in Sec.~\ref{sec:instabilities}. 
The physical properties of the possible two-component OPs are discussed in Sec.~\ref{sec:properties}.
Section \ref{sec:conclusion} contains our conclusions and further discussion, including a brief discussion of interband pairing that is found in some studies of SRO~\cite{Clepkens2021a, Kaeser2022}.
Finally, some details are left to Appendixes, including the derivation of the effective interactions in Appendix~\ref{sec:app_methods}, 
the effects of $V^{\NN}$ in Appendix~\ref{sec:app_NNU}, 
the more detailed analysis of the stability of $d_{x^2-y^2}$- and $g$-wave pairing in Appendix~\ref{sec:app_robustness}, 
and the general effects of longer-range interaction anisotropies in Appendix~\ref{sec:app_LRUanisotropy}. 

\section{Model and Method} \label{sec:model}

We consider the microscopic model Hamiltonian for the three conduction bands of SRO,
\begin{gather}
H= H_K + H_{\mathrm{int}},
\end{gather}
where $H_K$ is the kinetic energy part that gives rise to the normal state FSs, 
and $H_{\mathrm{int}}$ is the interaction.

$H_{K}$ can be written in the basis 
$\Psi(\vec{k})=[c_{\vec{k},1,\uparrow};c_{\vec{k},2,\uparrow};c_{\vec{k},3,\downarrow};
c_{\vec{k},1,\downarrow}; c_{\vec{k},2,\downarrow};c_{\vec{k},3,\uparrow}]^T$, so that it is block-diagonal.
\begin{gather} 
\label{eq:Hblock}
\hat{H}_K(\vec{k})=
\begin{pmatrix}
H_{\uparrow\uparrow}(\vec{k}) & 0 \\
0 & H_{\downarrow\downarrow}(\vec{k})
\end{pmatrix},
\end{gather}
where $\{1, 2, 3\}=\{d_{yz}, d_{xz}, d_{xy}\}$ orbitals, and $c^{\dagger} (c)$ is the electron creation (annihilation) operator.
\begin{align}
H_{s s} (\vec{k})= 
\begin{pmatrix}
\epsilon_{yz, \vec{k}}            & g_{\vec{k}} + i s \eta           & -s \eta -i\eta^{B_{2g}}_\vec{k}\\
g_{\vec{k}} - i s \eta     & \epsilon_{xz, \vec{k}}                     & i \eta +s\eta^{B_{2g}}_\vec{k}\\
- s \eta + i\eta^{B_{2g}}_\vec{k}                 & - i \eta + s\eta^{B_{2g}}_\vec{k}           & \epsilon_{xy, \vec{k}}
\end{pmatrix}, 
\label{eq:HK}
\end{align} 
with $s= 1~(-1)$ for spin $\uparrow~(\downarrow)$. 
$\epsilon_{yz (xz), \vec{k}} = -2t \cos{k_{y(x)}} -2t^{\perp}\cos{k_{x(y)}} - \mu$ and 
$\epsilon_{xy, \vec{k}} = -2t^{\prime}(\cos{k_x}+\cos{k_y})-4t^{\prime\prime}\cos{k_x}\cos{k_y}-\mu_c$ describe intra-orbital hoppings; $g_{\vec{k}} =-4t^{\prime\prime\prime} \sin{k_x} \sin{k_y}$ is the hopping between $d_{xz}$ and $d_{yz}$ orbitals. 
$\eta$ is the atomic SOC, and $\eta^{B_{2g}}_{\vec{k}}=4\eta_{B_{2g}}\sin{k_x}\sin{k_y}$ is the non-local SOC in the $B_{2g}$ channel. 
Diagonalizing $\hat{H}_{K}$ gives three doubly degenerate energy bands labeled by band index, $\{\alpha,\beta, \gamma\}$, and pseudo-spin, $\sigma = \uparrow ~(\downarrow)$. 
The band parameters are: $(t, t^{\perp}, t^{\prime\prime\prime}, t^{\prime}, t^{\prime\prime}, \mu, \mu_c) = (1, 0.11, 0.05, 0.8, 0.32, 1.05, 1.1)t$ that capture the overall band structure and FS sheets of SRO. 
For now, the magnitudes of $\eta$ and $\eta_{B_{2g}}$ are left undetermined and will be suitability varied to analyze the effects of SOC. 
The resulting FSs for two different values of the SOC parameters are shown in Fig.~\ref{fig:FS} in Appendix~\ref{sec:app_robustness}.

The interaction Hamiltonian (with on-site and longer-range interactions) is,
\begin{subequations}
\begin{align}
H_{\mathrm{int}} = &\frac{U}{2} \sum_{i,a} n_{ia\uparrow} n_{ia\downarrow} 
+ \frac{U^\prime}{2} \sum_{i,a\ne b,s,s^\prime} n_{ias} n_{ibs^\prime} \notag\\
& + \frac{J}{2} \sum_{i,a\ne b, s,s^\prime} c^\dagger_{i a s } c^\dagger_{i b s^\prime} c_{i a s^\prime} c_{i b s} \notag\\
& + \frac{J^\prime}{2} \sum_{i, a\ne b, s\ne s^\prime} c^\dagger_{i a  s} c^\dagger_{i a s^\prime}  c_{i b s^\prime}  c_{i b s}  \label{eq:Kanamori_onsite} \\
& + \sum_{i, \delta = \{\pm\hat{x}, \pm\hat{y}\}, a, b, s, s^{\prime}} \frac{V^{\NN}_{ab,\delta}}{2}n_{i,a,s} n_{i+\delta , b,s^{\prime}}\notag\\
& + \sum_{i, \delta = \{\pm\hat{x}\pm\hat{y}\}, a, b, s, s^{\prime}} \frac{V^{\NNN}_{ab,\delta}}{2}n_{i,a,s} n_{i+\delta , b,s^{\prime}}, \label{eq:Kanamori_NN}
\end{align}
\label{eq:Kanamori_total}
\end{subequations}
where, $n_{i,a,s}\equiv c^{\dagger}_{i,a,s} c_{i,a,s}$ is the spin and orbital resolved electron density operator at site $i$. 
Eq.~\eqref{eq:Kanamori_onsite} describes the on-site interaction, where
$U$ ($U^\prime$) is the intra (inter)-orbital repulsive Hubbard interaction, 
$J$ is the Hund's coupling, and $J^\prime$ is the pair hopping. 
Eq.~\eqref{eq:Kanamori_NN} describes the longer-range interactions, where $V^{\NN}_{ab,\delta} ~(V^{\NNN}_{ab,\delta})$ is the NN (NNN) Coulomb repulsion. 

For simplicity, we take $J^\prime=J$ and $U^\prime=U-2J$ ($SO(3)$ symmetry)~\cite{Georges2013} and ignore the $d_{xy/z}$ anisotropy due to hybridization with oxygen orbitals~\cite{Zhang2016,Mravlje2011}, but we will briefly comment on the effect of this anisotropy in Sec.~\ref{sec:conclusion}. 
$V^{\NN}_{ab,\delta}$ and $V^{\NNN}_{ab,\delta}$ are $t_{2g}$ orbital-dependent, and their orbital-anisotropies are defined as, 
\begin{align} 
\alpha_{ab,\delta} &\equiv \frac{ V_{ab,\delta}^{\NN}} {V^{\NN}_{11,\hat{x}}} -1 \equiv  \frac{ V_{ab,\delta}^{\NN}} {V^{\NN}}-1,\label{eq:anisotropy_Vnn}\\
\beta_{ab,\delta} &\equiv  \frac{ V_{ab,\delta}^{\NNN} } { V^{\NNN}_{11, \hat{x}+\hat{y}} } -1 \equiv \frac{ V_{ab,\delta}^{\NNN} } { V^{\NNN}}-1,
\label{eq:anisotropy_Vnnn}
\end{align} 
where $V^{\NN}_{11,\hat{x}}$ ($V^{\NNN}_{11,\hat{x}+\hat{y}}$) is the intra-orbital interaction between two NN (NNN)  $d_{yz}$ orbitals with $\delta=\hat{x}$ ($\delta=\hat{x}+\hat{y}$). $\{\alpha_{ab,\delta}\}= 0$ ($\{\beta_{ab,\delta}\}= 0$) describes the orbital-isotropic $V^{\NN}$ ($V^{\NNN}$) case. 
From rotation symmetry in the $t_{2g}$ orbital space, there are 6 free orbital-anisotropy parameters: $\alpha_{33}$, $\alpha_{23, \pm \hat{x}}$, $\alpha_{12}$, $\beta_{33}$, $\beta_{13}$ and $\beta_{12}$. Here, and in the following, we drop the subscript $\delta$ in $\alpha_{33,\delta}$, $\alpha_{12,\delta}$ and $\beta_{ab,\delta}$ as these parameters are $\delta$ independent. 
Following from symmetry: 
\begin{subequations}
\begin{align}
\alpha_{22,\pm \hat{y}} &= \alpha_{11,\pm \hat{x}} = 0, \\
\alpha_{33} &= \alpha_{22,\pm \hat{x}} = \alpha_{11,\pm \hat{y}},\\
\alpha_{12} &= \alpha_{23,\pm \hat{y}} = \alpha_{13,\pm \hat{x}}, \\
\alpha_{23,\pm \hat{x}} &= \alpha_{13,\pm \hat{y}}, 
\end{align}
\end{subequations} 
and
\begin{subequations}
\begin{align}
\beta_{11} &= \beta_{22} = 0, \\
\beta_{13} &= \beta_{23}.
\end{align}
\end{subequations} 

To study the superconducting instabilities, we obtain effective pairing vertices within the RPA. Taking the static limit, the effective interaction in the orbital basis reads,

\begin{align} \label{eq:VeffOrb}
V_{\mathrm{eff}} = \frac{1}{4} \sum_{\vec{k}, \vec{k}^\prime} [\Gamma(\vec{k} ,\vec{k}^\prime)]^{\tilde{a}_1 \tilde{a}_2}_{\tilde{a}_3 \tilde{a}_4} c^\dagger_{\vec{k},\tilde{a}_1} c^\dagger_{-\vec{k},\tilde{a}_3} c_{-\vec{k}^\prime,\tilde{a}_4} c_{\vec{k}^\prime, \tilde{a}_2},
\end{align}
where $\tilde{a}_1 = \{a_1, s_1\}$ is a composite index that labels both orbital and spin, and

\begin{widetext}
\begin{subequations}
\begin{align}
[\Gamma(\vec{k} ,\vec{k}^\prime)]^{\tilde{a}_1 \tilde{a}_2}_{\tilde{a}_3 \tilde{a}_4}  = 
\sum_{\delta, \delta'}\sum_{i, j = \{1, 2\} }
&\left[ \begin{pmatrix}
e^{i\vec{k} \cdot \delta} & 0\\
0 & e^{i \vec{k} \cdot \delta }
\end{pmatrix}
\big[\widetilde{W} (\delta) \big]^{\tilde{a}_1 \tilde{a}_2}_{\tilde{a}_3 \tilde{a}_4}
\begin{pmatrix}
 e^{-i \vec{k}^{\prime} \cdot \delta} & 0\\
 0 & e^{i\vec{k}^{\prime} \cdot \delta}
\end{pmatrix} \right.\\
&-\begin{pmatrix}
e^{i\vec{k} \cdot \delta} & 0\\
0 & e^{i \vec{k} \cdot \delta }
\end{pmatrix} 
\big[\widetilde{W} (\delta) 
\chi^{\mathrm{RPA}}(\vec{k}, \vec{k}'; \delta, \delta')
\widetilde{W}(\delta')\big] ^{\tilde{a}_1 \tilde{a}_2}_{\tilde{a}_3 \tilde{a}_4}
\begin{pmatrix}
 e^{-i \vec{k}' \cdot \delta'} & 0\\
 0 & e^{i \vec{k}' \cdot \delta' }
\end{pmatrix} \\
&+\left.  \begin{pmatrix}
e^{i\vec{k} \cdot \delta} & 0\\
0 & e^{i \vec{k} \cdot \delta }
\end{pmatrix}
\big[\widetilde{W} (\delta) 
\chi^{\mathrm{RPA}}(\vec{k}, - \vec{k}'; \delta, \delta') 
\widetilde{W}(\delta')\big] ^{\tilde{a}_1 \tilde{a}_4}_{\tilde{a}_3 \tilde{a}_2}
\begin{pmatrix}
 e^{i \vec{k}' \cdot \delta'} & 0\\
 0 & e^{-i \vec{k}' \cdot \delta' }
\end{pmatrix}\right]_{ij}.
\end{align}\label{eq:Gamma_RPA}
\end{subequations}
Here,  
\begin{align}\label{eq:chiRPA}
&\chi^{\mathrm{RPA}}(\vec{k}, \vec{k}'; \delta, \delta^{\prime}) = \frac{1}{1+ \chi(\vec{k}, \vec{k}'; \delta, \delta') \widetilde{W}(\delta^{\prime}) }\chi(\vec{k}, \vec{k}'; \delta, \delta^{\prime})
\end{align}
is a generalized $\delta$-dependent RPA particle-hole susceptibility matrix, 
with $\chi(\vec{k},  \vec{k}'; \delta, \delta^{\prime})$ the corresponding bare susceptibility, whose matrix element is 
\begin{align}
&\chi ^{\tilde{b}_1 \tilde{b}_2}_{\tilde{b}_3 \tilde{b}_4}(\vec{k}, \vec{k}'; \delta, \delta')
=\sum_{\vec{p}}\sum_{\alpha,\beta} \frac{n_F(\xi^\alpha_{ \vec{p} }) - n_F(\xi^\beta_{ \vec{p} -( \vec{k} - \vec{k}')}) } {\xi^\beta_{ \vec{p} -( \vec{k} - \vec{k}^{\prime})} - \xi^\alpha_{ \vec{p} }}
\mathcal{F}^{\tilde{b}_1 \tilde{b}_2}_{\tilde{b}_3 \tilde{b}_4} (\alpha, \beta; \vec{p},\vec{k} - \vec{k}')
\begin{pmatrix}
 e^{-i \vec{k}^{\prime} \cdot \delta +i\vec{k}\cdot \delta'} & e^{-i\vec{k}' \cdot \delta + i\vec{p}\cdot \delta'}\\
 e^{-i\vec{p} \cdot \delta + i\vec{k}\cdot \delta'} & e^{-i\vec{p} \cdot (\delta-\delta')}
\end{pmatrix}.
\label{eq:chi_0}
\end{align}
$\mathcal{F}^{\tilde{b}_1 \tilde{b}_2}_{\tilde{b}_3 \tilde{b}_4} (\alpha, \beta; \vec{p},\vec{q})$
is the form factor associated with the band-to-orbital transformations, 
\begin{align} 
\mathcal{F}^{\tilde{b}_1 \tilde{b}_2}_{\tilde{b}_3 \tilde{b}_4} (\alpha, \beta; \vec{p},\vec{q}) 
= \psi^\alpha_{\tilde{b}_2}(\vec{p} )  [\psi^\alpha_{\tilde{b}_3}(\vec{p} )]^*   [\psi^\beta_{\tilde{b}_1}(\vec{p} -\vec{q} )]^* \psi^\beta_{\tilde{b}_4}(\vec{p} - \vec{q}). 
\label{eq:Fo2b}
\end{align}
\end{widetext}
In these equations, $\alpha$ and $\beta$ are energy band labels (including the pseudospin). 
$\xi_{\vk}^{\alpha}$ is the $\alpha$-th band dispersion, 
$\psi_{\tilde{b}}^{\alpha} (\vec{k})$ is the corresponding matrix element of the orbital-to-band transformation, and $n_F$ is the Fermi-Dirac distribution function. 
$\widetilde{W}(\delta)$ is the bare interaction, $H_{\mathrm{int}}$, written in $\vec{k}$-space but with its $\vec{k}$-dependence peeled off and absorbed into the definition of the susceptibility $\chi$,
which reduces the computational complexity. (Similar methods have been introduced in Ref.~\onlinecite{Romer2021}.)
This is achieved by introducing a redundant $2\times 2$ subspace, indexed by $\{i,j\}$ in Eq.~\eqref{eq:Gamma_RPA}. 
More details can be found in Appendix~\ref{sec:app_methods}. 
$\widetilde{W}$ , $\chi$, and $\chi^{\mathrm{RPA}}$ are $N\times N$ matrices for given momenta  with $N = 6\times 6 \times 9\times 2$, 
where $6\times 6$ comes from the two sets of composite indices $\{\tilde{a}_1, \tilde{a}_2\}$, each of which consists of three orbitals $\otimes$ two spin species, nine from the label of neighboring sites $\delta = \{0, ~\hat{x}, ~\hat{y}, ~-\hat{x}, ~-\hat{y}, ~\hat{x}+\hat{y}, ~-\hat{x}+\hat{y}, ~-\hat{x}-\hat{y}, ~\hat{x}-\hat{y}\}$, and two from the additional subspace label $i= \{1,2\}$.
Eq.~\eqref{eq:chi_0} will be evaluated at low temperatures where $\chi$ is temperature independent and using a sufficiently large $\vk$-mesh in the first Brillouin zone~\cite{Wolf2018}.
Throughout this work, we choose $k_BT = 0.001t$ and a $512\times 512$ grid mesh for the integration. 

Transforming $V_{\mathrm{eff}}$ in Eq.~\eqref{eq:VeffOrb} to the band basis leads to 
\begin{align}\hspace{-0.2cm}
V_{\mathrm{eff}} = &\sum_{\vec{k}, \vec{k}^{\prime}} \sum_{\alpha, \beta} \Gamma^{\alpha\beta}(\vec{k}, \vec{k}^{\prime}) 
 c^{\dagger}_{\alpha}(\vec{k})c^{\dagger}_{\alpha}(-\vec{k}) c_{\beta }(-\vec{k}^\prime) c_{\beta}(\vec{k}^\prime) ,
\end{align}
where, 
\begin{align}
\Gamma^{\alpha\beta} &(\vec{k},\vec{k}^{\prime})  
= \frac{1}{4} \sum_{\tilde{a}_i} \Gamma^{\tilde{a}_1 \tilde{a}_2}_{\tilde{a}_3 \tilde{a}_4}(\vec{k},\vec{k}^\prime)  \notag\\
&\times [ \psi^{\alpha}_{\tilde{a}_1}(\vec{k}) ]^* [ \psi^{\alpha}_{\tilde{a}_3}(-\vec{k}) ]^* \psi^\beta_{\tilde{a}_4}(-\vec{k}^\prime)  \psi^\beta_{\tilde{a}_2}(\vec{k}^\prime).
\label{eq:Fo2b2}
\end{align}
Note that we have used $\Gamma$ for both the orbital- and band-basis effective interaction, which are distinguished by their indices. 
Projecting $\Gamma^{\alpha\beta}$ onto the FS, one can determine
the superconducting instabilities by solving the following BCS linearized gap equation~\cite{Scaffidi2014}: 
\begin{align}
\sum_{\beta} \int_{S_{ \beta}} \frac{d k^\prime_{\parallel}}{|S_{\beta}|}g(\vec{k}_{ \alpha}, \vec{k}^{\prime}_{ \beta}) \psi\left(\vec{k}^{\prime}_{ \beta}\right)=\lambda \psi\left(\vec{k}_{ \alpha}\right),
\label{eq:gap}
\end{align}
where
\begin{align}
g(\vec{k}_{ \alpha}, \vec{k}^{\prime}_{ \beta}) =
\sqrt{ \frac{\rho_{\alpha}~\bar{v}_{F, \alpha}}{v_{F}(\vec{k}_{\alpha})}}~ \Gamma^{\alpha\beta} (\vec{k}_{ \alpha}, \vec{k}^{\prime}_{  \beta}) \sqrt{ \frac{\rho_{\beta}~\bar{v}_{F,\beta}}{v_{F}(\vec{k}^{\prime}_{\beta})}},
\label{eq:gap_matrix}
\end{align}
and $\lambda = \rho V_{\mathrm{eff}}$, where $\rho$ is the density of states at the Fermi level.~\cite{Raghu2010}
In Eq.~\eqref{eq:gap}, all momenta are defined on the FS. $S_\beta$ is the FS of the $\beta$-th band, which is a one-dimensional contour for our two-dimensional calculations; $|S_\beta|$ is its corresponding area (or contour length). 
$\rho_{\alpha}$ is the density of states of the $\alpha$-th band, and the average of the norm of the Fermi velocity is given by,  $\bar{v}_{F,\alpha}^{-1} = \int_{S_\alpha} \frac{d k_{\parallel}}{|S_{\alpha}|} v_{F}^{-1}(\vec{k}_{\alpha})$. 
After discretizing Eq.~\eqref{eq:gap}, it becomes a matrix equation to be solved numerically. 
To get good convergence, we discretize the FS contours with $\sim$ 1000 equally spaced points.
Alternatively, one can take an easier method by discretizing the whole first Brillouin zone, but only keeping states that lie within a thin energy window from the Fermi level.
However, as pointed out in Ref.~\onlinecite{Cho2013}, a much larger number of points is then required for the same level of accuracy.

The critical temperature, $T_c$, is determined by the most negative eigenvalue, $\lambda$, through $T_{c} \sim \mathcal{W} e^{-1/|\lambda|}$, where $\mathcal{W}$ is of the order of the bandwidth.
The superconducting gap is 
\begin{equation}
\Delta (\vec{k}_{\alpha}) \propto  \sqrt{ \frac{v_{F}\left(\vec{k}_{\alpha}\right)}{\rho_{\alpha}\bar{v}_{F, \alpha}}} \psi(\vec{k}_{\alpha}), 
\end{equation}
where $\Delta (\vec{k}_{\alpha})$ can be written in the {\textit {pseudospin}} basis as
\begin{align}
\Delta(\vk_{\alpha}) &  =  
\begin{pmatrix}
\Delta_{\uparrow \uparrow}  & \Delta_{\uparrow \downarrow}  \\
\Delta_{\downarrow \uparrow}  & \Delta_{\downarrow \downarrow}
\end{pmatrix}
\end{align}
for a given $\vec{k}_\alpha$ point on one of the three FS sheets.  

\section{Pairing results in the presence of longer-range interactions} \label{sec:instabilities}
We first ignore the effects of non-local SOC, $\eta_{B_{2g}}=0$, but we include a sizable atomic SOC of $\eta/t = 0.2$. 
Similar calculations have been conducted in several theoretical works with only local interactions, where $s^{\prime}$-, $d_{x^2-y^2}$-, helical, or chiral pairing is obtained depending on microscopic details.~\cite{Zhang2018, Romer2019, Wang2020}
As in Ref.~\onlinecite{Romer2021}, we investigate the effects of orbital-independent NN Coulomb repulsions, $V^{\NN}$, in Appendix~\ref{sec:app_NNU}. 
We include a wide range of Hubbard-$U$ from weak to intermediate coupling, i.e. $U/t \in (10^{-4}, 1.1)$, as $U$ can strongly influence the leading pairing within the RPA.~\cite{Zhang2018,Wang2020, Romer2020b}
The Hund's coupling is set as $J/U= 0.2$ as obtained via constrained local-density approximation~\cite{Pchelkina2007} and the constrained RPA~\cite{Mravlje2011}.
The size of $V^{\NN}$ for SRO is not clear. For cuprates with identical crystal structures, $V^{\NN}/U$ is about~\cite{Hirayama2018} 0.2, and this value was used in Ref.~\onlinecite{Romer2021}.
As Ru 4d orbitals are more extended than Cu 3d orbitals, $V^{\NN}/U$ for SRO may be larger.
One finds $V^{\NN}/U \approx 0.38$ from integrals over Slater-type Ru d orbitals where screening effects and hybridizations between the Ru d and oxygen p orbitals are neglected.\cite{Hoggan2011}
By comparison, the same calculation for Cu $d_{x^2-y^2}$ orbitals gives $V^{\NN}/U\approx 0.22$, suggesting the effects of hybridization and screening in the cuprates essentially cancel each other.
Guided by this analysis, we perform calculations for SRO in the range of $V^{\NN}/U \in (0, 0.4)$.
We find that $V^{\NN}$ has little effect in stabilizing $g$-wave pairing and tends to destabilize the $d_{x^2-y^2}$-wave phase.
However, it favors helical pairing in the weak-$U$ regime, and $d_{xy}$-wave at intermediate values of $U$.
As a result, $s^{\prime}+id_{xy}$, $d_{x^2-y^2}+ id_{xy}$, or a mixed helical state can be obtained (at phase boundaries) in different regimes of the interaction parameter space. (See Fig.~\ref{fig:phase_etaUUk} in Appendix~\ref{sec:app_NNU}.)

Our results for intermediate-$U$ are in rough agreement with Ref.~\onlinecite{Romer2021}, except for the absence of the $d_{xy}$ phase there.
We find the $d_{xy}$ state may be overtaken by $s^{\prime}$ if we increase $T$ or decrease $N_{\mathrm{FS}}$, the number of patching points used to solve the linearized gap equation.
The sensitivity to temperature, even at relatively low temperatures, has been noted previously in the context of similar RPA calculations.\cite{Romer2015}
In all our calculations, we choose a sufficiently low temperature for the susceptibility calculations so that the results no longer change with decreasing temperature.

In this section, we focus on the superconducting instabilities in the presence of both $V^{\NN}$ and $V^{\NNN}$. We first ignore the effects of orbital anisotropies of $V^{\NN}$ and $V^{\NNN}$.
The ratio of $V^{\NNN}/V^{\NN}$ can be roughly estimated through integrals over Ru 4d orbitals, as discussed above for $V^{\NN}/U$, which gives $V^{\NNN}/V^{\NN}\sim 0.7$. 
This neglects hybridization and screening, the combined effect of which likely reduces $V^{\NNN}/V^{\NN}$.
Our calculations will focus on the range of $V^{\NNN}/V^{\NN}\in (0,~ 0.7)$. 

Non-zero $V^{\NN}$ and $V^{\NNN}$ produce a correction, $\delta\Gamma(\vk,\vkprime)$, to the effective pairing interaction.
For weak $V^{\NN}$ and $V^{\NNN}$, $\delta\Gamma(\vk,\vkprime)$ is dominated by the bare-$V^{\NN}$ and $V^{\NNN}$ contributions, 
which can be schematically written as, 
\begin{widetext}
\begin{subequations}
\begin{align}
\delta \Gamma^{(1)} (\vk,\vkprime)  &  \sim  V^{\NN} \left[\cos{(k_x-k^\prime_x)}+\cos{(k_y-\kprime_y)}\right]\mathcal{F}_{o\rightarrow b}(\vk,\vkprime) 
 +  2 V^{\NNN} \cos{(k_x-\kprime_x)}\cos{(k_y-\kprime_y)}\mathcal{F}_{o\rightarrow b}(\vk,\vkprime) \\
 & = \sum_{\Lambda, i} \left(g_{\Lambda, i}^{\NN} + g_{\Lambda, i }^{\NNN}  \right) [ \phi^{\Lambda, i}(\vk)]^*  \phi^{\Lambda, i}(\vkprime) ,
 \label{eq:V_basis} 
\end{align} 
\end{subequations}
\end{widetext}
where $\phi^{\Lambda, i}$denotes the $i$-th lattice harmonic of irrep. $\Lambda$ in the $D_{4h}$ group and $g^{\NN (\NNN)}_{\Lambda, i}$ is the corresponding pairing interaction strength. 
In the presence of SOC, $\mathcal{F}_{o\rightarrow b}$, the form factor associated with orbital-to-band transformation, carries nontrivial pseudospin structures [see Eq.~\eqref{eq:Fo2b2}], which are omitted here for a qualitative discussion. 

In the single-band case, $\mathcal{F}_{o\rightarrow b}(\vk,\vkprime) = 1$ and $\delta \Gamma^{(1)}(\vk,\vkprime)$ can be greatly simplified.
$g_{\Lambda, j}^{\NNN}$ is non-zero and repulsive only for NNN harmonics in the $\Lambda = \{ A_{1g}, B_{2g}, E_{u}\} = \{ s^{\prime}, d_{xy}, p\}$ irrep. with eigenfunctions: $\phi^{s^{\prime}, 2}=\cos{k_x}\cos{k_y}$,  $\phi^{d_{xy},2}= \sin{k_x}\sin{k_y}$, and $\phi^{p_{x(y)},2}= \cos{k_{y(x)}}\sin{k_{x(y)}}$.~\cite{Raghu2012} 
Similarly, as discussed in Appendix~\ref{sec:app_NNU} and in Ref.~\onlinecite{Raghu2012}, $g_{\Lambda, i}^{\NN}$ is repulsive for $\phi^{s^{\prime}, 1}=\cos k_x + \cos k_y$, $\phi^{d_{x^2-y^2}, 1}=\cos k_x -\cos k_y$, and $\phi^{p_{x/y}, 1}=\sin k_{x/y}$. 
In summary, $\delta \Gamma^{(1)}$ has repulsive components in all the pairing channels except for $g$-wave. 
In the multi-band model with SOC, our numerical results show that $\delta \Gamma^{(1)}$ remains repulsive as long as $J/U\le 1/3$ and also has small components in the $g$-wave channel. 

For sizable $V^{\NN}$ and $V^{\NNN}$, the second-order correction, $\delta\Gamma^{(2)}$, becomes important. 
$\delta\Gamma^{(2)}$ usually involves higher angular harmonics and can be attractive due to fluctuations.
For example, $V^{\NN}(\vk,\vkprime) \widetilde{\chi}(\vk-\vkprime)V^{\NNN}(\vk,\vkprime)$ is one of the second order correction terms from the bubble diagram in Fig.~\ref{fig:diagrams}(b) where $\widetilde{\chi}$ represents the bubble;
the expansion of this term into angular harmonics contains the $g$-wave component with basis functions such as $\phi^{g,4}(\vk) =\phi^{d_{xy},2} (\vk) \phi^{d_{x^2-y^2},1} (\vk) =  \sin{k_{x}}\sin{k_{y}}(\cos{k_{x}}-\cos{k_{y}})$ for the single band case.  
This argument applies even in the presence of multi-orbitals and SOC.
Thus, $g$-wave can be promoted by the combined effects of $V^{\NN}$ and $V^{\NNN}$. 

\begin{figure}[!htp]
\centering
\includegraphics[width=0.8\linewidth,trim={0mm 0mm 0mm 0mm},clip]{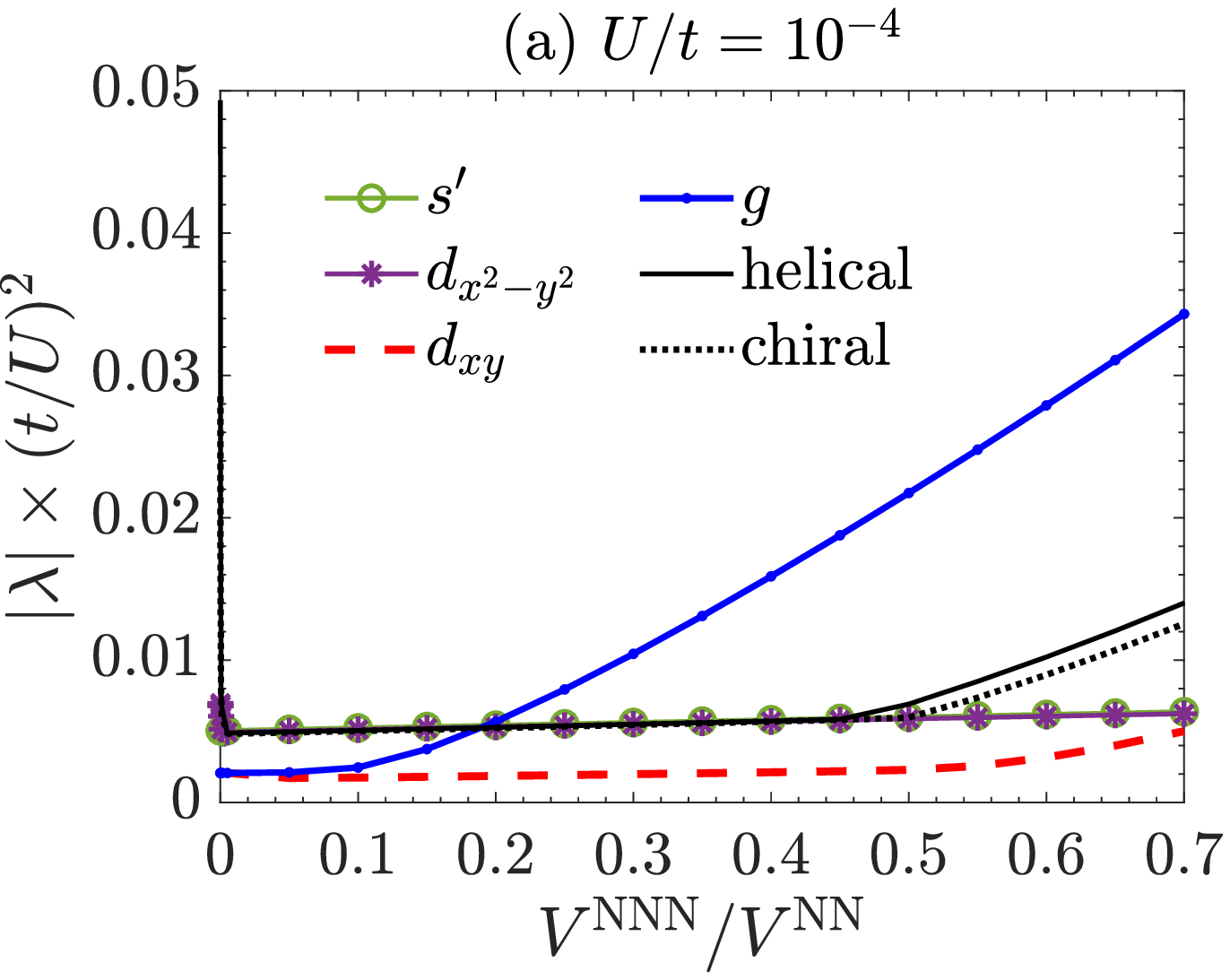}
\includegraphics[width=0.8\linewidth,trim={0mm 0mm 0mm 0mm},clip]{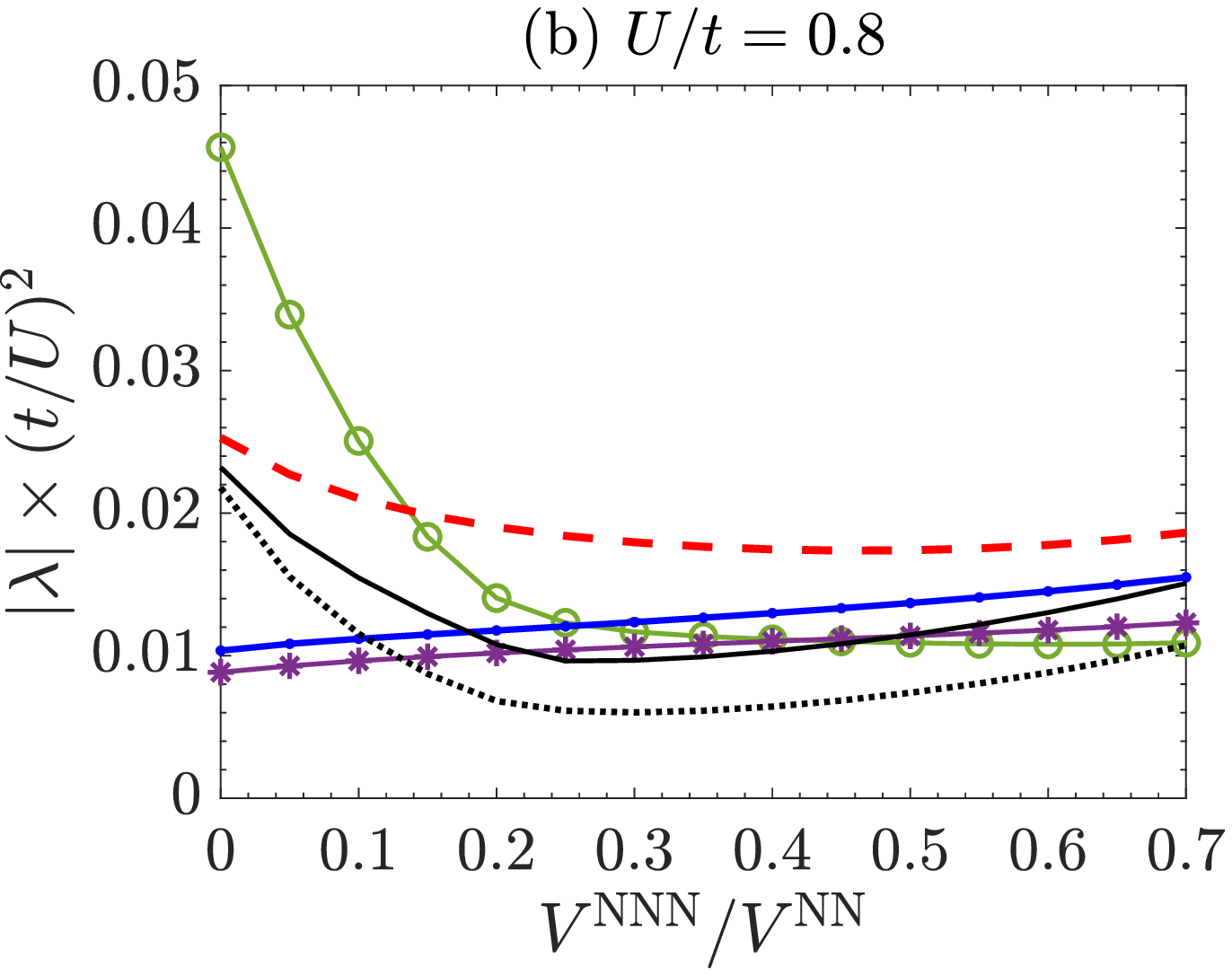}
\caption{Superconducting instabilities as a function of $V^{\NNN}/V^{\NN}$ for: (a) $U/t=10^{-4}$, and (b) $U/t=0.8$, where helical and $s^{\prime}$-wave is favored at $V^{\NNN}=0$, respectively. Only the largest eigenvalue (in magnitude) of each irrep. is shown. $\eta/t=0.2$, $\eta_{B_{2g}}=0$, $J/U=0.2$ ~ and $V^{\NN}/U=0.25$.}
\label{fig:lambda_Uk1Ukk}
\end{figure}

Figure~\ref{fig:lambda_Uk1Ukk} shows the effects of $V^{\NNN}$ on the leading superconducting instability in each irrep. in the case of $V^{\NN}/U$=0.25 for (a), $U/t=10^{-4}$ and (b), $U/t=0.8$, where helical and $s^{\prime}$-wave is favored without $V^{\NNN}/V^{\NN}$, respectively.
One sees that $g$-wave pairing is enhanced by $V^{\NNN}$.
In the weak-$U$ case, $g$-wave state becomes the leading order at $V^{\NNN}/V^{\NN}\gtrsim 0.2$, as other pairing channels are largely suppressed by the bare and repulsive $V^{\NN}$ and $V^{\NNN}$. 
An $s^{\prime}+ig$ pairing can be obtained close to the multi-critical point (i.e. $V^{\mathrm{NNN}}/V^{\mathrm{NN}} \approx 0.2$), where the $s^{\prime}$- and $g$-wave channels are near degenerate. 
For intermediate $U$, $g$-wave order becomes the first sub-leading pairing for a substantial range of $V^{\NNN}/V^{\NN}$, $V^{\NNN}/V^{\NN}\gtrsim 0.3$, whereas $d_{xy}$-wave pairing is dominant. 
We also find that $s^{\prime}$-wave pairing is significantly suppressed by $V^{\NNN}$-induced corrections at the RPA level, in contrast to the case studied in Ref.~\onlinecite{Romer2022}, where the suppression effect is moderate. 
In summary, $g$-wave is the leading or the first sub-leading pairing for a broad range of $U$, $V^{\NN}$ and $V^{\NNN}$ (not shown); while $d_{x^2-y^2}$-wave is not favored. 

\begin{figure}[htp]
\centering
\includegraphics[width=0.8\linewidth,trim={0mm 0mm 0mm 0mm},clip]{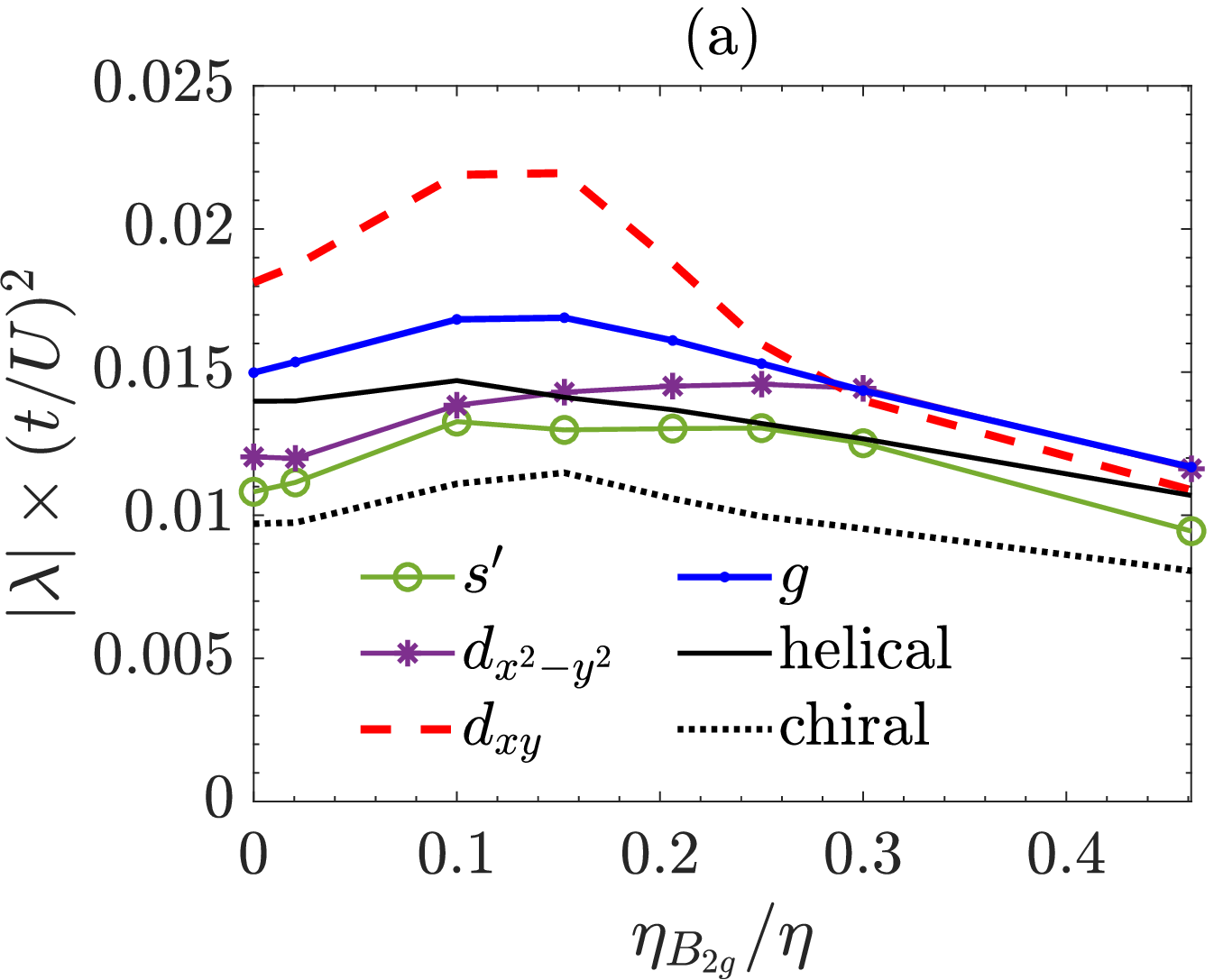}
\includegraphics[width=0.8\linewidth,trim={0mm 0mm 0mm 0mm},clip]{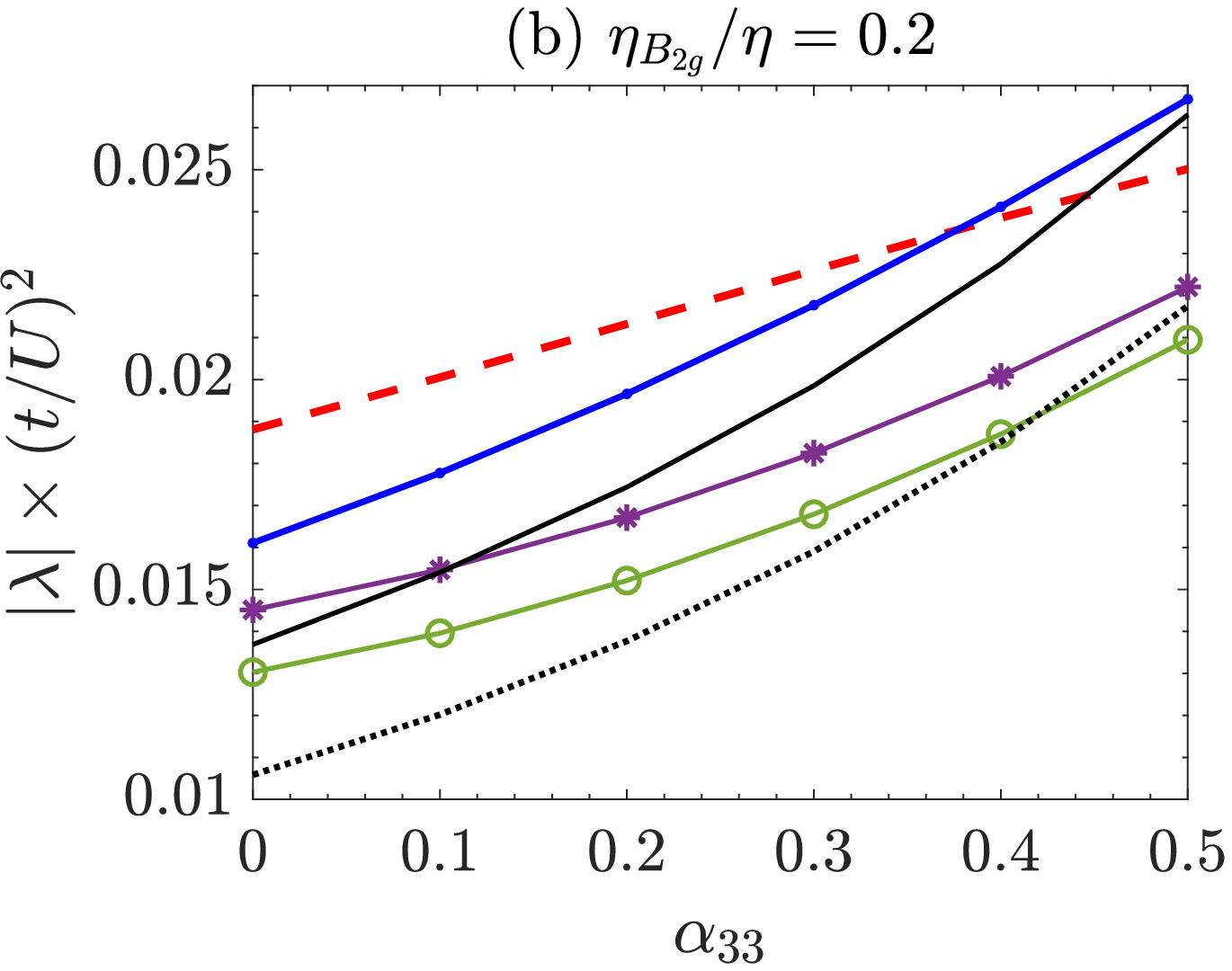}
\caption{(a) Effects of $\eta_{B_{2g}}/\eta$ on the superconducting instabilities in different channels for $U/t=0.8, V^{\NN}/U=0.25, V^{\NNN}/V^{\NN}=0.65$. (b) Evolutions of the leading superconducting instabilities as a function of anisotropy parameter $\alpha_{33}$ for $\eta_{B_{2g}}/\eta=0.2$ where $\alpha_{33}=0$ represents the isotropic longer-range interaction case in (a). Other anisotropy parameters are chosen as: $(\alpha_{23,\pm\hat{x}}, \alpha_{12}, \beta_{33}, \beta_{13}, \beta_{12}) = (1, 0.4, 0.33, 0.17, 0)\alpha_{33}$.}
\label{fig:lambda_mucetak}
\end{figure}

$\eta_{B_{2g}}$ can strongly impact the higher angular momentum pairings as it involves NNN Ru sites. 
Figure~\ref{fig:lambda_mucetak}~(a) shows the superconducting instabilities as a function of $\eta_{B_{2g}}/\eta$ in the intermediate $U$ case ($V^{\NN}/U=0.25$ and $V^{\NNN}/V^{\NN}=0.65$), where $g$-wave is the first sub-leading order at $\eta_{B_{2g}} = 0$. 
$\eta_{B_{2g}}/\eta$ is increased by decreasing $\eta$ linearly while increasing $\eta_{B_{2g}}$, so that the sum of $\eta$ and $\eta_{B_{2g}}$ is constant.~ \cite{Clepkens2021a}
The chemical potential for the $d_{xy}$ orbitals is adjusted, $\tilde{\mu}_{c} = \mu_{c}+\delta \mu_{c}$ to fit the ARPES data~\cite{Tamai2019}. 
In Fig.~\ref{fig:lambda_mucetak}~(a), we find that both $d_{x^2-y^2}$- and $g$-wave pairings are dominant and near/accidentally degenerate at $\eta_{B_{2g}}/\eta \gtrsim 0.3$. 
The very close overlap of these two states for $\eta_{B_{2g}}/\eta \gtrsim 0.3$ is accidental.  
In general, we find $d_{x^2-y^2}$- and $g$-wave states are the first two leading pairings in the range of $V^{\mathrm{NN}}/U\in (0.2,~0.3)$ and $V^{\mathrm{NNN}}/V^{\mathrm{NN}}>0.5$ (in Fig.~\ref{fig:lambda_etak0.3UkUkk} of Appendix~\ref{sec:app_robustness}). 
Also, for the weak-$U$ case, $d_{x^2-y^2}$ and $g$-wave are found to be the first two leading channels at a much smaller $\eta_{B_{2g}}/\eta$, $\eta_{B_{2g}}/\eta>0.15$ (see Fig.~\ref{fig:lambda_Uk0.0001mucetak} in Appendix~\ref{sec:app_robustness}).

We note that the required $\eta_{B_{2g}}/\eta$ for the presence of both $d_{x^2-y^2}$ and $g$-wave phase is much larger than the density functional theory (DFT) estimate $\sim 0.02$.~\cite{Clepkens2021a} 
However, it has been pointed out that SOC is underestimated in the DFT calculations and can be further enhanced by correlation effects.~\cite{Liu2008, Isobe2014, Zhang2016, Kim2018, Tamai2019} 
In addition, in the following, we will show that this value can be reduced by including longer-range interaction anisotropies. 

The $d_{x^2-y^2}+ig$ pairing stabilized by $\eta_{B_{2g}}$ is also observed in a recent study using a mean-field approach, although a much larger $\eta_{B_{2g}}/\eta \gtrsim 0.45$ is required there.~\cite{Clepkens2021a}
In addition, our RPA calculations find that non-zero $V^{\NN}$ and $V^{\NNN}$ are necessary to obtain the $d_{x^2-y^2}+ig$-wave phase, unlike in Ref.~\onlinecite{Clepkens2021a}. 

We further consider the effects of the longer-range interaction anisotropies. 
The magnitudes of the orbital-anisotropies, defined in Eqs.~\eqref{eq:anisotropy_Vnn} and \eqref{eq:anisotropy_Vnnn}, largely depend on the spread of the $d$-orbitals. 
We can roughly estimate the anisotropy parameters through integrals over Ru 4d Slater-type orbitals, where we find that the largest orbital anisotropy parameter, $\alpha_{33}$, is about $0.12$. As discussed above, such estimations do not include the hybridization and screening effects. These effects substantially enhance the interaction anisotropies in HgBa$_2$CuO$_4$. For example, the NN interaction for Cu $d_{x^2-y^2}$-orbitals is about 25$\%$ larger than that for $d_{3z^2-r^2}$-orbitals according to Ref.~\onlinecite{Hirayama2019}, while a direct Slater integral gives only $3\%$. 
Similarly, the hybridization and screening effects may also enhance the estimates in SRO. 
In comparison to HgBa$_2$CuO$_4$, the orbitals are larger (which should increase the hybridization) but the Ru-O bonds are substantially less anisotropic in different crystal directions (which decreases the enhancement).  Consequently, in the absence of a detailed calculation, we treat the anisotropy as a variable parameter.
Figure~\ref{fig:lambda_mucetak}(b) shows the superconducting instabilities as a function of $\alpha_{33}$ in the case of $\eta_{B_{2g}}/\eta=0.2$. 
The relative magnitudes of other parameters are chosen as  $(\alpha_{23,\pm\hat{x}}, \alpha_{12}, \beta_{33}, \beta_{13}, \beta_{12}) = (1, 0.4, 0.33, 0.17, 0)\alpha_{33}$, based on rough estimates through Ru $t_{2g}$ Slater-type orbitals integrals. (Details can be found in Appendix~\ref{sec:app_LRUanisotropy}.) 
The $t_{2g}$ orbital-anisotropy increases the stability of the $g$-wave so that it becomes the leading order for $\alpha_{33}\gtrsim 0.36$. 
We note that the required $\alpha_{33}$ to stabilize the $g$-wave is much larger than its Slater orbital estimate, 0.12. 
 However, this does not need to be an obstruction since we also found that the required $\alpha_{33}$ can be much smaller in some parts of the parameter space, for example, with larger $V^{\mathrm{NN}}$ and $V^{\mathrm{NNN}}$ and/or in the weak-$U$ regime (not shown). Furthermore, as discussed above, the actual $\alpha_{33}$ is expected to be larger than our simple estimate.
Although $d_{x^2-y^2}$-wave pairing is not favored, it is promoted relative to the $d_{xy}$-channel, suggesting that the $d_{x^2-y^2}$ phase can be enlarged by orbital anisotropies. 
More discussion on the anisotropy effects are shown in Appendix~\ref{sec:app_LRUanisotropy}, where, in particular, we find that $\alpha_{33}$ helps to stabilize $g$-wave pairing, and $\beta_{33}$ favors $d_{x^2-y^2}$-wave for finite-$U$. 

\section{Properties of the stable pairing states} \label{sec:properties}

\begin{figure*}[!htp]
\centering
\includegraphics[width=0.345\linewidth,trim={0mm 0mm 10mm 0mm},clip]{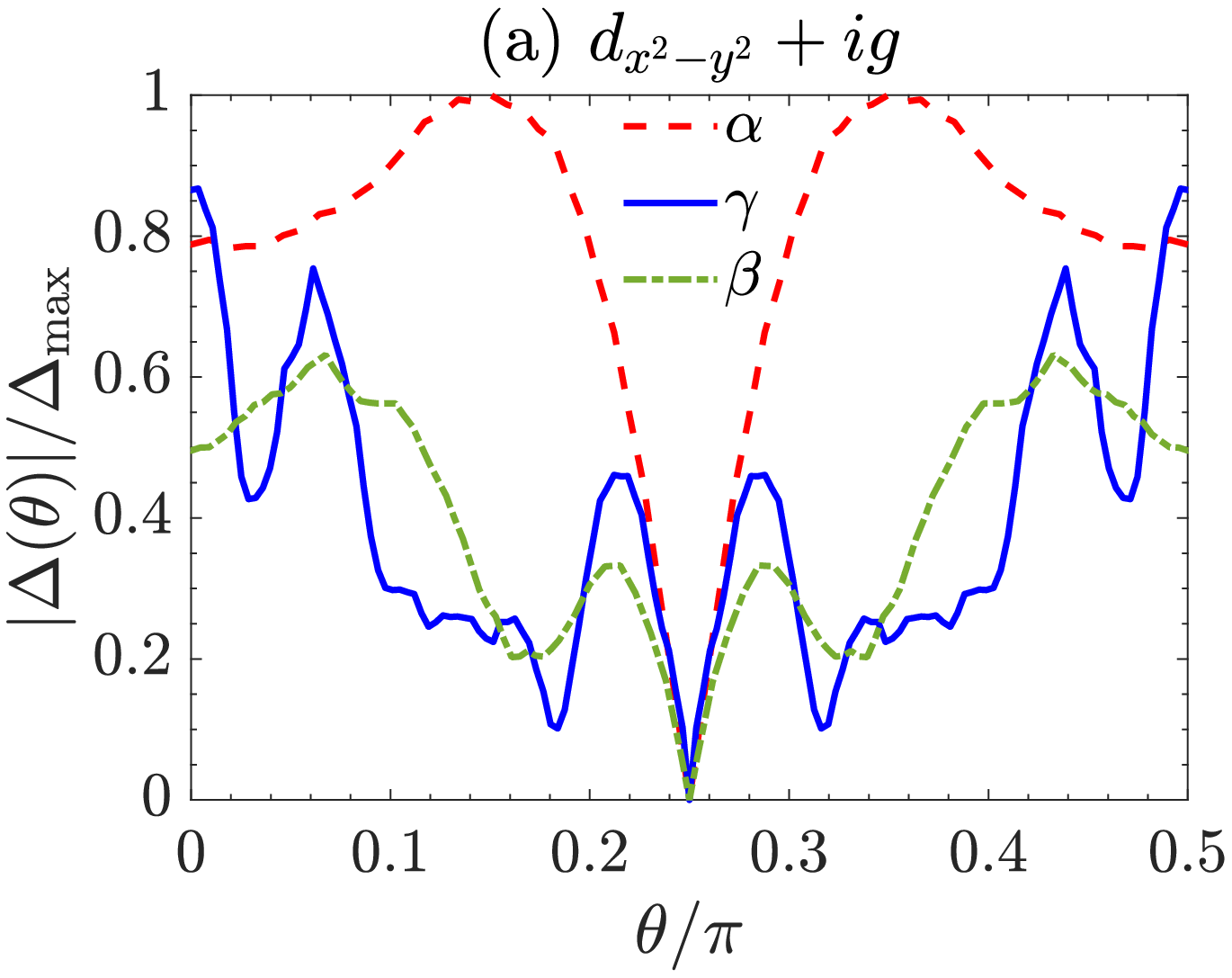}
\includegraphics[width=0.32\linewidth,trim={9mm 0mm 10mm 0mm},clip]{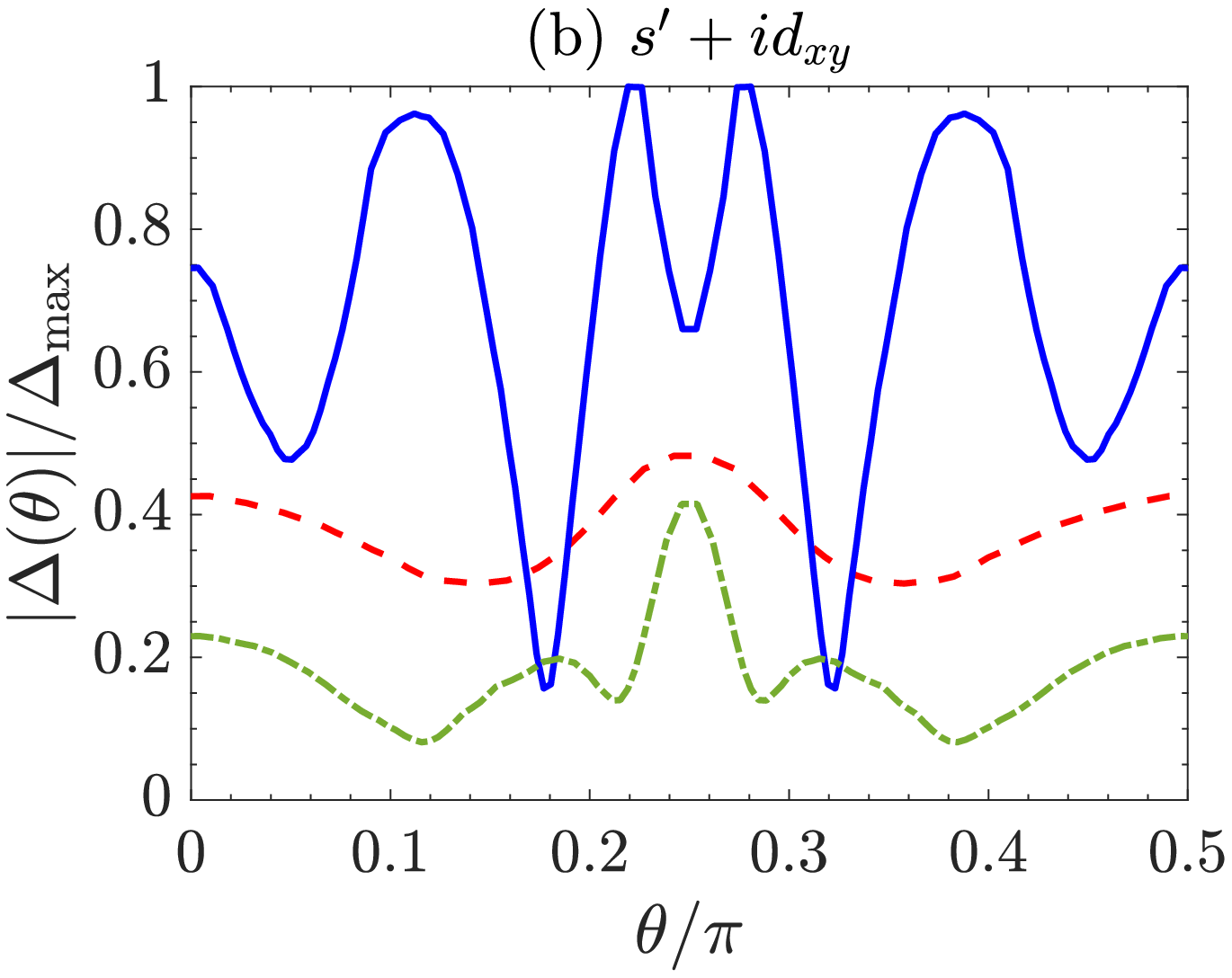}
\includegraphics[width=0.32\linewidth,trim={9mm 0mm 10mm 0mm},clip]{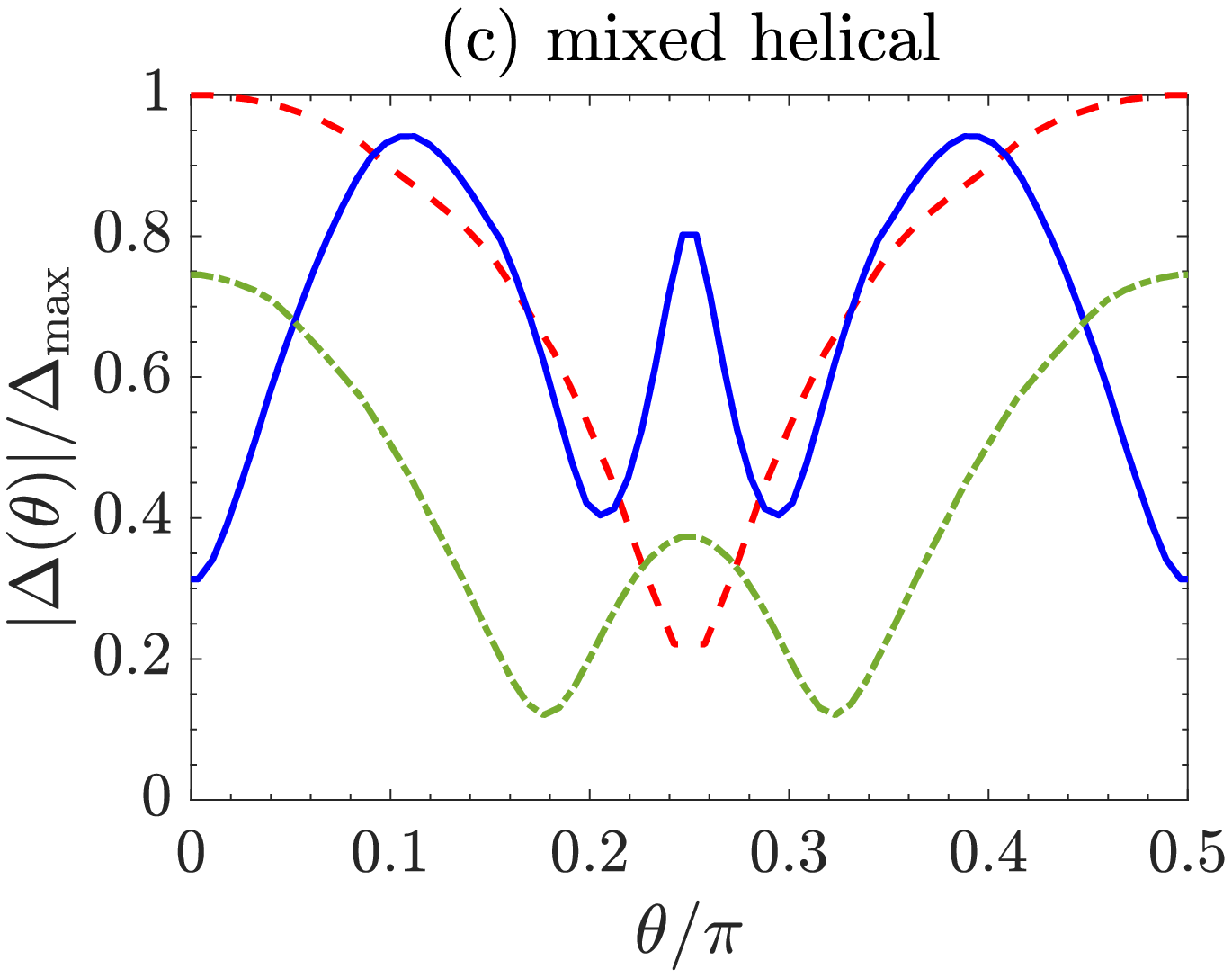}
\caption{Gap function profiles of three TRSB solutions (a) $d_{x^2-y^2}+ig$, (b) $s^{\prime}+id_{xy}$ and (c) mixed-helical, along the three FS contours in one quadrant of the first BZ. The FS contour of each band is parameterized by the angle of the vector $\vec{k}_F= k_{F}(\theta)(\cos{\theta}, \sin{\theta})$. Angle $\theta$ is measured from $(\pi, \pi)$ for the $\alpha$ band, and from
$(0, 0)$ for $\beta$ and $\gamma$ bands. 
The three states are obtained at phase boundaries where their two respective components are degenerate.
The band/ interaction parameters for the phase boundaries are: $(\eta_{B_{2g}}/\eta, U/t, V^{\NN}/U, V^{\NNN}/V^{\NN}) =$ (a) $(0.3, 0.8, 0.25, 0.65)$, (b) $(0, 0.8, 0.05, 0)$, and (c) (0, 0.0001, 0.15, 0).
Note, for the non-unitary mixed-helical pairing, because $|\Delta_{\downarrow \downarrow}| \ll |\Delta_{\uparrow \uparrow}|$, only the latter is shown. }
\label{fig:OP_TRSB}
\end{figure*}

In Sec.\ref{sec:instabilities}, we show that $d_{x^2-y^2}$- and $g$-wave pairing can be favored in SRO by the effects of longer-range interactions and $\eta_{B_{2g}}$. 
Consequently, at certain parameters, $d_{x^2-y^2}+ig$ pairing can be realized. 
In this section, we explore the gap structure, spin susceptibility, and spontaneous edge current of $d_{x^2-y^2}+ig$ pairing using the stable OP configurations found at the phase boundary: $(\eta_{B_{2g}}/\eta, U/t, V^{\NN}/U, V^{\NNN}/V^{\NN}) =$ (0.3, 0.8, 0.25, 0.65), to see if it can be compatible with experiments on SRO. 
In addition, we also compare these properties of  $d_{x^2-y^2}+ig$ pairing to those of two other recently proposed pairings, $s^{\prime}+id_{xy}$ and mixed helical pairing, which are obtained at $(\eta_{B_{2g}}/\eta, U/t, V^{\NN}/U, V^{\NNN}/V^{\NN}) =$ (0, 0.8, 0.05, 0) and (0, 0.0001, 0.15, 0), respectively. 
The stability of the latter two TRSB pairing candidates is discussed in Appendix~\ref{sec:app_NNU} with nonzero $V^{\NN}$.
The $s^{\prime}+id_{xy}$ can be obtained for a finite-$U$, as in Ref.~\onlinecite{Romer2021},  while the mixed helical pairings are realized in the weak-$U$ limit. 
We also find that the splitting between helical pairings in $B_{1u}$ and $B_{2u}$ (or $A_{1u}$ and $A_{2u}$) is rather small throughout $V^{\NN}\in(0, 0.3)$ in the weak-$U$ limit as shown in  Fig.~\ref{fig:lambda_Uk} (in Appendix~\ref{sec:app_NNU}). 
The result at $V^{\NN}=0$, i.e. with only on-site interactions, is consistent with previous studies both in 2D~\cite{Wang2020} and in 3D~\cite{Roising2019}. 

\subsection{Gap structure}

The gaps are of similar size on all bands and exhibit strong gap anisotropy with multiple nodes or near-nodes on the FS for all three pairings (shown in Fig.~\ref{fig:OP_TRSB}).
We find $|\Delta|_{\mathrm{min}}/|\Delta|_{\mathrm{max}}\sim 10\%$~\footnote{$|\Delta|_{\mathrm{max (min)}}$ is the maximum (minimum) gap magnitude over all three bands.} in the realized $s^{\prime}+id_{xy}$ and mixed helical states.
Since the experiment estimate of $|\Delta|_{\mathrm{min}}/|\Delta|_{\mathrm{max}}$ is $\lesssim 3\%$,~\cite{Lupien2001, Hassinger2017} further fine-tuning of the interaction parameters is needed to make the $s^{\prime}+id_{xy}$ and the mixed helical states compatible with the experiments.
In agreement with the previous studies,~\cite{Scaffidi2014, Zhang2018, Romer2019, Romer2021} the locations of the minima are slightly off the $k_x = \pm k_y$ diagonal lines and are robust against the change of interaction parameters. 
Thus, future experiments on the precise location of the nodes or near-nodes can help in identifying the pairing symmetry.

\subsection{Spin susceptibility and Knight shift} \label{subsec:knightshift}

\begin{figure}[!htp]
\centering
\includegraphics[width=0.8\linewidth,trim={0mm 0mm 0mm 0mm},clip]{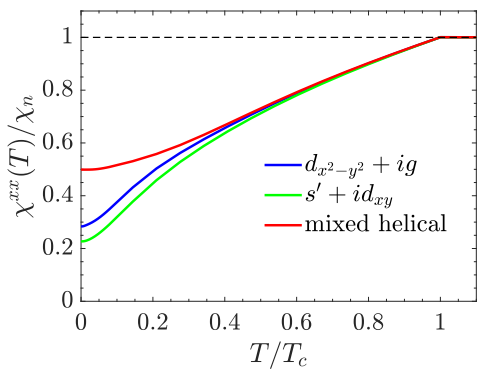}
\caption{The temperature dependence of the spin susceptibility (normalized by the normal state value $\chi_{n}$) for the three OP pairings, calculated in the presence of a small Zeeman field in the $x$-direction and with Fermi-liquid corrections~\cite{Leggett1965, Ishida2020} included. 
We set $k_BT_c = 0.005t$, and the maximum magnitude of the gap is $|\Delta|_{\text{max}}=0.015t$. These calculations follow those in Ref.~\onlinecite{Roising2019}.}
\label{fig:knight_fit}
\end{figure}

Recent NMR measurements reveal a substantial in-plane Knight shift drop below $T_c$,~\cite{Pustogow2019, Ishida2020} which is most straightforwardly explained by spin-singlet pairings. 
It has been argued that spin-triplet helical pairings could also be consistent with the susceptibility drop.~\cite{Huang2021} 

Figure~\ref{fig:knight_fit} shows the calculated spin susceptibility as a function of temperature for the three pairings. The residual $\chi(T=0)$ is roughly similar for all three pairings due to SOC, which mixes spin-singlet and triplet states. 
$\chi(T=0)/\chi_{n}$ is about $28\%$, $23\%$, $50\%$ for the $d_{x^2-y^2}+ig$, $s^{\prime}+id_{xy}$, and mixed helical pairing, respectively. 
Taking into account experimental precision along with vortex and disorder effects, the $s^{\prime}+id_{xy}$ and $d_{x^2-y^2}+ig$ pairings are in better agreement with the experiments. 
It might be difficult to clearly distinguish these two pairings in Knight shift measurements, especially if the $s^{\prime}+id_{xy}$ had extremely deep gap minima, i.e. $|\Delta|_{\mathrm{min}}/|\Delta|_{\mathrm{max}} < 3\%$, which is expected to increase the residual spin susceptibility.
Nevertheless, the calculated $\chi(T=0)/\chi_{n}$ is much higher than the upper bound of 10$\%$ suggested by the experiments~\cite{Chronister2021}. 
We note that the experimental interpretations are complicated by the difficulties in disentangling the orbital and spin contributions. 
Our results for the $s^{\prime}+id_{xy}$ pairing are consistent with Ref.~\onlinecite{Romer2021}. 

\subsection{Spontaneous edge current}\label{subsec:edgecurrent}

\begin{figure}[!htp]
\centering
\includegraphics[width=0.8\linewidth,trim={0mm 0mm 0mm 0mm},clip]{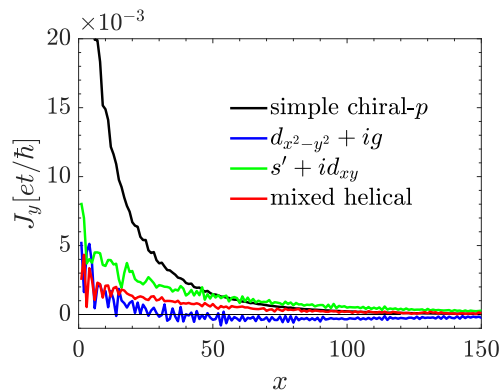}
\caption{Distribution of zero-temperature spontaneous edge current for the three pairings compared with simple chiral $p$-wave. A superconducting region of width $L_S = 800$ sites was taken with surface along (1,0,0). We set $k_BT_c = 0.005t$ and the maximum magnitude of the gap is $|\Delta|_{\text{max}}=0.015t$. These calculations follow those in Ref.~\onlinecite{Scaffidi2015}, and lattice constant is defined to be unity.}
\label{fig:current}
\end{figure}

A TRSB superconducting state may support finite spontaneous edge currents which are expected to be detected by high sensitivity magnetic scanning microscopy. 
Experiments on SRO show no evidence for such edge currents, suggesting that the current is either absent or too tiny to be resolved.~\cite{Kirtley2007, Hicks2010, Curran2014} 
It's been pointed out that the spontaneous edge current can be dramatically reduced by gap anisotropies ~\cite{Scaffidi2015,Huang2015}, indicating that the three pairings may be reconciled with the null results in experiments, although often fine-tuning is required. 

The spontaneous edge current for the $d_{x^2-y^2}+ig$ and $s^{\prime}+id_{xy}$ pairings is sensitive to the edge orientation: the current is generally finite at (1, 0, 0) surfaces and vanishes at (1, 1, 0) surfaces.
For the mixed helical pairing, the current is independent of the edge orientation. 
As shown in Fig.\ref{fig:current}, the predicted edge currents at the (1, 0, 0) surfaces, $J_{y}(x)$, for the three pairings are much smaller than the simple chiral $p$-wave case. 
In addition, there is a sign change in $J_{y}(x)$ for the $d_{x^2-y^2}+ig$ pairing, which significantly reduces the total integrated edge current, $I_{y} = \int dx J_{y}(x)$. 
In particular, this current is compatible with the experiments,~\cite{Kirtley2007} since  $I_{y}^{ d_{x^2-y^2}+ig} / I_{y}^{\mathrm{chiral-}p} \approx 0.6\% $, where $I^{\mathrm{chiral-}p}_{y}$ is the simple chiral $p$-wave result~\cite{Stone2004}. This current ratio is $19\%$ and $36\%$ for the $s^{\prime}+id_{xy}$ and mixed helical pairings, respectively. 
The larger current reduction in the $d_{x^2-y^2}+ig$  state is a result of the higher angular harmonics in the gap functions and should be robust since it comes from an intrinsic property of the bulk superconducting state.
We note that the $s^{\prime}+id_{xy}$ and mixed helical pairings may also support edge currents smaller than the measurable limit, however, this would need fine-tuning. 
Our results for the mixed helical pairing are in rough agreement with Ref.\onlinecite{Huang2021}. 

\section{Conclusions} \label{sec:conclusion}
Within the RPA, we find that both $d_{x^2-y^2}$- and $g$-wave can be stabilized by the effects of longer-range interactions and $\eta_{B_{2g}}$, and that accidentally / near degenerate $d_{x^2-y^2}+ig$ can be stable for a specific range of parameters. 
We also calculate the physical properties of the realized $d_{x^2-y^2}+ig$ state and compare it with the realized $s^{\prime}+id_{xy}$~\cite{Romer2021} and mixed helical pairings~\cite{Huang2021}. We find that, although the $s^{\prime}+id_{xy}$ pairing is as competitive as the $d_{x^2-y^2}+ig$, and is better than the mixed helical pairing, in explaining the spin susceptibility data~\cite{Pustogow2019, Ishida2020}, the $d_{x^2-y^2}+ig$ state is more compatible with experimental evidence of the existence of nodes/near-nodes~\cite{Hassinger2017, Sharma2020} and the absence of spontaneous edge current~\cite{Kirtley2007, Hicks2010, Curran2014} than the other two proposals.

Since $T_c \sim e^{-1/|\lambda|}$, 
the TRSB phase would occur only when the interaction parameters are tuned essentially right at the phase boundaries, where $T_{c}^{d_{x^2-y^2}} \approx T_{c}^{g}$.
As pointed out in Refs.~\onlinecite{Yuan2021, Willa2021}, the relative stability of the $d_{x^2-y^2}$- and $g$-wave states can be sensitive to local strains, so that in the presence of such strains, one may have coexisting domains of $g$-wave and $d_{x^2-y^2}$-wave order.
In this case, time-reversal symmetry can be broken at domain walls between different strain regions.~\cite{Yuan2021}
Ref.~\onlinecite{Willa2021} shows that strain-inhomogeneities can couple a single-component primary OP, e.g. $d_{x^2-y^2}~ (g)$ -wave, to a subleading pairing state, e.g. $g~ (d_{x^2-y^2})$-wave, and break time-reversal symmetry. 
As perfect degeneracy of the $d_{x^2-y^2}$ and $g$-wave is not required, the $d_{x^2-y^2}+ig$ state is expected to be stabilized in a broader parameter regime compared to a homogeneous $d_{x^2-y^2}+ig$ state. 
Recent studies show that the inhomogeneous TRSB domain walls may provide a route to explaining the observation of half-quantum vortices~\cite{Yuan2021}, heat capacity~\cite{Li2021} and ultrasound attenuation measurements~\cite{Ghosh2022} on SRO. 
Our calculations suggest that even modest strains may be sufficient for such a scenario since we find $d_{x^2-y^2}$ and $g$-wave to be the first two leading orders over a range of parameters. 

The RPA formalism we employed here is limited to small values of the Coulomb interaction, $U\lesssim \mathcal{O}(t)$, it neglects correlation effects away from the FS, and it neglects many diagrams in calculating the effective interaction. 
Thus, the RPA approach may not adequately capture the physics of SRO with typical estimates of $U \sim \mathcal{O}(eV) \sim 10t$.~\cite{Mravlje2011, Vaugier2012, Behrmann2012, Zhang2016} 
A recent study~\cite{Romer2020b} compares the weak coupling RG, RPA, and dynamical cluster approximation (DCA) approaches within the single band Hubbard model, and it finds good agreement among these approaches, in the region where they can be compared, suggesting a smooth crossover in pairing states within the Hubbard model from weak to strong coupling.
While such an analysis has not been done for the multiband case, it suggests that the RPA approach can provide valuable insight into superconducting pairings in SRO. 
We believe that some of the observations about the effects of interaction-anisotropy and $\eta_{B_{2g}}$ in our RPA calculations will survive at strong coupling since they are independent of the strength of $U$.
Also, beyond the weak-$U$ limit, correlation effects away from the FS can be important to the superconducting instabilities. 
For example, the RPA approach may underestimate the stability of $d_{x^2-y^2}$ pairing by not adequately accounting for the effect of the van Hove singularity away from, but near, the FS of the $\gamma$ band.
Thus, the functional RG approach~\cite{Wang2013, Tsuchiizu2015} with longer-range interactions would be of interest since it can be employed in the sizable $U$-regime and it treats states away from the FS more accurately.

The superconducting pairings discussed above are classified according to the irrep. of the point group of the lattice in the band basis. 
Recently, it was argued that the above analysis of the pairing function is insufficient.~\cite{Kaba2019, Ramires2019, Huang2019} 
Instead, some recent studies of SRO focus on the orbital basis approach and propose specific interorbital pairings.~\cite{Clepkens2021a, Kaeser2022} 
Ref.~\onlinecite{Kaeser2022} finds an odd frequency inter-orbital singlet pairing is favored by solving the linearized Eliashberg equation. 
Using a mean-field approach, Ref.~\onlinecite{Clepkens2021a} finds stable inter-orbital $d_{x^2-y^2}+ig$ pairing at $J/U>1/3$. 
Transforming these inter-orbital pairings into the band basis, one finds that they both support substantial interband pairing away from the FS.
By contrast, the approach we take ignores interband pairing.  

The relative size of intraband and interband pairing in SRO is an open question. However, since the minimum energy difference between electron states on different bands and with opposite momentum is much larger than the superconducting pairing energy in SRO, one might expect interband pairing to be small. 
For example, interband pairing might be reduced by finite frequency effects that are usually ignored in weak-coupling formalisms.
In any case, one does not expect interband pairing to significantly impact the relative stability of different pairing states close to $T_c$, since interband pairing does not open up a gap anywhere on the FS. 
In particular, interband pairing only contributes in order $\Delta^2/E_F$ to the pairing gap at the FSs. 
Nevertheless, interband pairing can be important for some properties, including the polar Kerr effect~\cite{Xia2006}, which has been measured at high frequencies where all superconducting contributions are of order $(\Delta/E_F)^2$ or smaller. 
While the approach of Ref.~\onlinecite{Kaeser2022} could, in principle, address the size of interband pairing, the calculations are restricted to temperatures $\gtrsim$ 300$K$, not only well above $T_c$, but also above the temperature where a well-defined FS emerges. 
One would likely need to go to much lower temperatures to reliably capture the relative size of intraband and interband pairing. 
Since the presence of substantial interband pairing could impact the interpretation of some experiments, this is an interesting avenue for future studies. 

Lastly, we comment on the effects of $d_{xy/z}$ anisotropy we have neglected throughout this paper.
In the crystal field with $D_{4h}$ symmetry, the Ru $t_{2g}$ states split into an $e_g$ doublet ($d_{xz}$, $d_{yz}$) and a $b_{2g}$ singlet ($d_{xy}$). 
As the RuO bond in the $c$-direction is elongated, $d_{xy}$ orbitals are more spread out than $(d_{xz}, d_{yz})$ orbitals, suggesting stronger interactions for $d_{xy}$ orbitals. 
This anisotropy would slightly suppress $g$-wave pairing. 
However, it won't significantly affect our key results, as it is very small in SRO, i.e. the spread of the $d_{xy}$ orbital is larger than $(d_{xz}, d_{yz})$ by about $7\%$.~\cite{Mravlje2011, Zhang2016} 

\section{Acknowledgments}
We would like to thank Sung-Sik Lee, Erik S. S{\o}rensen, and Wen Huang for useful discussions. 
This research is supported by the Natural Sciences and Engineering Research Council. 
This work was made possible by the facilities of the Shared Hierarchical Academic Research Computing Network (SHARCNET:www.sharcnet.ca) and Compute/Calcul Canada. 
Z.W. is supported by James Franck Institute at the University of Chicago. 

\appendix \label{sec:appendix}
\begin{appendix}

\section{Susceptibility and effective interaction} \label{sec:app_methods}
In this section, we show the derivation of the effective interaction in the Cooper pair channel that takes into account on-site, NN, and NNN interactions. 
The bare interaction Hamiltonian is defined in Eq.~\eqref{eq:Kanamori_total}, and it can be rewritten in a more compact form as, 

\begin{widetext} 
\begin{align}
H_{\mathrm{int}} = \frac{1}{4} &\sum_{i, \delta, {\tilde{a}_i}} \left( \big[ W_1(\delta)\big] ^{\tilde{a}_1 \tilde{a}_2}_{\tilde{a}_3 \tilde{a}_4} c^{\dagger}_{i, \tilde{a}_1}c^{\dagger}_{i+\delta, \tilde{a}_3}c_{i+\delta, \tilde{a}_4}c_{i, \tilde{a}_2} 
+ \big[W_2(\delta)\big] ^{\tilde{a}_1 \tilde{a}_2}_{\tilde{a}_3 \tilde{a}_4}
c^{\dagger}_{i, \tilde{a}_1}c^{\dagger}_{i+\delta, \tilde{a}_3}c_{i, \tilde{a}_4}c_{i+\delta, \tilde{a}_2} \right).
 \label{eq:Hint_matrix}
\end{align} 
$\tilde{a}_{j} = \{a_j, s_j\}$ is a composite index that labels both orbital ($a_j$) and spin ($s_j$). 
$[W_1(\delta)]^{\tilde{a}_1, \tilde{a}_2}_{\tilde{a}_3, \tilde{a}_4}$ has the following non-zero elements, 
\begin{subequations}
\begin{align}
&[W_1(0)]^{a s, a s}_{a \bar{s} , a \bar{s}}  = U, \quad
[W_1(0)]^{a s, a s}_{b \bar{s}, b \bar{s}}  = U^{\prime},  \quad
[W_1(0)]^{a s, b s}_{a \bar{s}, b \bar{s}}  = J^{\prime}, \quad
[W_1(0)]^{a s, b s}_{b \bar{s}, a \bar{s}}  =J, \quad
[W_1(0)]^{a s, a s}_{b s, b s}  = U^{\prime}-J, \\
&[W_1(\delta\neq 0)]^{a s, a s}_{a \bar{s}, a \bar{s}} = [W_1(\delta\neq 0)]^{a s, a s}_{a s, a s}  = V_{a a,\delta},  \quad\quad\quad
[W_1(\delta\neq 0)]^{a s, a s}_{b \bar{s}, b \bar{s}} = [W_1(\delta\neq 0)]^{a s, a s}_{b s, b s} = V_{a b,\delta},
\end{align}
\end{subequations}
where $a\ne b$ and $\bar{s}=-s$. 
$W_2(\delta)$ is related to $W_1(\delta)$ such that the whole interaction matrix coefficient is antisymmetric with respect to exchanges of indices of two creation or annihilation operators in $H_{\mathrm{int}}$: 
\begin{align}\label{eq:antisymm}
 [W_2(\delta)]^{\tilde{a}_1, \tilde{a}_4}_{\tilde{a}_3, \tilde{a}_2} 
= [W_2(\delta)]_{\tilde{a}_1, \tilde{a}_4}^{\tilde{a}_3, \tilde{a}_2} 
= -[W_1(\delta)]^{\tilde{a}_1, \tilde{a}_2}_{\tilde{a}_3, \tilde{a}_4} 
= -[W_1(\delta)]_{\tilde{a}_1, \tilde{a}_2}^{\tilde{a}_3, \tilde{a}_4}.
\end{align}
$[W_1(\delta)]^{\tilde{a}_1, \tilde{a}_2}_{\tilde{a}_3, \tilde{a}_4}$ (and, similarly, $[W_2(\delta)]^{\tilde{a}_1, \tilde{a}_2}_{\tilde{a}_3, \tilde{a}_4}$) is a $36\times 36$ matrix, for each value of $\delta$, with $(\tilde{a}_1,  \tilde{a}_2)$ its row index and $(\tilde{a}_3, \tilde{a}_4)$ the column index. 

By Fourier transformation of the interaction we get, 
\begin{align}
H_{\mathrm{int}}(\vk) =\frac{1}{4}\sum_{\vec{k}_i,\tilde{a}_i}\sum_{\delta} &\big[ W(\vec{k}_1, \vec{k}_2; \vec{k}_3, \vec{k}_4, \delta)\big]  ^{\tilde{a}_1 \tilde{a}_2}_{\tilde{a}_3 \tilde{a}_4} ~ c^{\dagger}_{\vec{k}_1, \tilde{a}_1} c^{\dagger}_{\vec{k}_3, \tilde{a}_3}c_{\vec{k}_4, \tilde{a}_4}c_{\vec{k}_2, \tilde{a}_2},
 \label{eq:Hint_matrix_k}
\end{align}
where 
\begin{align}
\big[ W(\vec{k}_1, \vec{k}_2; \vec{k}_3, \vec{k}_4, \delta)\big]  ^{\tilde{a}_1 \tilde{a}_2}_{\tilde{a}_3 \tilde{a}_4}
=\sum_{i, j = \{1, 2\} }
\left[ \begin{pmatrix}
e^{i\vec{k}_1 \cdot \delta} & 0\\
0 & e^{i \vec{k}_1 \cdot \delta }
\end{pmatrix}
\underbrace{\begin{pmatrix}
\big[ W_1(\delta)\big] ^{\tilde{a}_1 \tilde{a}_2}_{\tilde{a}_3 \tilde{a}_4} & 0\\
0 & \big[ W_2(\delta)\big] ^{\tilde{a}_1 \tilde{a}_2}_{\tilde{a}_3 \tilde{a}_4}
\end{pmatrix}}
_{[\widetilde{W} (\delta) ]^{\tilde{a}_1 \tilde{a}_2}_{\tilde{a}_3 \tilde{a}_4}}
\begin{pmatrix}
 e^{-i \vec{k}_2 \cdot \delta} & 0\\
 0 & e^{-i\vec{k}_4 \cdot \delta}
\label{eq:Wkij}
\end{pmatrix} \right]_{ij}.
\end{align}
Here, we introduce an additional but redundant $2\times 2$ subspace, and the sum is taken over all the matrix elements in this subspace. 
In the form of Eq.~\eqref{eq:Wkij}, the momenta dependence of the bare interaction is factored out, which can facilitate our derivation of the RPA effective interactions, as will become clear in the following. 

In the presence of longer-range interactions, the bare interaction depends on the momentum transfer, $\vec{k}_1-\vec{k}_2$ and $\vec{k}_1-\vec{k}_4$. 
As a result, in deriving the effective interaction, the interaction vertex in the higher-order diagrams involves the internal loop momentum $\vec{p}$, unlike the on-site interaction case where the bare interaction is momentum independent~\cite{Scaffidi2014, Zhang2018, Romer2019}. 
The additional $\vec{p}$-dependence poses a challenge for writing diagrammatic contributions to the RPA effective interaction as a simple geometric and algebraic sum. 
This computational complexity can be reduced by factoring out the $\vec{p}$-dependence in the interaction vertex and absorbing it into the definition of particle-hole susceptibility as in Ref.~\onlinecite{Romer2021}. 
We follow a similar approach to that in Ref.~\onlinecite{Romer2021} except that we start with the antisymmetrized bare interaction. 
This parametrization of the interaction will prove convenient in the following derivation of the RPA effective interaction. 

\begin{figure*}[htp]
\centering
\includegraphics[width=0.85\linewidth]{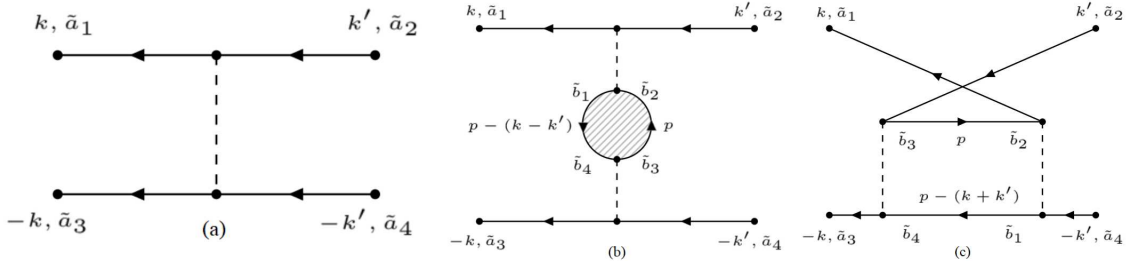}
\caption{The first and second order diagrams which contribute to the effective interaction in the Cooper pair channel. Note that each interaction line carries four joint composite indices $\tilde{a}_i = (a_i, s_i)$.
The internal momentum label, $p$, is a short-hand notation for frequency and momentum, both of which need to be summed over; on the other hand, for the external momenta, $k$ and $k^\prime$,
we only consider zero frequency, i. e. the retardation effect in the effective pairing interaction is neglected.} 
\label{fig:diagrams}
\end{figure*}

The exchange of the spin and orbital fluctuations can induce attractions responsible for superconductivity, even if the bare interaction is repulsive. 
To take into account this effect, one calculates the effective electron-electron interaction, $[ \Gamma (\vec{k}, \vec{k}^{\prime}) ]^{\ta_1 \ta_2}_{\ta_3 \ta_4} $ in Eq.~\eqref{eq:VeffOrb}, 
by summing up one-particle irreducible diagrams of different orders in the bare interaction, Eq.~\eqref{eq:Hint_matrix_k}. 
The first-order contribution [shown in Fig.~\ref{fig:diagrams}~(a)] is
\begin{align}
 [ \Gamma (\vec{k}, \vec{k}^{\prime})^{(1)} ]^{\ta_1 \ta_2}_{\ta_3 \ta_4} 
= \sum_{\delta}\sum_{i, j = \{1, 2\} }
\left[ \begin{pmatrix}
e^{i\vec{k} \cdot \delta} & 0\\
0 & e^{i \vec{k} \cdot \delta }
\end{pmatrix}
\big[\widetilde{W} (\delta) \big]^{\tilde{a}_1 \tilde{a}_2}_{\tilde{a}_3 \tilde{a}_4}
\begin{pmatrix}
 e^{-i \vec{k}^{\prime} \cdot \delta} & 0\\
 0 & e^{i\vec{k}^{\prime} \cdot \delta}
\end{pmatrix} \right]_{ij}.
\label{eq:Gamma_0}
\end{align}

The two second-order diagrams are shown in Fig.~\ref{fig:diagrams}~(b-c). The contribution of the bubble diagram is expressed as, 
\begin{align}
[ \Gamma (\vec{k}, \vec{k}^{\prime})^{(\mathrm{2, bubble})} ]^{\ta_1 \ta_2}_{\ta_3 \ta_4}
= & - \sum_{\delta, \delta'} \sum_{\tilde{b}_i}\sum_{\vec{p}}\big[ W(\vec{k}, \vec{k}'; \vec{p}-(\vec{k}-\vec{k}'), \vec{p}, \delta)\big] ^{\tilde{a}_1 \tilde{a}_2}_{\tilde{b}_1 \tilde{b}_2} 
 \sum_{\alpha,\beta} \frac{n_F(\xi^\alpha_{ \vec{p} }) - n_F(\xi^\beta_{ \vec{p} -( \vec{k} - \vec{k}')}) } {\xi^\beta_{ \vec{p} -( \vec{k} - \vec{k}^{\prime})} - \xi^\alpha_{ \vec{p} }} \notag\\
&\times\mathcal{F}^{\tilde{b}_1 \tilde{b}_2}_{\tilde{b}_3 \tilde{b}_4} (\alpha, \beta; \vec{p},\vec{k} - \vec{k}')
 \big[ W(\vec{p}, \vec{p}-(\vec{k}-\vec{k}') ; -\vec{k}, -\vec{k}', \delta')\big] _{\tilde{a}_3 \tilde{a}_4}^{\tilde{b}_3 \tilde{b}_4} 
\label{eq:bubbleline1}\\
= &  -   \sum_{\delta, \delta'}\sum_{i, j = \{1, 2\} }
\left[ \begin{pmatrix}
e^{i\vec{k} \cdot \delta} & 0\\
0 & e^{i \vec{k} \cdot \delta }
\end{pmatrix} 
\big[\widetilde{W} (\delta) 
\chi( \vec{k}, \vec{k}'; \delta, \delta')
\widetilde{W}(\delta')\big] ^{\tilde{a}_1 \tilde{a}_2}_{\tilde{a}_3 \tilde{a}_4}
\begin{pmatrix}
 e^{-i \vec{k}' \cdot \delta'} & 0\\
 0 & e^{i \vec{k}' \cdot \delta' }
\end{pmatrix} \right]_{ij}.
\label{eq:Gamma_2_bubble}
\end{align} 
$\big[ \chi(\vec{k}, \vec{k}'; \delta, \delta')\big]^{\tilde{b}_1 \tilde{b}_2}_{\tilde{b}_3 \tilde{b}_4}$ is the bare susceptibility defined in Eq.~\eqref{eq:chi_0}.
$\xi_{\vk}^{\alpha (\beta)}$ is the $\alpha (\beta)$-band dispersion, 
$n_F$ is the Fermi-Dirac distribution function, 
and $\mathcal{F}^{\tilde{b}_1 \tilde{b}_2}_{\tilde{b}_3 \tilde{b}_4} (\alpha, \beta; \vec{p},\vec{q})$ is the form factor associated with the band-to-orbital transformations given in Eq.~\eqref{eq:Fo2b}. 
From Eq.~\eqref{eq:bubbleline1} to \eqref{eq:Gamma_2_bubble}, the $\vec{p}$ dependence in the interaction vertices, $\big[ W(\vec{k}, \vec{k}'; \vec{p}-(\vec{k}-\vec{k}'), \vec{p}, \delta)\big]$ and $\big[ W(\vec{p}, \vec{p}-(\vec{k}-\vec{k}') ; -\vec{k}, -\vec{k}', \delta')\big]$, is factorized and absorbed into the integrand of 
$[\chi( \vec{k}, \vec{k}'; \delta, \delta')]$. 

Similarly, the ladder diagram contribution is, 
\begin{align}
[ \Gamma (\vec{k}, \vec{k}')^{(\mathrm{2, ladder})}  ]^{\ta_1 \ta_2}_{\ta_3 \ta_4}=
 \sum_{\delta, \delta'}\sum_{i, j = \{1, 2\} }
\left[ \begin{pmatrix}
e^{i\vec{k} \cdot \delta} & 0\\
0 & e^{i \vec{k} \cdot \delta }
\end{pmatrix}
\big[\widetilde{W} (\delta) \chi( \vec{k}, -\vec{k}'; \delta, \delta') \widetilde{W}(\delta')\big] ^{\tilde{a}_1 \tilde{a}_4}_{\tilde{a}_3 \tilde{a}_2}
\begin{pmatrix}
 e^{i \vec{k}' \cdot \delta'} & 0\\
 0 & e^{-i \vec{k}' \cdot \delta' }
\end{pmatrix} \right]_{ij}.
\label{eq:Gamma_2_ladder}
\end{align}
Notice that $[ \Gamma (\vec{k}, \vec{k}')^{(\mathrm{2, ladder})}  ]^{\ta_1 \ta_2}_{\ta_3 \ta_4} = - [ \Gamma (\vec{k}, -\vec{k}')^{(\mathrm{2, bubble})}  ]^{\ta_1 \ta_4}_{\ta_3 \ta_2}$. 
As a result, the effective interaction at the order of $(U/t)^{2}$ satisfies the same anti-symmetric property as the bare interaction in Eq.~\eqref{eq:Hint_matrix_k}.
The effective interaction vertex at the RPA level [in Eq.~\eqref{eq:Gamma_RPA}] is obtained by summing up the bare interaction in Eq.~\eqref{eq:Gamma_0}, and a geometric series of the bubble and ladder diagrams. The latter contribution takes a form similar to Eqs.~\eqref{eq:Gamma_2_bubble} and \eqref{eq:Gamma_2_ladder}, except that the susceptibility $\chi$ is replaced by $\chi^{\mathrm{RPA}}$,
given in Eq.~\eqref{eq:chiRPA}. Note that, not only the usual particle-hole bubble and ladder contributions but the vertex corrections consisting of admixtures of the bubble and ladder vertices~\cite{Altmeyer2016}, all of which are summed to infinite order, are included, since the bare interaction is anti-symmetrized. 
\end{widetext}

\section{Superconducting instabilities in the presence of NN Coulomb repulsion} \label{sec:app_NNU}

In this section, we explore the superconducting instabilities in the presence of the on-site Kanamori-Hubbard interaction, $U$, and NN Coulomb repulsion, $V^{\NN}$. 
To deduce the general behavior, we perform calculations from weak to intermediate strength of $U$. The effects of $\eta_{B_{2g}}$ and longer-range anisotropies are neglected here.

For comparison, we first briefly summarize the results with $U$ only.~\cite{Zhang2018, Wang2020, Romer2019, Romer2020a}
It has been pointed out that the interplay of the bare-$U$ interaction and fluctuations mediated interactions is nontrivial in determining the leading superconducting pairing within the RPA.~\cite{Zhang2018, Romer2020a} 
In a multi-orbital model with SOC, the bare-$U$ interaction is repulsive in the even-parity $s^{\prime}$-, $d_{x^2-y^2}$-, and $d_{xy}$-wave channels, but it does not affect $g$-wave or odd-parity pairings. 
On the other hand, fluctuation-induced interactions favor $s^{\prime}$- and $d_{x^2-y^2}$-wave pairings.~\cite{Zhang2018} 
Thus, as $U$ crosses from the weak- to the intermediate-coupling regime and the bare $U$ becomes relatively less important, the dominant pairing changes from a helical to an $s^{\prime}$- or $d_{x^2-y^2}$-wave.~\cite{Zhang2018,Wang2020}

When $V^{\NN}$ is taken into account, it produces a correction, $\delta \Gamma(\vk,\vkprime)$, to the effective pairing interaction. 
At the bare-$V^{\NN}$ level, $\delta \Gamma^{(1)}(\vk,\vkprime)$ has the following schematic form,
\begin{align}
\delta  \Gamma^{(1)}
&\sim V^{\NN} \left[\cos{(k_x-\kprime_x)}+\cos{(k_y-\kprime_y)}\right]\mathcal{F}_{o\rightarrow b}(\vk,\vkprime) \notag\\
&=  \sum_{\Lambda} \sum_{i} g_{ \Lambda, i}^{\NN} \;  [ \phi^{ \Lambda, i}(\vk) ]^* \phi^{ \Lambda, i}(\vkprime),
\label{eq:Vnn_basis_app}
\end{align} 
where $\phi^{\Lambda, i}$ is the $i$-th lattice harmonic of irrep. $\Lambda$ in the $D_{4h}$ group, with $g_{\Lambda, i}^{\NN}$ the corresponding pairing interaction strength. 
Note that, in the presence of SOC, $\mathcal{F}_{o\rightarrow b}(\vk,\vkprime)$, the form factor associated with the orbital-to-band transformation, is in general a complex matrix in the pseudospin subspace (whose dependence is omitted here for a qualitative discussion). 
In the single band Hubbard model, as discussed in Refs.~\onlinecite{Raghu2012, Wolf2018}, $\mathcal{F}_{o\rightarrow b}(\vk,\vkprime) = 1$; 
$g_{\Lambda, i}^{\NN}$ is repulsive in the $s$-, $d_{x^2-y^2}$- and $p$-wave channels, while it is zero for both the $d_{xy}$- and $g$-wave channels. 
In the multi-orbital case with a finite SOC, our numerical results show that $\delta \Gamma^{(1)}$ also contains small repulsive components in the $d_{xy}$- and $g$-wave channels, induced by the nontrivial form factor $\mathcal{F}_{o\rightarrow b}(\vk,\vkprime)$. 

Higher-order contributions to $\delta \Gamma(\vk,\vkprime)$ due to $V^{\NN}$ are again induced by particle-hole fluctuations in spin, charge and orbitals, and they can be either attractive or repulsive.  
They can make significant contributions to certain otherwise suppressed channels.
For example, expanding the second-order correction term in the form of $V^{\NN}(\vk,\vkprime) \widetilde{\chi}(\vk,\vkprime)V^{\NN}(\vk,\vkprime)$, where
$\widetilde{\chi}$ represents the bubble in Fig.~\ref{fig:diagrams}(b), into different harmonic channels shows that this term has a substantial weight in the $d_{xy}$-wave channel. 

\begin{figure}[htp]
\centering
\includegraphics[width=0.8\linewidth,trim={0mm 0mm 0mm 0mm},clip]{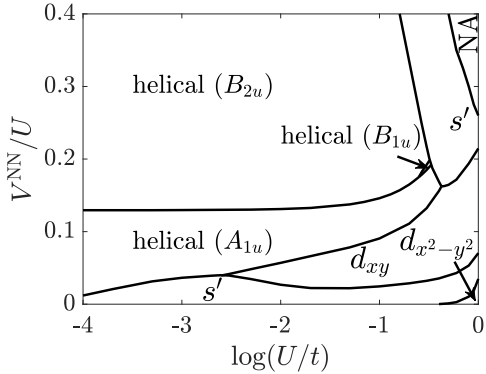}
\caption{Leading superconducting instability phase diagram as a function of $\log_{10}(U/t)$ and orbital-isotropic $V^{\NN}$. ‘NA’ corresponds to the regime where the RPA susceptibility diverges and where RPA breaks down. $\eta/t=0.2, \eta_{B_{2g}}/\eta = 0$, and $J/U=0.2$.}
\label{fig:phase_etaUUk}
\end{figure}

Figure~\ref{fig:phase_etaUUk} shows the phase diagram for the leading superconducting instability as a function of the dimensionless interaction parameters $U/t$ and $V^{\NN}/U$. 
$V^{\NN}$ stabilizes they helical state in the weak-$U$ regime and $d_{xy}$-wave pairing at intermediate $U$. 
As a result, TRSB $s^{\prime}+id_{x^2-y^2}$, $s^{\prime}+id_{xy}$, $d_{x^2-y^2}+id_{xy}$, mixed helical, or mixed parity $s^{\prime}+ip$ pairing, can be obtained at the phase boundaries.
However, $g$-wave pairing is not favored. 
The phase diagram is roughly robust against the change of $\eta$ and $J/U$.
In the following, we discuss two limiting $U$ cases in detail. 

\begin{figure}[!hpt]
\centering
\includegraphics[width=0.51\linewidth,trim={0mm 0mm 9mm 0mm},clip]{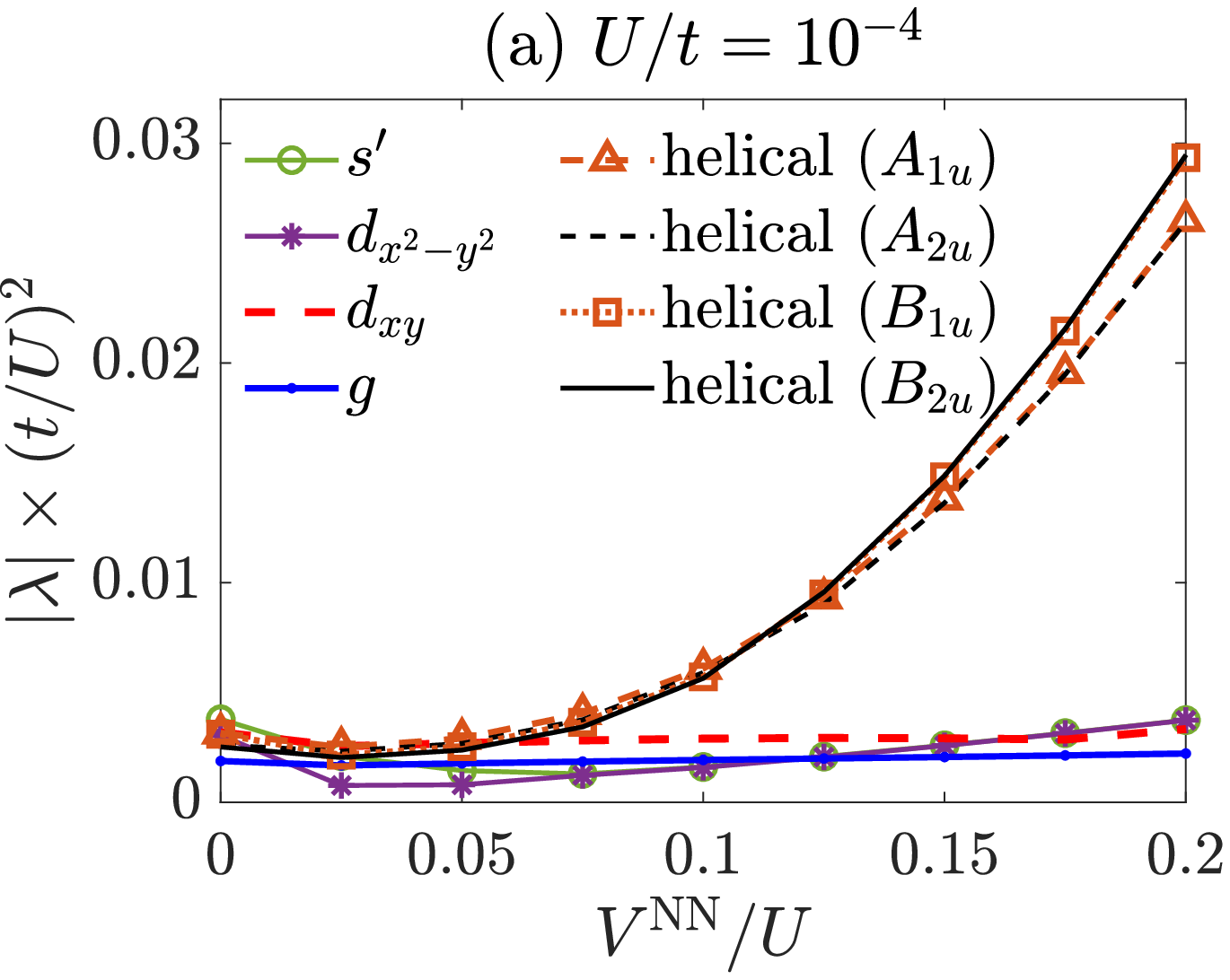}
\includegraphics[width=0.47\linewidth,trim={10mm 0mm 8mm 0mm},clip]{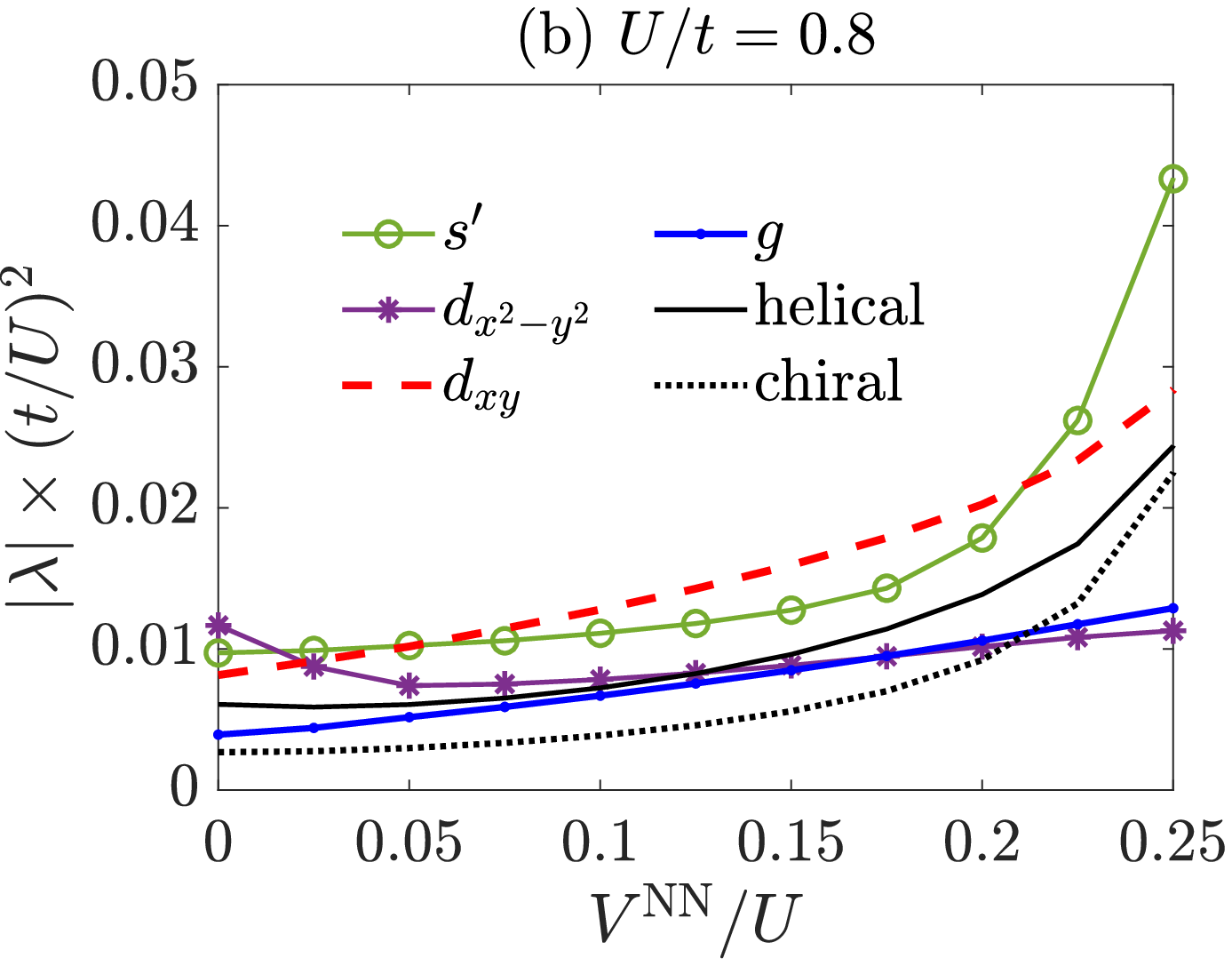}
\caption{Evolution of the largest eigenvalue (in magnitude) of the linearized gap equation, Eq.~\eqref{eq:gap}, in selected leading irrep. as a function of $V^{\NN}/U$ for (a) $U/t=10^{-4}$ and (b) $U/t=0.8$. (Some sub-leading irreps. are not shown.)}
\label{fig:lambda_Uk}
\end{figure}

We first consider the weak-$U$ limit and take $U/t=10^{-4}$, where $s^{\prime}$-wave is leading for $V^{\NN}=0$. 
The evolutions of the superconducting instabilities as a function of $V^{\NN}$ are shown in Fig.~\ref{fig:lambda_Uk}.
The leading eigenvalue in all the pairing channels shown is slightly suppressed at small $V^{\NN}$; helical pairings are promoted when $V^{\NN}/U \gtrsim 0.05$. 
In the latter case, the corresponding helical gap functions we obtained are similar to those obtained in the absence of $V^{\NN}$ in the previous studies\cite{Scaffidi2014, Tsuchiizu2015, Zhang2018}. 

A noticeable feature in Fig.~\ref{fig:lambda_Uk} (a) is that, independent of $V^{\NN}/U$, the splitting between helical pairings in the $B_{1u}$ and $B_{2u}$ (or $A_{1u}$ and $A_{2u}$) irrep. is rather small, making it reasonable to consider accidentally degenerate helical pairings $B_{1u}+iB_{2u}$ (or $A_{1u}+iA_{2u}$). 
These pairings are proposed in Ref.~\onlinecite{Huang2021} to explain some observations in SRO, including the intrinsic Hall and Kerr effects, the absence of observable spontaneous edge current, and the substantial Knight shift drop using simple gap functions without any microscopic details. 
We revisit the $B_{1u}+iB_{2u}$ state obtained at $V^{\NN}/U = 0.15$ in Sec.~\ref{sec:properties} to see if it can reconcile with the experiments. 

Figure~\ref{fig:lambda_Uk} (b) shows an intermediate $U$ case, $U/t=0.8$, where $d_{x^2-y^2}$ is slightly leading without $V^{\NN}$. 
The leading superconducting instabilities in most of the pairing channels, including $g$-wave, are enhanced, due to the enhancement of $\chi^{\RPA}$. 
Either the $s^{\prime}$- or $d_{xy}$-solution dominates over other channels depending on the value of $V^{\NN}$. 
Similar results were recently reported in Ref.~\onlinecite{Romer2021}, where $s^{\prime}+id_{xy}$ pairing is proposed. 
We discuss the properties of the $s^{\prime}+id_{xy}$ pairing at $V^{\NN}/U = 0.05$ in  Sec.~\ref{sec:properties}. 

\section{Stability of $d_{x^2-y^2}$ and $g$-wave pairing in the presence of longer-range interactions and $\eta_{B_{2g}}$}
\label{sec:app_robustness}

\begin{figure}[htp]
\centering
\includegraphics[width=0.45\linewidth,trim={10mm 0mm 25mm 0mm},clip]{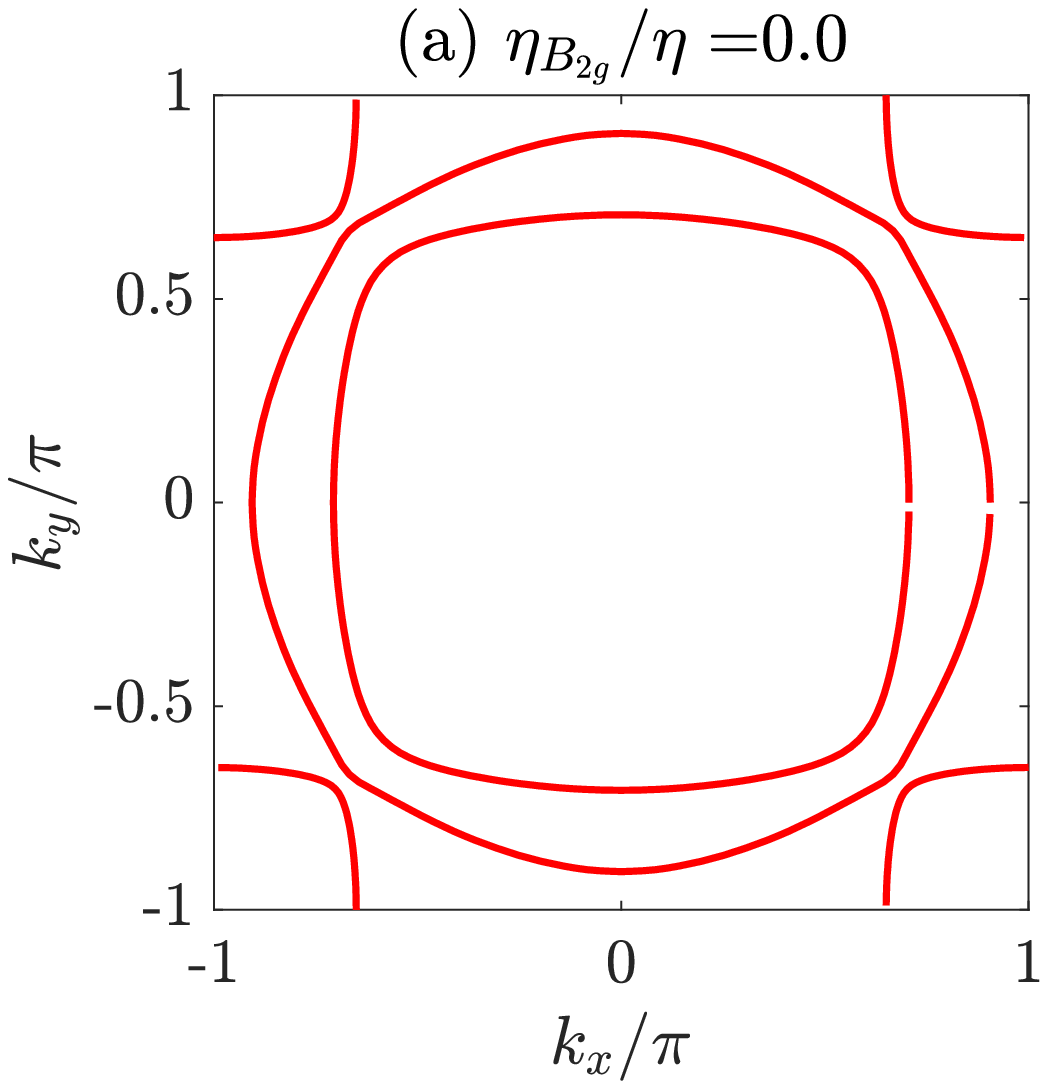}
\includegraphics[width=0.45\linewidth,trim={10mm 0mm 25mm 0mm},clip]{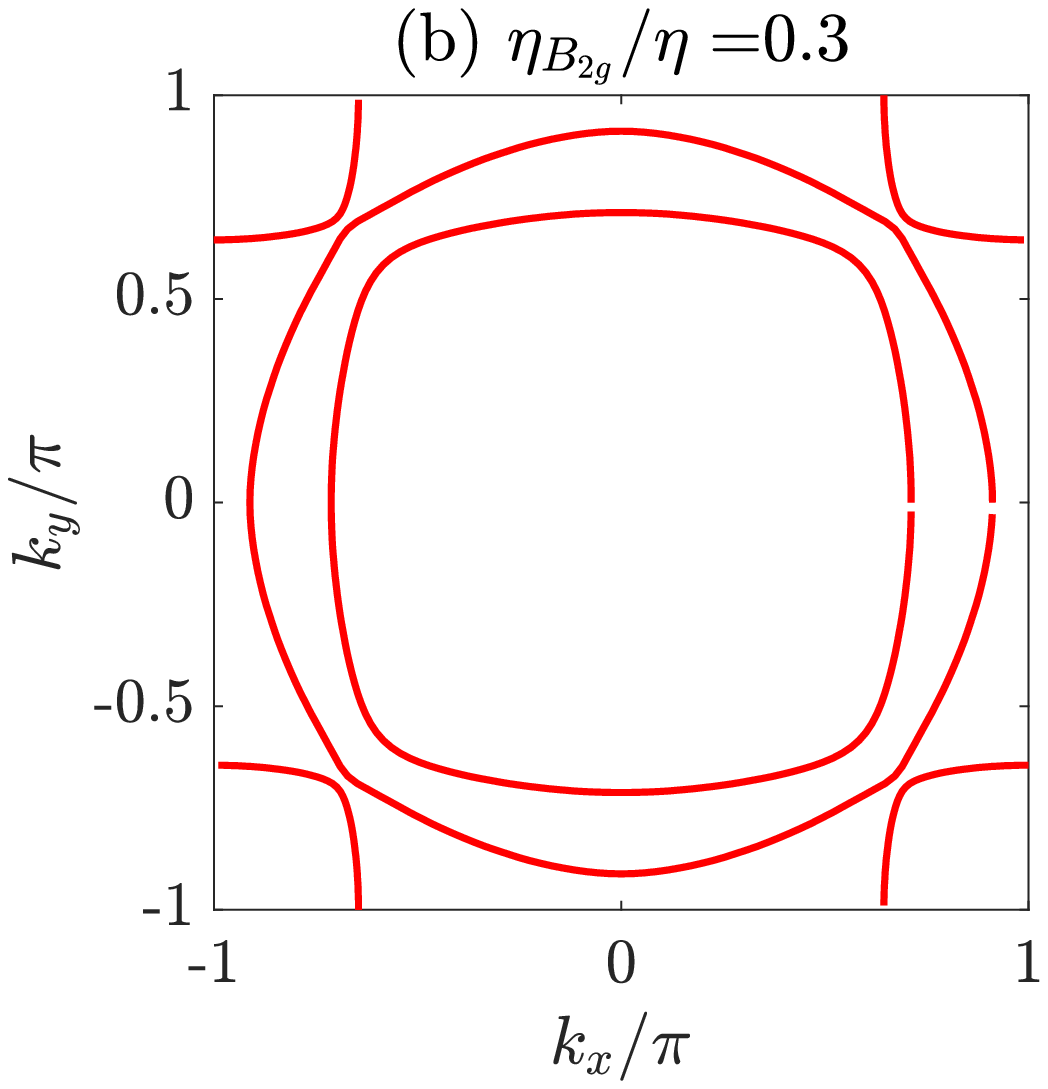}
\caption{Fermi surfaces for the tight-binding model given in Eq.~\eqref{eq:HK} with SOC parameters: $(\eta,~ \eta_{B_{2g}})$ = (a) $(0.2,~0)t$ and (b) $(0.154,~ 0.046) t$. $\tilde{\mu}_c $ is adjusted to $1.14t$ for (b).}
\label{fig:FS}
\end{figure}

In Sec.~\ref{sec:instabilities}, we show that both $d_{x^2-y^2}$- and $g$-wave pairing can be stabilized at $\eta_{B_{2g}}/\eta \gtrsim 0.3$ in the intermediate $U$ case ($U=0.8, V^{\NN}/U=0.25, V^{\NNN}/V^{\NN}=0.65$). (See Fig.~\ref{fig:lambda_mucetak}~(a).) 
In this appendix, we show that this is a robust result relevant for a large region of parameter space. The resulting FSs for two distinct SOC parameter values are shown in Fig.~\ref{fig:FS}.
The orbital anisotropies, which can further promote the $d_{x^2-y^2}$- and $g$-wave phase, are neglected here. 

\begin{figure}[htp]
\centering
\includegraphics[width=0.5\linewidth,trim={0mm 0mm 10mm 0mm},clip]{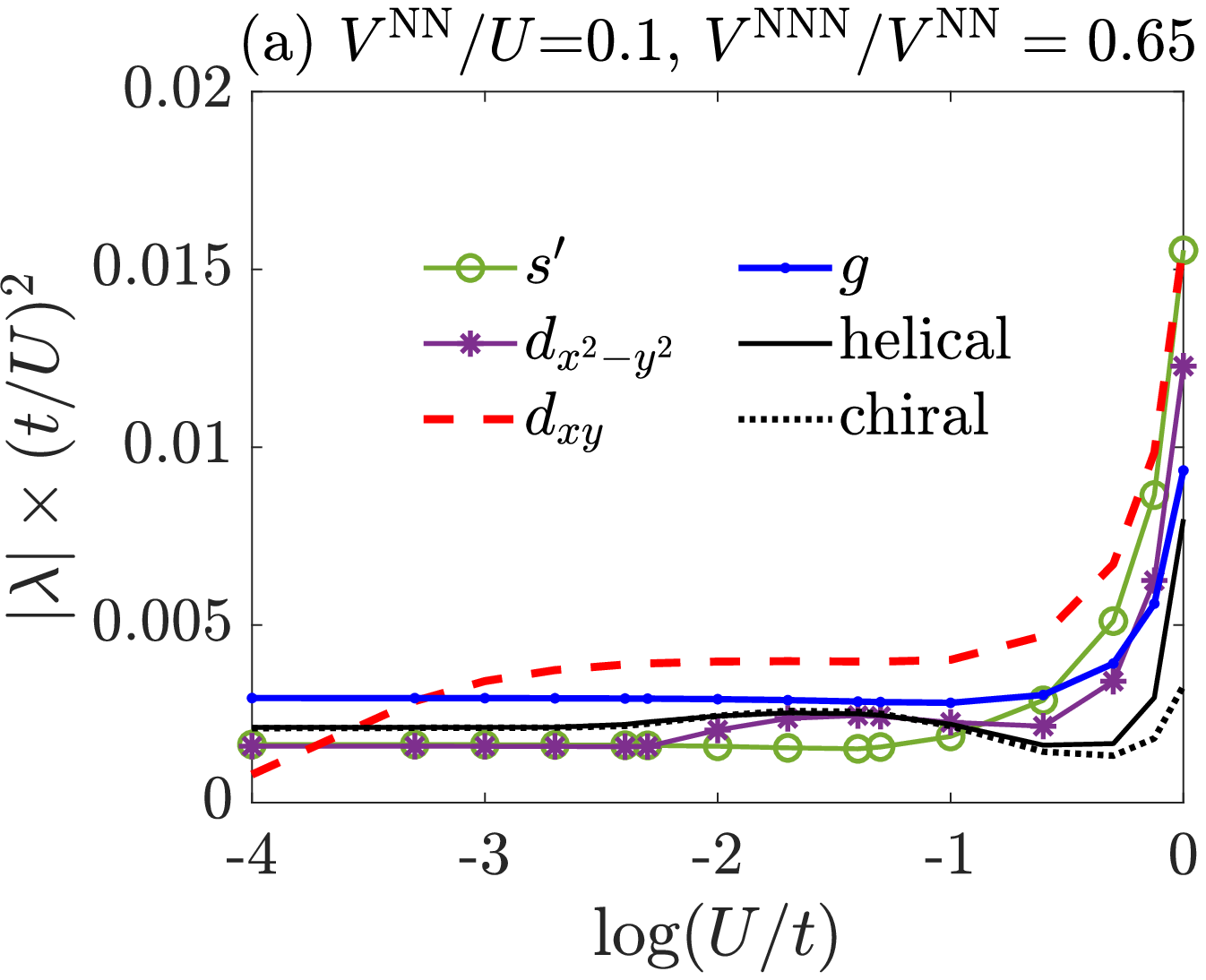}
\includegraphics[width=0.46\linewidth,trim={10mm 0mm 10mm 0mm},clip]{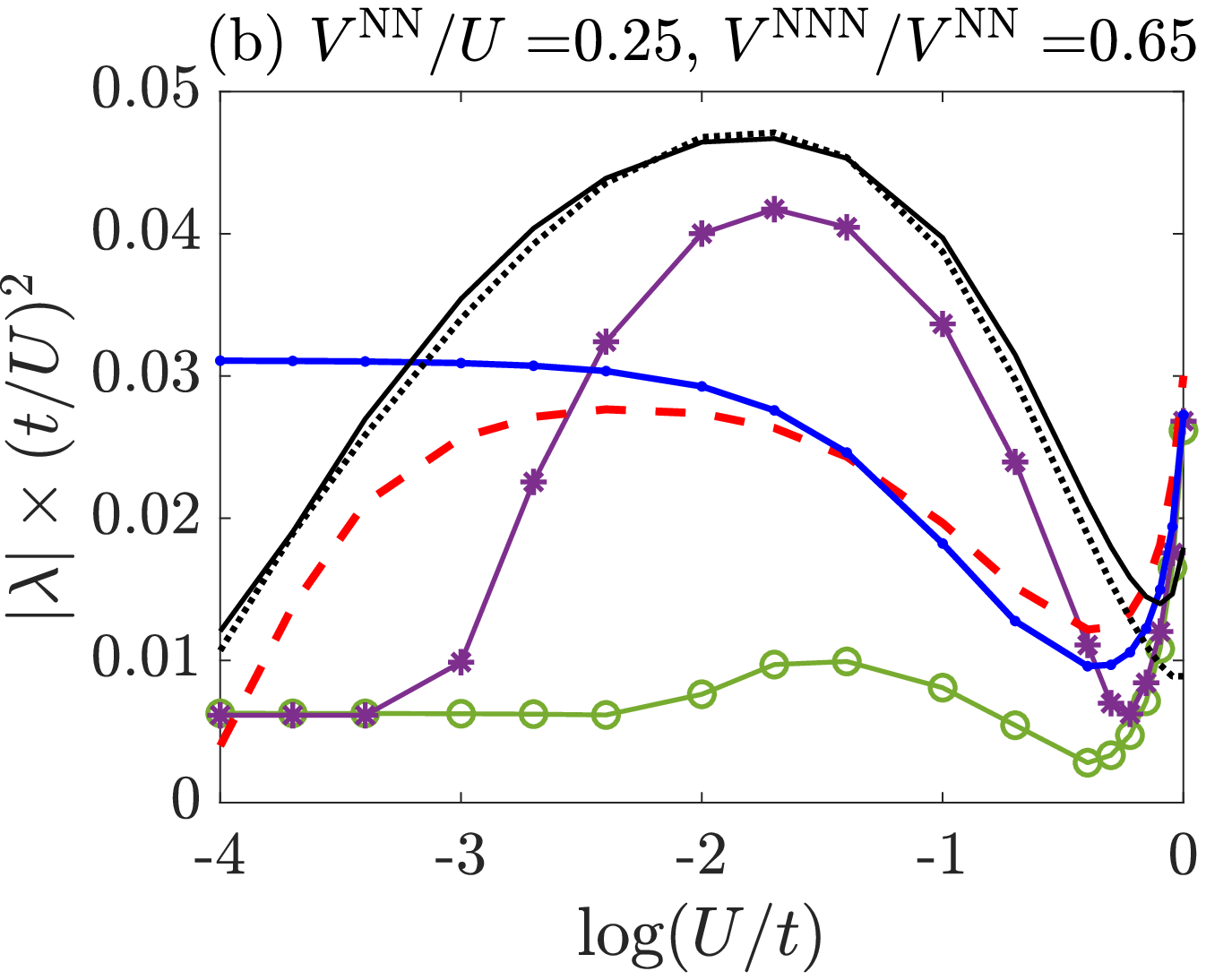}
\caption{Leading superconducting instabilities as a function of $\log(U/t)$ for (a), $V^{\NN}/U = 0.1$ and (b), $V^{\NN}/U = 0.25$. $V^{\NNN}/V^{\NN} = 0.65$.}
\label{fig:lambda_U}
\end{figure}

Before we focus on a detailed case, we first give a general picture of the evolutions of the $g$-wave superconducting instability for a wide range of $U$ in the presence of sizable $V^{\mathrm{NNN}}$ (in Fig.~\ref{fig:lambda_U}). 
We find that $g$-wave pairing becomes dominant in the weak-U limit when longer-range interactions are included and is less favored for finite $U$, as observed in Fig.~\ref{fig:lambda_Uk1Ukk}. 
However, the dependence of the leading pairing and of the $g$-wave state on $U$ is non-monotonic. 
Another important piece of information we can get from Fig.~\ref{fig:lambda_U} is that, for a given $U$, the $g$-wave can always become the leading or the first sub-leading pairing in the parameter space of $V^{\NN}$ and $V^{\NNN}$. For the latter case, it can be further promoted by the effects of $\eta_{B_{2g}}$ and interaction-anisotropies, as discussed in the main text.

\begin{figure}[htp]
\centering
\includegraphics[width=0.52\linewidth,trim={0mm 0mm 10mm 0mm},clip]{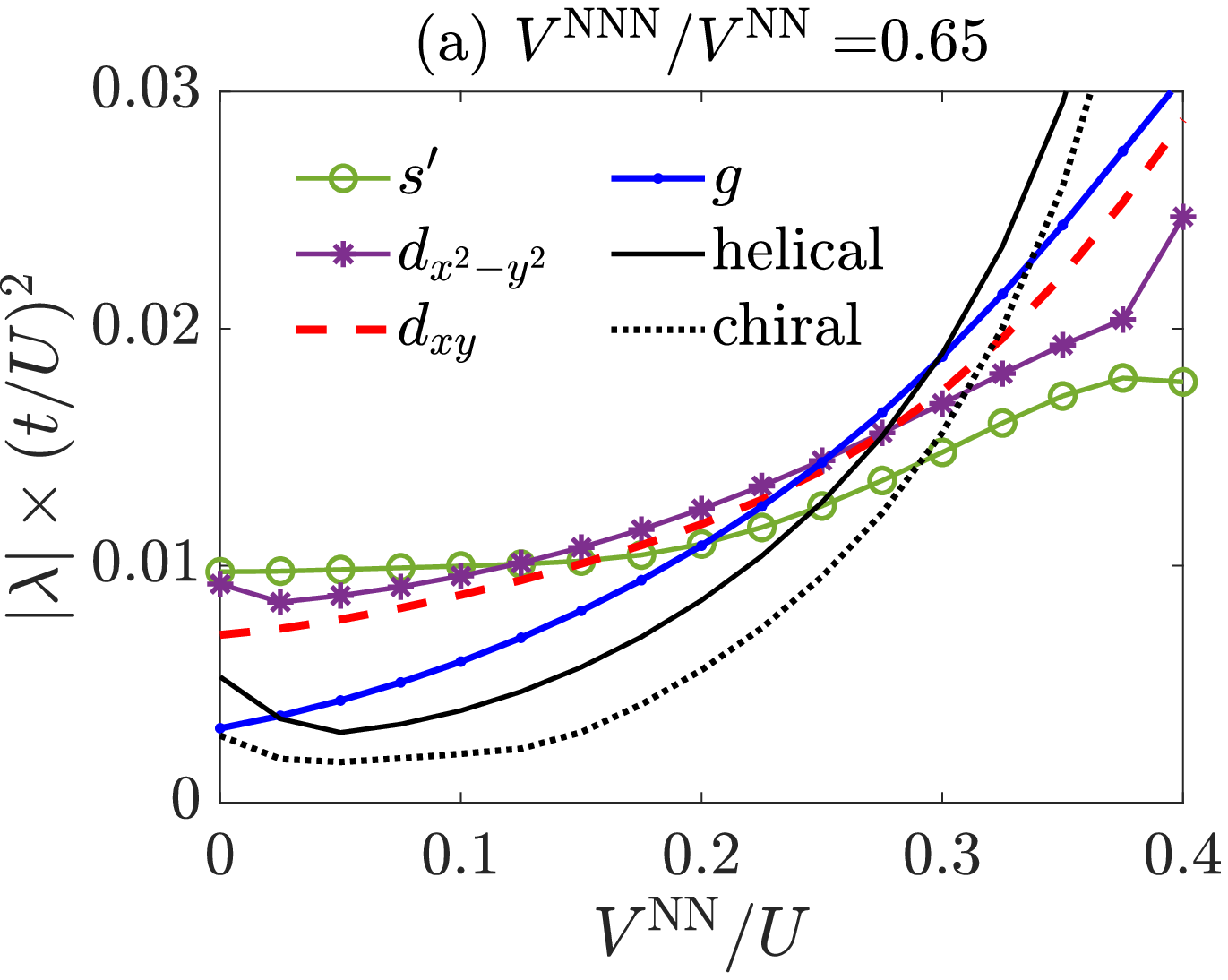}
\includegraphics[width=0.468\linewidth,trim={10.5mm 0mm 13mm 0mm},clip]{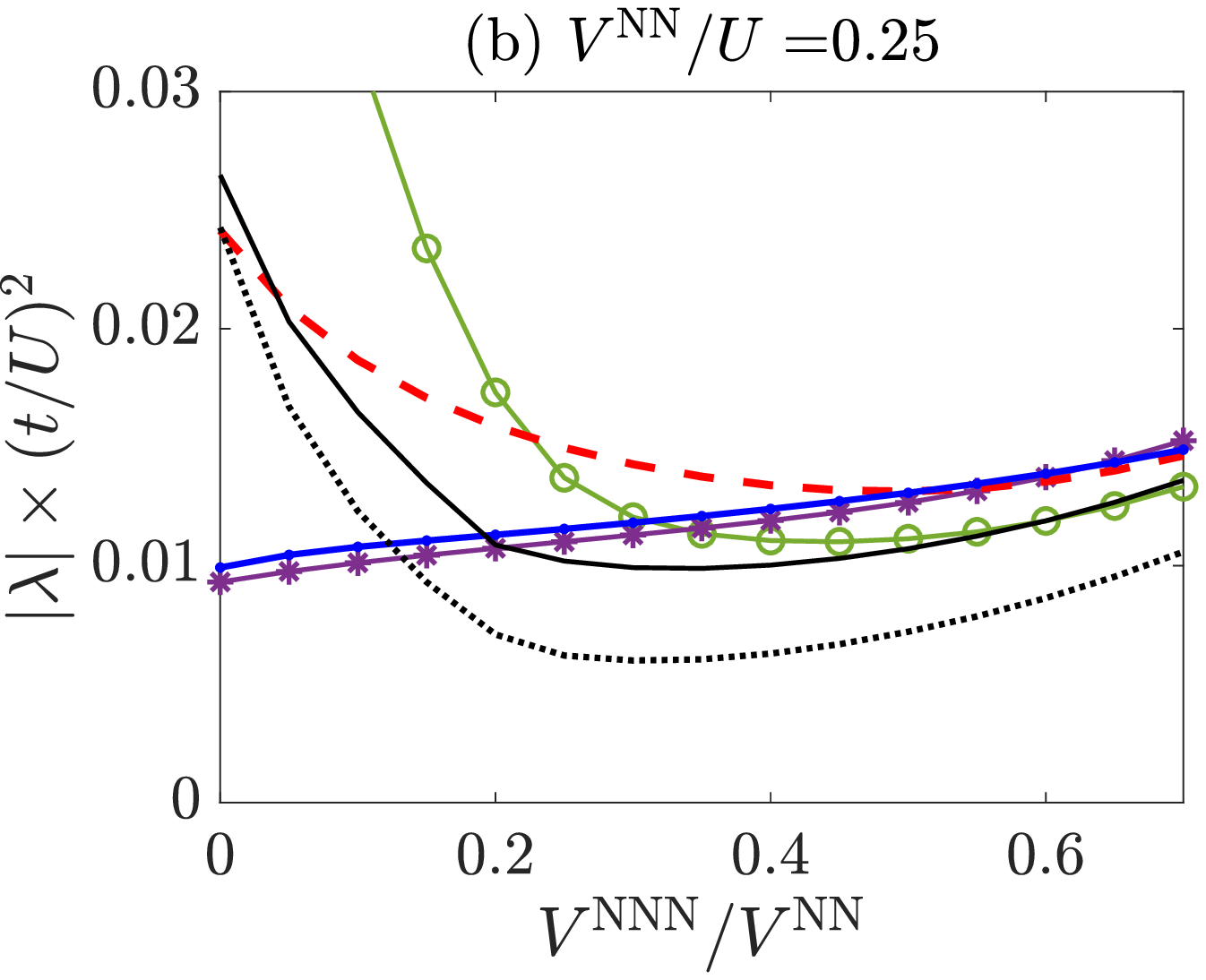}
\caption{Leading superconducting instabilities as (a) $V^{\NN}/U$ and (b) $V^{\NNN}/V^{\NN}$ is varied. $\eta_{B_{2g}}/\eta = 0.3$. The case at $V^{\NN}/U=0.25, V^{\NNN}/V^{\NN}=0.65$ is shown in Fig.~\ref{fig:lambda_mucetak}(a).}
\label{fig:lambda_etak0.3UkUkk}
\centering
\includegraphics[width=0.65\linewidth,trim={0mm 0mm 0mm 0mm},clip]{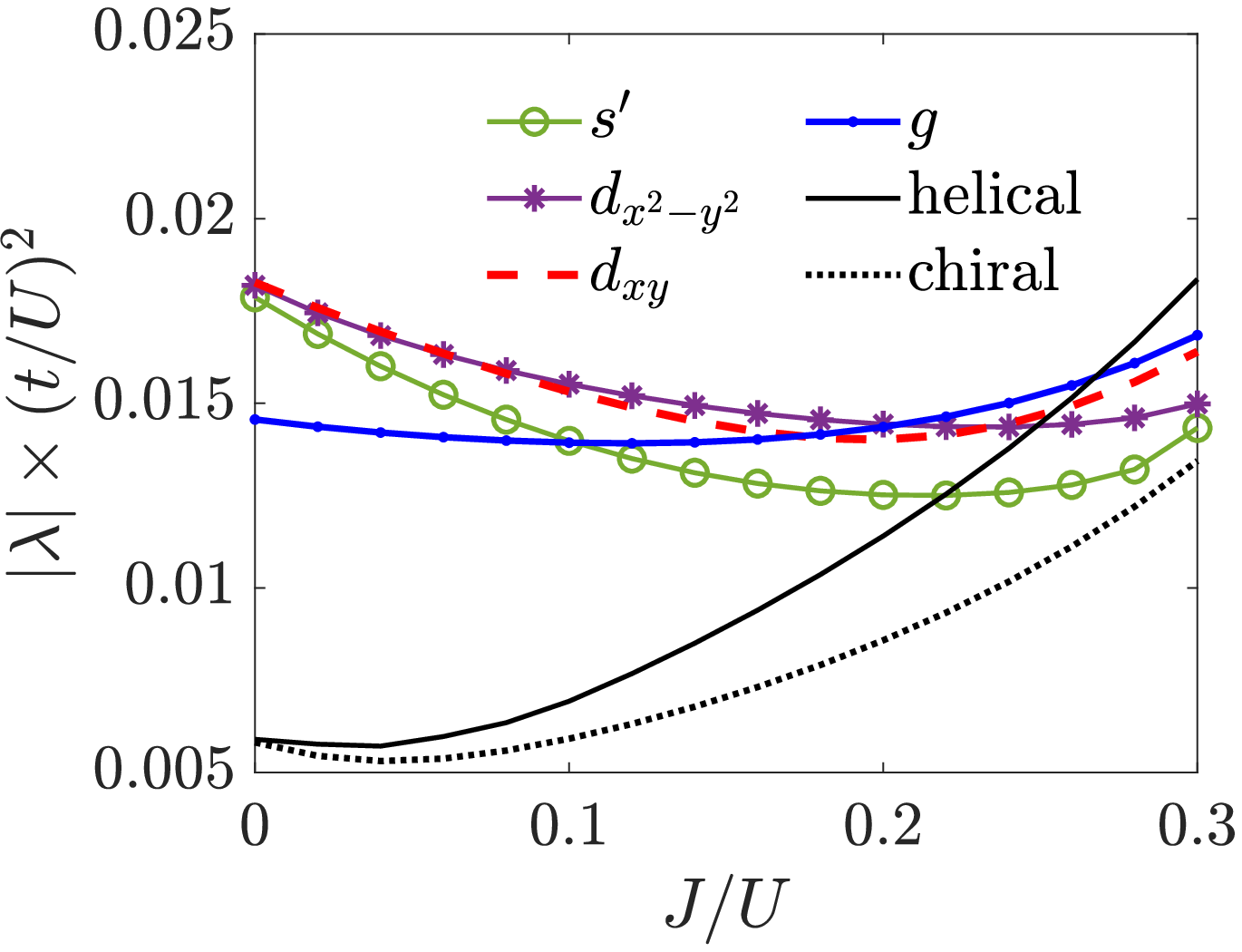}
\caption{Superconducting instabilities v.s. $J/U$ in the $\eta_{B_{2g}}/\eta=0.3$ case ($U/t=0.8$, $V^{NN}/U=0.25$, $V^{\NNN}/V^{\NN} = 0.65$).}
\label{fig:lambda_Uk1Ukk0.65J}
\end{figure}

We then consider the case at $\eta_{B_{2g}}/\eta = 0.3$ where the $d_{x^2-y^2}$ state is slightly dominant and the $g$-wave is the first subleading channel. 
As shown in Fig.~\ref{fig:lambda_etak0.3UkUkk} and \ref{fig:lambda_Uk1Ukk0.65J}, $d_{x^2-y^2}$- and $g$-wave are the first two leading pairings in the range of $V^{\NN}/U \in (0.2, 0.3)$, $V^{\NNN}/V^{\NN}>0.5$ and $J/U\in (0.16, 0.24)$. 
The $d_{x^2-y^2}$ and $g$-wave phase can be larger by increasing $\eta_{B_{2g}}$ or including longer-range interaction anisotropies.

\begin{figure}[htp]
\centering
\includegraphics[width=0.65\linewidth,trim={0mm 0mm 0mm 0mm},clip]{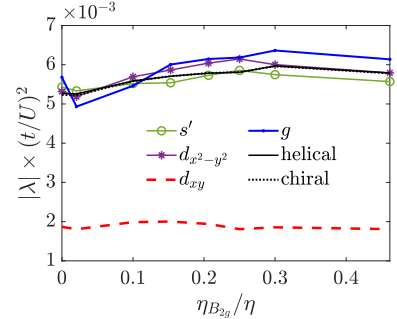}
\caption{Effects of $\eta_{B_{2g}}/\eta$ on the superconducting instabilities in different channels for $U/t=0.0001$. $V^{NN}/U=0.25$, $V^{\NNN}/V^{\NN} = 0.2$.}
\label{fig:lambda_Uk0.0001mucetak}
\end{figure}

Figure~\ref{fig:lambda_Uk0.0001mucetak} shows the effects of $\eta_{B_{2g}}/\eta$ in the weak-$U$ case ($U/t=10^{-4}, V^{\NN}/U=0.25, V^{\NNN}/V^{\NN}=0.2$), where the $g$-wave is slightly dominant at $\eta_{B_{2g}}/\eta=0$.
We find that $d_{x^2-y^2}$- and $g$-wave become the first two leading pairings at $\eta_{B_{2g}}/\eta \gtrsim 0.15$. 

\section{Effects of orbital-anisotropies of the longer-range interactions}  \label{sec:app_LRUanisotropy}
Orbital anisotropies of the longer-range interactions can significantly impact the leading superconducting instabilities. 
In particular, they help to stabilize the $g$- and $d_{x^2-y^2}$-wave state [in Fig.\ref{fig:lambda_mucetak} (b)]. 
In this appendix, we show that anisotropy parameters $\alpha_{33}$ and $\beta_{33}$ [defined in Eq.\eqref{eq:anisotropy_Vnn} and \eqref{eq:anisotropy_Vnnn}] are crucial to stabilizing the $g$- and $d_{x^2-y^2}$-wave state, respectively. The effects of $\eta_{B_{2g}}$ are not included.

\begin{table}[htp]
\begin{tabular}{c|c||c|c|c|c|c|c}
\hline
\hline
\multirow{2}{*}{Parameter} & \multirow{2}{*}{Estimate} & \multicolumn{6}{c}{Effects}\\
\cline{3-8} 
& &  $~~s^{\prime}~~$ & $d_{x^2-y^2}$ & $~d_{xy}~$ & $~~g~~$ & helical & chiral\\
\hline
$\alpha_{33}$ & 0.12 & $\uparrow\uparrow$ & $\uparrow\uparrow$ & $\uparrow$ & $\uparrow\uparrow$ & $\uparrow\uparrow$ & $\uparrow\uparrow$\\
\hline
$\alpha_{23,x}$ & 0.11 & $\uparrow$ & $\uparrow$ & $\uparrow$ & $\uparrow$ & $\uparrow$ & $\uparrow$ \\
\hline
$\alpha_{12}$ & 0.05 & $-$ & $-$ &  $\downarrow$ & $-$ & $\uparrow$ & $\uparrow$\\
\hline
\hline
$\beta_{33}$ &  0.04 & $\uparrow\uparrow$ & $\uparrow\uparrow$ & $-$ & $-$ & $\uparrow$ & $\uparrow$\\
\hline
$\beta_{13}$ & 0.02 & $-$ & $-$ & $\uparrow$ &  $\uparrow$ & $\uparrow\uparrow$ & $\uparrow\uparrow$\\
\hline
$\beta_{12}$ & 0.002 & $-$ & $-$ & $\uparrow$ &  $\uparrow$ &  $\uparrow$ &  $\uparrow$\\
\hline
\hline
\end{tabular}
\caption{Effects of orbital-anisotropy parameters on different pairing channels (last six columns) for finite-$U$. $\uparrow (~\downarrow$ or $-)$ means that the eigenvalue (magnitude) of that pairing channel is enhanced (suppressed or barely changed). $\uparrow\uparrow$ indicates that the eigenvalue of that channel is more enhanced than those with $\uparrow$. The second column gives the estimates of the parameters through integrals over Ru 4d-$t_{2g}$ Slater-type orbitals.}
\label{table:anisotropy}
\end{table}

Rough estimates of the orbital-anisotropy parameters through integrals over Ru 4d-$t_{2g}$ Slater-type orbitals are shown in Table. \ref{table:anisotropy}. 
The largest anisotropy parameter, $\alpha_{33}$, which quantifies the NN interactions for $d_{xy}$-orbitals (or $d_{xz}$-orbitals along the $x$-direction) relative to that for $d_{xz}$-orbitals along the $y$-direction, is about $0.12$.
As discussed in Sec.\ref{sec:instabilities}, this value is underestimated because the hybridization effects are neglected. 
However, Table.~\ref{table:anisotropy} can still indicate how those parameters compare with each other. 
For example, $(\alpha_{23,x}, \alpha_{12} , \beta_{33} , \beta_{13}, \beta_{12}) \approx (1, 0.4, 0.33, 0.17, 0)\alpha_{33}$ and this is the ratio used in the main text.

\begin{figure}[!htp]
\centering
\includegraphics[width=0.51\linewidth,trim={0mm 0mm 10mm 0mm},clip]{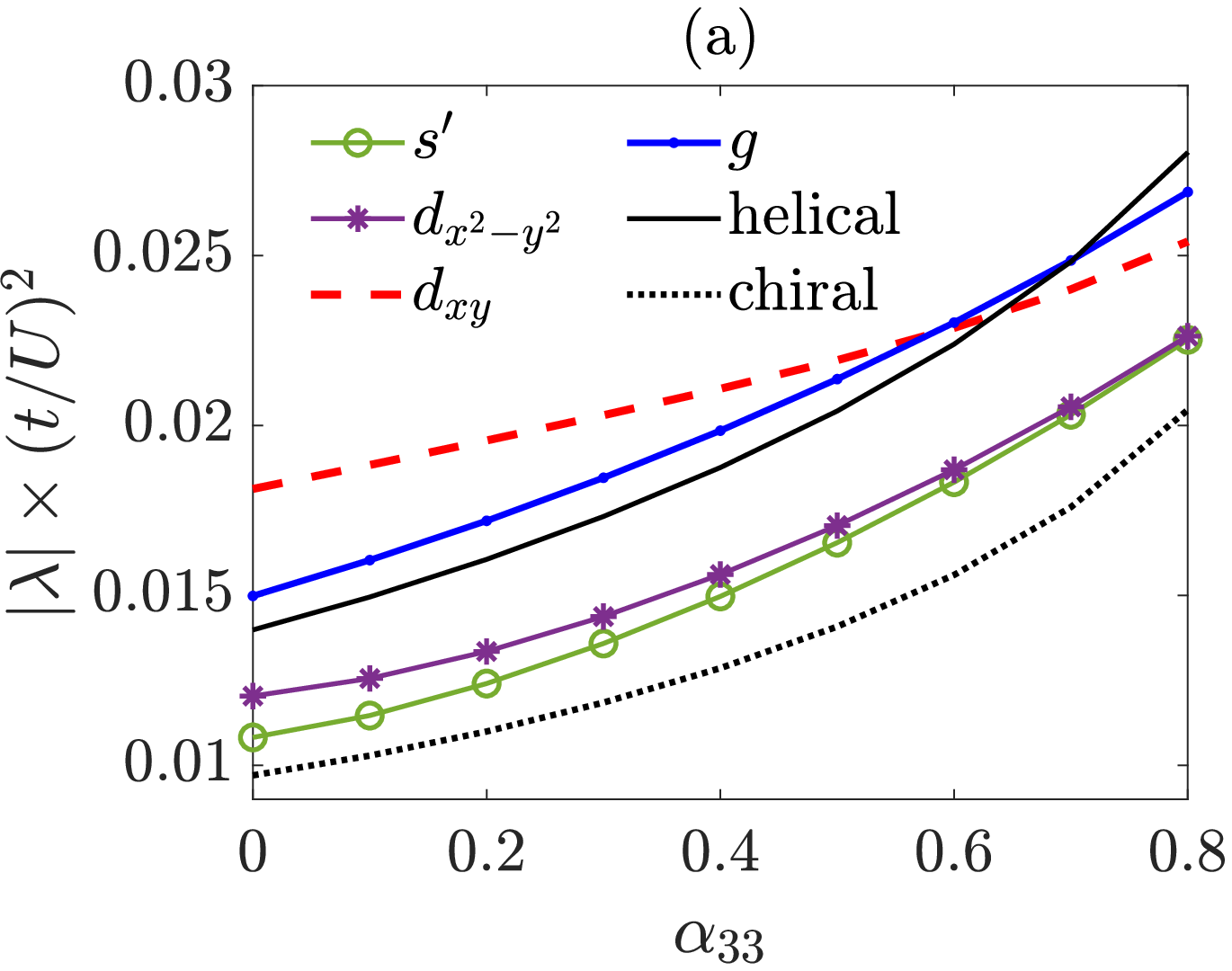}
\includegraphics[width=0.475\linewidth,trim={10mm 0mm 10mm 0mm},clip]{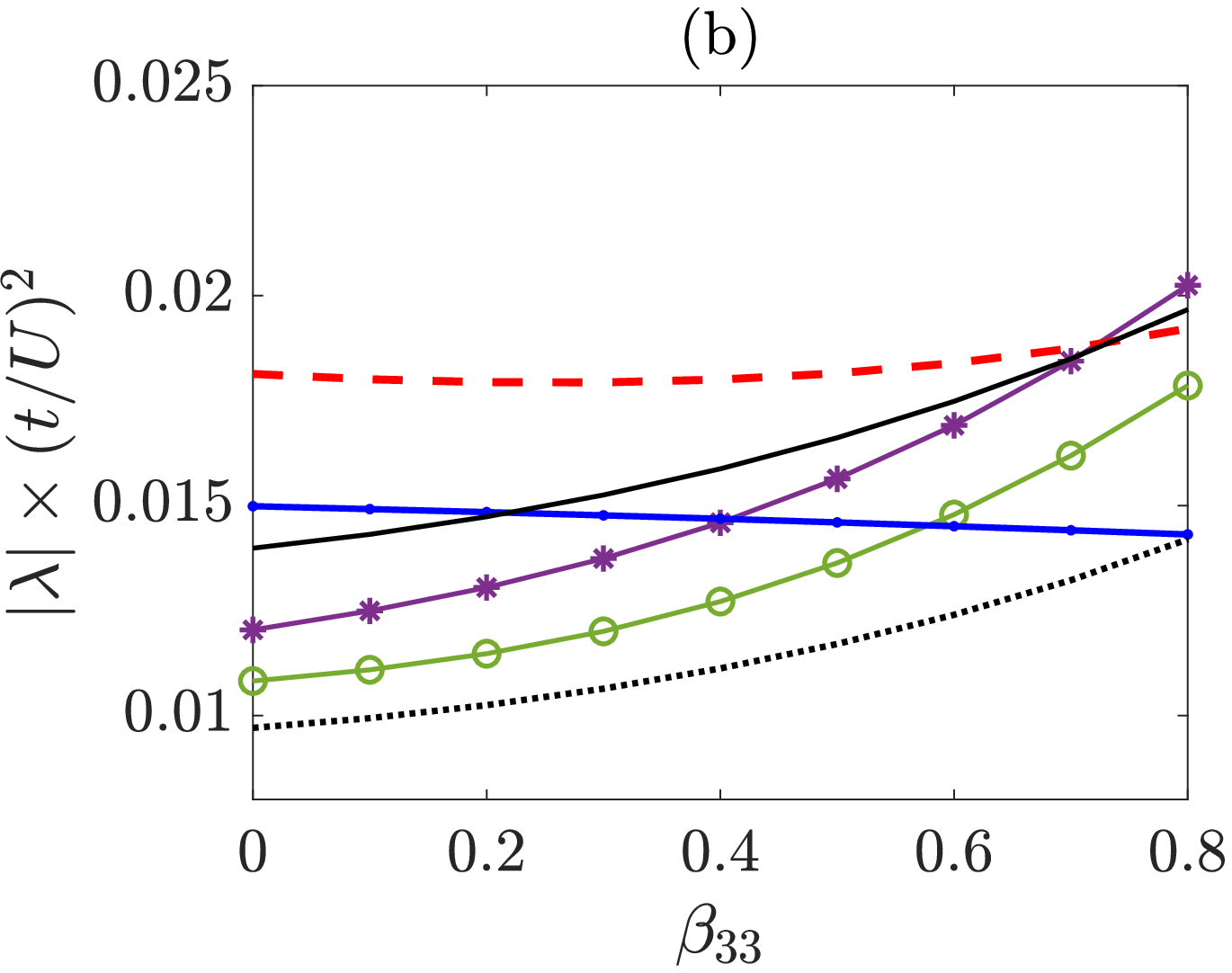}
\caption{Evolutions of the superconducting instability as a function of (a) $\alpha_{33}$ and (b) $\beta_{33}$. $U/t = 0.8, V^{\NN}/U = 0.25, V^{\NNN}/V^{\NN}=0.65$. Other anisotropy parameters are set to zero. $\alpha_{33} (\beta_{33})=0$ describes the orbital-isotropic case in Fig.~\ref{fig:lambda_Uk1Ukk}. }
\label{fig:lambda_a33b33}
\end{figure}

Table.~\ref{table:anisotropy} shows the effects of the anisotropy parameters on different pairing channels for finite-$U$. See caption for explanations of notation (arrows and dash). 
For example, $\alpha_{33}$ tends to significantly enhance all the pairing channels except  $d_{xy}$. 
As a consequence, $d_{xy}$ is surpassed by $g$-wave or helical at large $\alpha_{33}$ [shown in Fig.~\ref{fig:lambda_a33b33}(a)]. 
In addition, $\beta_{33}$ is critical to stabilizing $d_{x^2-y^2}$-pairing. 
As shown in Fig.~\ref{fig:lambda_a33b33}(b), $d_{x^2-y^2}$-wave pairing is favored at $\beta_{33}>0.7$. 
In the weak-$U$ limit, the anisotropies tend to favor the $g$-wave state (not shown), because the enhancement effects on other pairing channels, as discussed above, are generally canceled out by the bare interactions. 

\end{appendix}


\begin{thebibliography}{66}%
\makeatletter
\providecommand \@ifxundefined [1]{%
 \@ifx{#1\undefined}
}%
\providecommand \@ifnum [1]{%
 \ifnum #1\expandafter \@firstoftwo
 \else \expandafter \@secondoftwo
 \fi
}%
\providecommand \@ifx [1]{%
 \ifx #1\expandafter \@firstoftwo
 \else \expandafter \@secondoftwo
 \fi
}%
\providecommand \natexlab [1]{#1}%
\providecommand \enquote  [1]{``#1''}%
\providecommand \bibnamefont  [1]{#1}%
\providecommand \bibfnamefont [1]{#1}%
\providecommand \citenamefont [1]{#1}%
\providecommand \href@noop [0]{\@secondoftwo}%
\providecommand \href [0]{\begingroup \@sanitize@url \@href}%
\providecommand \@href[1]{\@@startlink{#1}\@@href}%
\providecommand \@@href[1]{\endgroup#1\@@endlink}%
\providecommand \@sanitize@url [0]{\catcode `\\12\catcode `\$12\catcode
  `\&12\catcode `\#12\catcode `\^12\catcode `\_12\catcode `\%12\relax}%
\providecommand \@@startlink[1]{}%
\providecommand \@@endlink[0]{}%
\providecommand \url  [0]{\begingroup\@sanitize@url \@url }%
\providecommand \@url [1]{\endgroup\@href {#1}{\urlprefix }}%
\providecommand \urlprefix  [0]{URL }%
\providecommand \Eprint [0]{\href }%
\providecommand \doibase [0]{http://dx.doi.org/}%
\providecommand \selectlanguage [0]{\@gobble}%
\providecommand \bibinfo  [0]{\@secondoftwo}%
\providecommand \bibfield  [0]{\@secondoftwo}%
\providecommand \translation [1]{[#1]}%
\providecommand \BibitemOpen [0]{}%
\providecommand \bibitemStop [0]{}%
\providecommand \bibitemNoStop [0]{.\EOS\space}%
\providecommand \EOS [0]{\spacefactor3000\relax}%
\providecommand \BibitemShut  [1]{\csname bibitem#1\endcsname}%
\let\auto@bib@innerbib\@empty
\bibitem [{\citenamefont {Luke}\ \emph {et~al.}(1998)\citenamefont {Luke},
  \citenamefont {Fudamoto}, \citenamefont {Kojima}, \citenamefont {Larkin},
  \citenamefont {Merrin}, \citenamefont {Nachumi}, \citenamefont {Uemura},
  \citenamefont {Maeno}, \citenamefont {Mao}, \citenamefont {Mori},
  \citenamefont {Nakamura},\ and\ \citenamefont {Sigrist}}]{Luke1998}%
  \BibitemOpen
  \bibfield  {author} {\bibinfo {author} {\bibfnamefont {G.~M.}\ \bibnamefont
  {Luke}}, \bibinfo {author} {\bibfnamefont {Y.}~\bibnamefont {Fudamoto}},
  \bibinfo {author} {\bibfnamefont {K.~M.}\ \bibnamefont {Kojima}}, \bibinfo
  {author} {\bibfnamefont {M.~I.}\ \bibnamefont {Larkin}}, \bibinfo {author}
  {\bibfnamefont {J.}~\bibnamefont {Merrin}}, \bibinfo {author} {\bibfnamefont
  {B.}~\bibnamefont {Nachumi}}, \bibinfo {author} {\bibfnamefont {Y.~J.}\
  \bibnamefont {Uemura}}, \bibinfo {author} {\bibfnamefont {Y.}~\bibnamefont
  {Maeno}}, \bibinfo {author} {\bibfnamefont {Z.~Q.}\ \bibnamefont {Mao}},
  \bibinfo {author} {\bibfnamefont {Y.}~\bibnamefont {Mori}}, \bibinfo {author}
  {\bibfnamefont {H.}~\bibnamefont {Nakamura}}, \ and\ \bibinfo {author}
  {\bibfnamefont {M.}~\bibnamefont {Sigrist}},\ }\href {\doibase 10.1038/29038}
  {\bibfield  {journal} {\bibinfo  {journal} {Nature}\ }\textbf {\bibinfo
  {volume} {394}},\ \bibinfo {pages} {558} (\bibinfo {year}
  {1998})}\BibitemShut {NoStop}%
\bibitem [{\citenamefont {Grinenko}\ \emph {et~al.}(2021)\citenamefont
  {Grinenko}, \citenamefont {Ghosh}, \citenamefont {Sarkar}, \citenamefont
  {Orain}, \citenamefont {Nikitin}, \citenamefont {Elender}, \citenamefont
  {Das}, \citenamefont {Guguchia}, \citenamefont {Br{\"u}ckner}, \citenamefont
  {Barber} \emph {et~al.}}]{Grinenko2021}%
  \BibitemOpen
  \bibfield  {author} {\bibinfo {author} {\bibfnamefont {V.}~\bibnamefont
  {Grinenko}}, \bibinfo {author} {\bibfnamefont {S.}~\bibnamefont {Ghosh}},
  \bibinfo {author} {\bibfnamefont {R.}~\bibnamefont {Sarkar}}, \bibinfo
  {author} {\bibfnamefont {J.-C.}\ \bibnamefont {Orain}}, \bibinfo {author}
  {\bibfnamefont {A.}~\bibnamefont {Nikitin}}, \bibinfo {author} {\bibfnamefont
  {M.}~\bibnamefont {Elender}}, \bibinfo {author} {\bibfnamefont
  {D.}~\bibnamefont {Das}}, \bibinfo {author} {\bibfnamefont {Z.}~\bibnamefont
  {Guguchia}}, \bibinfo {author} {\bibfnamefont {F.}~\bibnamefont
  {Br{\"u}ckner}}, \bibinfo {author} {\bibfnamefont {M.~E.}\ \bibnamefont
  {Barber}},  \emph {et~al.},\ }\href
  {https://doi.org/10.1038/s41567-021-01182-7} {\bibfield  {journal} {\bibinfo
  {journal} {Nature Physics}\ }\textbf {\bibinfo {volume} {17}},\ \bibinfo
  {pages} {748} (\bibinfo {year} {2021})}\BibitemShut {NoStop}%
\bibitem [{\citenamefont {Xia}\ \emph {et~al.}(2006)\citenamefont {Xia},
  \citenamefont {Maeno}, \citenamefont {Beyersdorf}, \citenamefont {Fejer},\
  and\ \citenamefont {Kapitulnik}}]{Xia2006}%
  \BibitemOpen
  \bibfield  {author} {\bibinfo {author} {\bibfnamefont {J.}~\bibnamefont
  {Xia}}, \bibinfo {author} {\bibfnamefont {Y.}~\bibnamefont {Maeno}}, \bibinfo
  {author} {\bibfnamefont {P.~T.}\ \bibnamefont {Beyersdorf}}, \bibinfo
  {author} {\bibfnamefont {M.~M.}\ \bibnamefont {Fejer}}, \ and\ \bibinfo
  {author} {\bibfnamefont {A.}~\bibnamefont {Kapitulnik}},\ }\href {\doibase
  10.1103/PhysRevLett.97.167002} {\bibfield  {journal} {\bibinfo  {journal}
  {Phys. Rev. Lett.}\ }\textbf {\bibinfo {volume} {97}},\ \bibinfo {pages}
  {167002} (\bibinfo {year} {2006})}\BibitemShut {NoStop}%
\bibitem [{\citenamefont {Kidwingira}\ \emph {et~al.}(2006)\citenamefont
  {Kidwingira}, \citenamefont {Strand}, \citenamefont {Van~Harlingen},\ and\
  \citenamefont {Maeno}}]{Kidwingira2006}%
  \BibitemOpen
  \bibfield  {author} {\bibinfo {author} {\bibfnamefont {F.}~\bibnamefont
  {Kidwingira}}, \bibinfo {author} {\bibfnamefont {J.~D.}\ \bibnamefont
  {Strand}}, \bibinfo {author} {\bibfnamefont {D.~J.}\ \bibnamefont
  {Van~Harlingen}}, \ and\ \bibinfo {author} {\bibfnamefont {Y.}~\bibnamefont
  {Maeno}},\ }\href {\doibase 10.1126/science.1133239} {\bibfield  {journal}
  {\bibinfo  {journal} {Science}\ }\textbf {\bibinfo {volume} {314}},\ \bibinfo
  {pages} {1267} (\bibinfo {year} {2006})}\BibitemShut {NoStop}%
\bibitem [{\citenamefont {Ghosh}\ \emph {et~al.}(2021)\citenamefont {Ghosh},
  \citenamefont {Shekhter}, \citenamefont {Jerzembeck}, \citenamefont
  {Kikugawa}, \citenamefont {Sokolov}, \citenamefont {Brando}, \citenamefont
  {Mackenzie}, \citenamefont {Hicks},\ and\ \citenamefont
  {Ramshaw}}]{Ghosh2021}%
  \BibitemOpen
  \bibfield  {author} {\bibinfo {author} {\bibfnamefont {S.}~\bibnamefont
  {Ghosh}}, \bibinfo {author} {\bibfnamefont {A.}~\bibnamefont {Shekhter}},
  \bibinfo {author} {\bibfnamefont {F.}~\bibnamefont {Jerzembeck}}, \bibinfo
  {author} {\bibfnamefont {N.}~\bibnamefont {Kikugawa}}, \bibinfo {author}
  {\bibfnamefont {D.~A.}\ \bibnamefont {Sokolov}}, \bibinfo {author}
  {\bibfnamefont {M.}~\bibnamefont {Brando}}, \bibinfo {author} {\bibfnamefont
  {A.~P.}\ \bibnamefont {Mackenzie}}, \bibinfo {author} {\bibfnamefont {C.~W.}\
  \bibnamefont {Hicks}}, \ and\ \bibinfo {author} {\bibfnamefont {B.~J.}\
  \bibnamefont {Ramshaw}},\ }\href {\doibase 10.1038/s41567-020-1032-4}
  {\bibfield  {journal} {\bibinfo  {journal} {Nature Physics}\ }\textbf
  {\bibinfo {volume} {17}},\ \bibinfo {pages} {199} (\bibinfo {year}
  {2021})}\BibitemShut {NoStop}%
\bibitem [{\citenamefont {Benhabib}\ \emph {et~al.}(2021)\citenamefont
  {Benhabib}, \citenamefont {Lupien}, \citenamefont {Paul}, \citenamefont
  {Berges}, \citenamefont {Dion}, \citenamefont {Nardone}, \citenamefont
  {Zitouni}, \citenamefont {Mao}, \citenamefont {Maeno}, \citenamefont
  {Georges}, \citenamefont {Taillefer},\ and\ \citenamefont
  {Proust}}]{Benhabib2021}%
  \BibitemOpen
  \bibfield  {author} {\bibinfo {author} {\bibfnamefont {S.}~\bibnamefont
  {Benhabib}}, \bibinfo {author} {\bibfnamefont {C.}~\bibnamefont {Lupien}},
  \bibinfo {author} {\bibfnamefont {I.}~\bibnamefont {Paul}}, \bibinfo {author}
  {\bibfnamefont {L.}~\bibnamefont {Berges}}, \bibinfo {author} {\bibfnamefont
  {M.}~\bibnamefont {Dion}}, \bibinfo {author} {\bibfnamefont {M.}~\bibnamefont
  {Nardone}}, \bibinfo {author} {\bibfnamefont {A.}~\bibnamefont {Zitouni}},
  \bibinfo {author} {\bibfnamefont {Z.~Q.}\ \bibnamefont {Mao}}, \bibinfo
  {author} {\bibfnamefont {Y.}~\bibnamefont {Maeno}}, \bibinfo {author}
  {\bibfnamefont {A.}~\bibnamefont {Georges}}, \bibinfo {author} {\bibfnamefont
  {L.}~\bibnamefont {Taillefer}}, \ and\ \bibinfo {author} {\bibfnamefont
  {C.}~\bibnamefont {Proust}},\ }\href {\doibase 10.1038/s41567-020-1033-3}
  {\bibfield  {journal} {\bibinfo  {journal} {Nature Physics}\ }\textbf
  {\bibinfo {volume} {17}},\ \bibinfo {pages} {194} (\bibinfo {year}
  {2021})}\BibitemShut {NoStop}%
\bibitem [{\citenamefont {Ghosh}\ \emph {et~al.}(2022)\citenamefont {Ghosh},
  \citenamefont {Kiely}, \citenamefont {Shekhter}, \citenamefont {Jerzembeck},
  \citenamefont {Kikugawa}, \citenamefont {Sokolov}, \citenamefont
  {Mackenzie},\ and\ \citenamefont {Ramshaw}}]{Ghosh2022}%
  \BibitemOpen
  \bibfield  {author} {\bibinfo {author} {\bibfnamefont {S.}~\bibnamefont
  {Ghosh}}, \bibinfo {author} {\bibfnamefont {T.~G.}\ \bibnamefont {Kiely}},
  \bibinfo {author} {\bibfnamefont {A.}~\bibnamefont {Shekhter}}, \bibinfo
  {author} {\bibfnamefont {F.}~\bibnamefont {Jerzembeck}}, \bibinfo {author}
  {\bibfnamefont {N.}~\bibnamefont {Kikugawa}}, \bibinfo {author}
  {\bibfnamefont {D.~A.}\ \bibnamefont {Sokolov}}, \bibinfo {author}
  {\bibfnamefont {A.~P.}\ \bibnamefont {Mackenzie}}, \ and\ \bibinfo {author}
  {\bibfnamefont {B.~J.}\ \bibnamefont {Ramshaw}},\ }\href {\doibase
  10.1103/PhysRevB.106.024520} {\bibfield  {journal} {\bibinfo  {journal}
  {Phys. Rev. B}\ }\textbf {\bibinfo {volume} {106}},\ \bibinfo {pages}
  {024520} (\bibinfo {year} {2022})}\BibitemShut {NoStop}%
\bibitem [{\citenamefont {Pustogow}\ \emph {et~al.}(2019)\citenamefont
  {Pustogow}, \citenamefont {Luo}, \citenamefont {Chronister}, \citenamefont
  {Su}, \citenamefont {Sokolov}, \citenamefont {Jerzembeck}, \citenamefont
  {Mackenzie}, \citenamefont {Hicks}, \citenamefont {Kikugawa}, \citenamefont
  {Raghu}, \citenamefont {Bauer},\ and\ \citenamefont {Brown}}]{Pustogow2019}%
  \BibitemOpen
  \bibfield  {author} {\bibinfo {author} {\bibfnamefont {A.}~\bibnamefont
  {Pustogow}}, \bibinfo {author} {\bibfnamefont {Y.}~\bibnamefont {Luo}},
  \bibinfo {author} {\bibfnamefont {A.}~\bibnamefont {Chronister}}, \bibinfo
  {author} {\bibfnamefont {Y.-S.}\ \bibnamefont {Su}}, \bibinfo {author}
  {\bibfnamefont {D.}~\bibnamefont {Sokolov}}, \bibinfo {author} {\bibfnamefont
  {F.}~\bibnamefont {Jerzembeck}}, \bibinfo {author} {\bibfnamefont
  {A.}~\bibnamefont {Mackenzie}}, \bibinfo {author} {\bibfnamefont
  {C.}~\bibnamefont {Hicks}}, \bibinfo {author} {\bibfnamefont
  {N.}~\bibnamefont {Kikugawa}}, \bibinfo {author} {\bibfnamefont
  {S.}~\bibnamefont {Raghu}}, \bibinfo {author} {\bibfnamefont
  {E.}~\bibnamefont {Bauer}}, \ and\ \bibinfo {author} {\bibfnamefont
  {S.}~\bibnamefont {Brown}},\ }\href {\doibase 10.1038/s41586-019-1596-2}
  {\bibfield  {journal} {\bibinfo  {journal} {Nature}\ }\textbf {\bibinfo
  {volume} {574}},\ \bibinfo {pages} {1} (\bibinfo {year} {2019})}\BibitemShut
  {NoStop}%
\bibitem [{\citenamefont {Ishida}\ \emph {et~al.}(2020)\citenamefont {Ishida},
  \citenamefont {Manago}, \citenamefont {Kinjo},\ and\ \citenamefont
  {Maeno}}]{Ishida2020}%
  \BibitemOpen
  \bibfield  {author} {\bibinfo {author} {\bibfnamefont {K.}~\bibnamefont
  {Ishida}}, \bibinfo {author} {\bibfnamefont {M.}~\bibnamefont {Manago}},
  \bibinfo {author} {\bibfnamefont {K.}~\bibnamefont {Kinjo}}, \ and\ \bibinfo
  {author} {\bibfnamefont {Y.}~\bibnamefont {Maeno}},\ }\href {\doibase
  10.7566/JPSJ.89.034712} {\bibfield  {journal} {\bibinfo  {journal} {Journal
  of the Physical Society of Japan}\ }\textbf {\bibinfo {volume} {89}},\
  \bibinfo {pages} {034712} (\bibinfo {year} {2020})}\BibitemShut {NoStop}%
\bibitem [{\citenamefont {Hassinger}\ \emph {et~al.}(2017)\citenamefont
  {Hassinger}, \citenamefont {Bourgeois-Hope}, \citenamefont {Taniguchi},
  \citenamefont {Ren\'e~de Cotret}, \citenamefont {Grissonnanche},
  \citenamefont {Anwar}, \citenamefont {Maeno}, \citenamefont
  {Doiron-Leyraud},\ and\ \citenamefont {Taillefer}}]{Hassinger2017}%
  \BibitemOpen
  \bibfield  {author} {\bibinfo {author} {\bibfnamefont {E.}~\bibnamefont
  {Hassinger}}, \bibinfo {author} {\bibfnamefont {P.}~\bibnamefont
  {Bourgeois-Hope}}, \bibinfo {author} {\bibfnamefont {H.}~\bibnamefont
  {Taniguchi}}, \bibinfo {author} {\bibfnamefont {S.}~\bibnamefont {Ren\'e~de
  Cotret}}, \bibinfo {author} {\bibfnamefont {G.}~\bibnamefont
  {Grissonnanche}}, \bibinfo {author} {\bibfnamefont {M.~S.}\ \bibnamefont
  {Anwar}}, \bibinfo {author} {\bibfnamefont {Y.}~\bibnamefont {Maeno}},
  \bibinfo {author} {\bibfnamefont {N.}~\bibnamefont {Doiron-Leyraud}}, \ and\
  \bibinfo {author} {\bibfnamefont {L.}~\bibnamefont {Taillefer}},\ }\href
  {\doibase 10.1103/PhysRevX.7.011032} {\bibfield  {journal} {\bibinfo
  {journal} {Phys. Rev. X}\ }\textbf {\bibinfo {volume} {7}},\ \bibinfo {pages}
  {011032} (\bibinfo {year} {2017})}\BibitemShut {NoStop}%
\bibitem [{\citenamefont {Sharma}\ \emph {et~al.}(2020)\citenamefont {Sharma},
  \citenamefont {Edkins}, \citenamefont {Wang}, \citenamefont {Kostin},
  \citenamefont {Sow}, \citenamefont {Maeno}, \citenamefont {Mackenzie},
  \citenamefont {Davis},\ and\ \citenamefont {Madhavan}}]{Sharma2020}%
  \BibitemOpen
  \bibfield  {author} {\bibinfo {author} {\bibfnamefont {R.}~\bibnamefont
  {Sharma}}, \bibinfo {author} {\bibfnamefont {S.~D.}\ \bibnamefont {Edkins}},
  \bibinfo {author} {\bibfnamefont {Z.}~\bibnamefont {Wang}}, \bibinfo {author}
  {\bibfnamefont {A.}~\bibnamefont {Kostin}}, \bibinfo {author} {\bibfnamefont
  {C.}~\bibnamefont {Sow}}, \bibinfo {author} {\bibfnamefont {Y.}~\bibnamefont
  {Maeno}}, \bibinfo {author} {\bibfnamefont {A.~P.}\ \bibnamefont
  {Mackenzie}}, \bibinfo {author} {\bibfnamefont {J.~C.~S.}\ \bibnamefont
  {Davis}}, \ and\ \bibinfo {author} {\bibfnamefont {V.}~\bibnamefont
  {Madhavan}},\ }\href {\doibase 10.1073/pnas.1916463117} {\bibfield  {journal}
  {\bibinfo  {journal} {Proceedings of the National Academy of Sciences}\
  }\textbf {\bibinfo {volume} {117}},\ \bibinfo {pages} {5222} (\bibinfo {year}
  {2020})}\BibitemShut {NoStop}%
\bibitem [{\citenamefont {R\o{}ising}\ \emph {et~al.}(2019)\citenamefont
  {R\o{}ising}, \citenamefont {Scaffidi}, \citenamefont {Flicker},
  \citenamefont {Lange},\ and\ \citenamefont {Simon}}]{Roising2019}%
  \BibitemOpen
  \bibfield  {author} {\bibinfo {author} {\bibfnamefont {H.~S.}\ \bibnamefont
  {R\o{}ising}}, \bibinfo {author} {\bibfnamefont {T.}~\bibnamefont
  {Scaffidi}}, \bibinfo {author} {\bibfnamefont {F.}~\bibnamefont {Flicker}},
  \bibinfo {author} {\bibfnamefont {G.~F.}\ \bibnamefont {Lange}}, \ and\
  \bibinfo {author} {\bibfnamefont {S.~H.}\ \bibnamefont {Simon}},\ }\href
  {\doibase 10.1103/PhysRevResearch.1.033108} {\bibfield  {journal} {\bibinfo
  {journal} {Phys. Rev. Research}\ }\textbf {\bibinfo {volume} {1}},\ \bibinfo
  {pages} {033108} (\bibinfo {year} {2019})}\BibitemShut {NoStop}%
\bibitem [{\citenamefont {R\o{}mer}\ \emph {et~al.}(2022)\citenamefont
  {R\o{}mer}, \citenamefont {Maier}, \citenamefont {Kreisel}, \citenamefont
  {Hirschfeld},\ and\ \citenamefont {Andersen}}]{Romer2022}%
  \BibitemOpen
  \bibfield  {author} {\bibinfo {author} {\bibfnamefont {A.~T.}\ \bibnamefont
  {R\o{}mer}}, \bibinfo {author} {\bibfnamefont {T.~A.}\ \bibnamefont {Maier}},
  \bibinfo {author} {\bibfnamefont {A.}~\bibnamefont {Kreisel}}, \bibinfo
  {author} {\bibfnamefont {P.~J.}\ \bibnamefont {Hirschfeld}}, \ and\ \bibinfo
  {author} {\bibfnamefont {B.~M.}\ \bibnamefont {Andersen}},\ }\href {\doibase
  10.1103/PhysRevResearch.4.033011} {\bibfield  {journal} {\bibinfo  {journal}
  {Phys. Rev. Research}\ }\textbf {\bibinfo {volume} {4}},\ \bibinfo {pages}
  {033011} (\bibinfo {year} {2022})}\BibitemShut {NoStop}%
\bibitem [{\citenamefont {Suh}\ \emph {et~al.}(2020)\citenamefont {Suh},
  \citenamefont {Menke}, \citenamefont {Brydon}, \citenamefont {Timm},
  \citenamefont {Ramires},\ and\ \citenamefont {Agterberg}}]{Suh2020}%
  \BibitemOpen
  \bibfield  {author} {\bibinfo {author} {\bibfnamefont {H.~G.}\ \bibnamefont
  {Suh}}, \bibinfo {author} {\bibfnamefont {H.}~\bibnamefont {Menke}}, \bibinfo
  {author} {\bibfnamefont {P.~M.~R.}\ \bibnamefont {Brydon}}, \bibinfo {author}
  {\bibfnamefont {C.}~\bibnamefont {Timm}}, \bibinfo {author} {\bibfnamefont
  {A.}~\bibnamefont {Ramires}}, \ and\ \bibinfo {author} {\bibfnamefont
  {D.~F.}\ \bibnamefont {Agterberg}},\ }\href {\doibase
  10.1103/PhysRevResearch.2.032023} {\bibfield  {journal} {\bibinfo  {journal}
  {Phys. Rev. Research}\ }\textbf {\bibinfo {volume} {2}},\ \bibinfo {pages}
  {032023(R)} (\bibinfo {year} {2020})}\BibitemShut {NoStop}%
\bibitem [{\citenamefont {Clepkens}\ \emph {et~al.}(2021)\citenamefont
  {Clepkens}, \citenamefont {Lindquist}, \citenamefont {Liu},\ and\
  \citenamefont {Kee}}]{Clepkens2021a}%
  \BibitemOpen
  \bibfield  {author} {\bibinfo {author} {\bibfnamefont {J.}~\bibnamefont
  {Clepkens}}, \bibinfo {author} {\bibfnamefont {A.~W.}\ \bibnamefont
  {Lindquist}}, \bibinfo {author} {\bibfnamefont {X.}~\bibnamefont {Liu}}, \
  and\ \bibinfo {author} {\bibfnamefont {H.-Y.}\ \bibnamefont {Kee}},\ }\href
  {\doibase 10.1103/PhysRevB.104.104512} {\bibfield  {journal} {\bibinfo
  {journal} {Phys. Rev. B}\ }\textbf {\bibinfo {volume} {104}},\ \bibinfo
  {pages} {104512} (\bibinfo {year} {2021})}\BibitemShut {NoStop}%
\bibitem [{\citenamefont {Yuan}\ \emph {et~al.}(2021)\citenamefont {Yuan},
  \citenamefont {Berg},\ and\ \citenamefont {Kivelson}}]{Yuan2021}%
  \BibitemOpen
  \bibfield  {author} {\bibinfo {author} {\bibfnamefont {A.~C.}\ \bibnamefont
  {Yuan}}, \bibinfo {author} {\bibfnamefont {E.}~\bibnamefont {Berg}}, \ and\
  \bibinfo {author} {\bibfnamefont {S.~A.}\ \bibnamefont {Kivelson}},\ }\href
  {\doibase 10.1103/PhysRevB.104.054518} {\bibfield  {journal} {\bibinfo
  {journal} {Phys. Rev. B}\ }\textbf {\bibinfo {volume} {104}},\ \bibinfo
  {pages} {054518} (\bibinfo {year} {2021})}\BibitemShut {NoStop}%
\bibitem [{\citenamefont {Willa}\ \emph {et~al.}(2021)\citenamefont {Willa},
  \citenamefont {Hecker}, \citenamefont {Fernandes},\ and\ \citenamefont
  {Schmalian}}]{Willa2021}%
  \BibitemOpen
  \bibfield  {author} {\bibinfo {author} {\bibfnamefont {R.}~\bibnamefont
  {Willa}}, \bibinfo {author} {\bibfnamefont {M.}~\bibnamefont {Hecker}},
  \bibinfo {author} {\bibfnamefont {R.~M.}\ \bibnamefont {Fernandes}}, \ and\
  \bibinfo {author} {\bibfnamefont {J.}~\bibnamefont {Schmalian}},\ }\href
  {\doibase 10.1103/PhysRevB.104.024511} {\bibfield  {journal} {\bibinfo
  {journal} {Phys. Rev. B}\ }\textbf {\bibinfo {volume} {104}},\ \bibinfo
  {pages} {024511} (\bibinfo {year} {2021})}\BibitemShut {NoStop}%
\bibitem [{\citenamefont {Kivelson}\ \emph {et~al.}(2020)\citenamefont
  {Kivelson}, \citenamefont {Yuan}, \citenamefont {Ramshaw},\ and\
  \citenamefont {Thomale}}]{Kivelson2020}%
  \BibitemOpen
  \bibfield  {author} {\bibinfo {author} {\bibfnamefont {S.~A.}\ \bibnamefont
  {Kivelson}}, \bibinfo {author} {\bibfnamefont {A.~C.}\ \bibnamefont {Yuan}},
  \bibinfo {author} {\bibfnamefont {B.}~\bibnamefont {Ramshaw}}, \ and\
  \bibinfo {author} {\bibfnamefont {R.}~\bibnamefont {Thomale}},\ }\href
  {\doibase 10.1038/s41535-020-0245-1} {\bibfield  {journal} {\bibinfo
  {journal} {npj Quantum Materials}\ }\textbf {\bibinfo {volume} {5}},\
  \bibinfo {pages} {43} (\bibinfo {year} {2020})}\BibitemShut {NoStop}%
\bibitem [{\citenamefont {Willa}(2020)}]{Willa2020}%
  \BibitemOpen
  \bibfield  {author} {\bibinfo {author} {\bibfnamefont {R.}~\bibnamefont
  {Willa}},\ }\href {\doibase 10.1103/PhysRevB.102.180503} {\bibfield
  {journal} {\bibinfo  {journal} {Phys. Rev. B}\ }\textbf {\bibinfo {volume}
  {102}},\ \bibinfo {pages} {180503(R)} (\bibinfo {year} {2020})}\BibitemShut
  {NoStop}%
\bibitem [{\citenamefont {Wang}\ \emph {et~al.}(2013)\citenamefont {Wang},
  \citenamefont {Platt}, \citenamefont {Yang}, \citenamefont {Honerkamp},
  \citenamefont {Zhang}, \citenamefont {Hanke}, \citenamefont {Rice},\ and\
  \citenamefont {Thomale}}]{Wang2013}%
  \BibitemOpen
  \bibfield  {author} {\bibinfo {author} {\bibfnamefont {Q.~H.}\ \bibnamefont
  {Wang}}, \bibinfo {author} {\bibfnamefont {C.}~\bibnamefont {Platt}},
  \bibinfo {author} {\bibfnamefont {Y.}~\bibnamefont {Yang}}, \bibinfo {author}
  {\bibfnamefont {C.}~\bibnamefont {Honerkamp}}, \bibinfo {author}
  {\bibfnamefont {F.~C.}\ \bibnamefont {Zhang}}, \bibinfo {author}
  {\bibfnamefont {W.}~\bibnamefont {Hanke}}, \bibinfo {author} {\bibfnamefont
  {T.~M.}\ \bibnamefont {Rice}}, \ and\ \bibinfo {author} {\bibfnamefont
  {R.}~\bibnamefont {Thomale}},\ }\href {\doibase 10.1209/0295-5075/104/17013}
  {\bibfield  {journal} {\bibinfo  {journal} {{EPL} (Europhysics Letters)}\
  }\textbf {\bibinfo {volume} {104}},\ \bibinfo {pages} {17013} (\bibinfo
  {year} {2013})}\BibitemShut {NoStop}%
\bibitem [{\citenamefont {Scaffidi}\ \emph {et~al.}(2014)\citenamefont
  {Scaffidi}, \citenamefont {Romers},\ and\ \citenamefont
  {Simon}}]{Scaffidi2014}%
  \BibitemOpen
  \bibfield  {author} {\bibinfo {author} {\bibfnamefont {T.}~\bibnamefont
  {Scaffidi}}, \bibinfo {author} {\bibfnamefont {J.~C.}\ \bibnamefont
  {Romers}}, \ and\ \bibinfo {author} {\bibfnamefont {S.~H.}\ \bibnamefont
  {Simon}},\ }\href {\doibase 10.1103/PhysRevB.89.220510} {\bibfield  {journal}
  {\bibinfo  {journal} {Phys. Rev. B}\ }\textbf {\bibinfo {volume} {89}},\
  \bibinfo {pages} {220510(R)} (\bibinfo {year} {2014})}\BibitemShut {NoStop}%
\bibitem [{\citenamefont {Zhang}\ \emph {et~al.}(2018)\citenamefont {Zhang},
  \citenamefont {Huang}, \citenamefont {Yang},\ and\ \citenamefont
  {Yao}}]{Zhang2018}%
  \BibitemOpen
  \bibfield  {author} {\bibinfo {author} {\bibfnamefont {L.-D.}\ \bibnamefont
  {Zhang}}, \bibinfo {author} {\bibfnamefont {W.}~\bibnamefont {Huang}},
  \bibinfo {author} {\bibfnamefont {F.}~\bibnamefont {Yang}}, \ and\ \bibinfo
  {author} {\bibfnamefont {H.}~\bibnamefont {Yao}},\ }\href {\doibase
  10.1103/PhysRevB.97.060510} {\bibfield  {journal} {\bibinfo  {journal} {Phys.
  Rev. B}\ }\textbf {\bibinfo {volume} {97}},\ \bibinfo {pages} {060510(R)}
  (\bibinfo {year} {2018})}\BibitemShut {NoStop}%
\bibitem [{\citenamefont {Gingras}\ \emph {et~al.}(2019)\citenamefont
  {Gingras}, \citenamefont {Nourafkan}, \citenamefont {Tremblay},\ and\
  \citenamefont {C\^ot\'e}}]{Gingras2019}%
  \BibitemOpen
  \bibfield  {author} {\bibinfo {author} {\bibfnamefont {O.}~\bibnamefont
  {Gingras}}, \bibinfo {author} {\bibfnamefont {R.}~\bibnamefont {Nourafkan}},
  \bibinfo {author} {\bibfnamefont {A.M.S.}\ \bibnamefont {Tremblay}}, \ and\
  \bibinfo {author} {\bibfnamefont {M.}~\bibnamefont {C\^ot\'e}},\ }\href
  {\doibase 10.1103/PhysRevLett.123.217005} {\bibfield  {journal} {\bibinfo
  {journal} {Phys. Rev. Lett.}\ }\textbf {\bibinfo {volume} {123}},\ \bibinfo
  {pages} {217005} (\bibinfo {year} {2019})}\BibitemShut {NoStop}%
\bibitem [{\citenamefont {R\o{}mer}\ \emph {et~al.}(2019)\citenamefont
  {R\o{}mer}, \citenamefont {Scherer}, \citenamefont {Eremin}, \citenamefont
  {Hirschfeld},\ and\ \citenamefont {Andersen}}]{Romer2019}%
  \BibitemOpen
  \bibfield  {author} {\bibinfo {author} {\bibfnamefont {A.~T.}\ \bibnamefont
  {R\o{}mer}}, \bibinfo {author} {\bibfnamefont {D.~D.}\ \bibnamefont
  {Scherer}}, \bibinfo {author} {\bibfnamefont {I.~M.}\ \bibnamefont {Eremin}},
  \bibinfo {author} {\bibfnamefont {P.~J.}\ \bibnamefont {Hirschfeld}}, \ and\
  \bibinfo {author} {\bibfnamefont {B.~M.}\ \bibnamefont {Andersen}},\ }\href
  {\doibase 10.1103/PhysRevLett.123.247001} {\bibfield  {journal} {\bibinfo
  {journal} {Phys. Rev. Lett.}\ }\textbf {\bibinfo {volume} {123}},\ \bibinfo
  {pages} {247001} (\bibinfo {year} {2019})}\BibitemShut {NoStop}%
\bibitem [{\citenamefont {Wang}\ \emph {et~al.}(2019)\citenamefont {Wang},
  \citenamefont {Zhang}, \citenamefont {Zhang},\ and\ \citenamefont
  {Wang}}]{Wang2019}%
  \BibitemOpen
  \bibfield  {author} {\bibinfo {author} {\bibfnamefont {W.-S.}\ \bibnamefont
  {Wang}}, \bibinfo {author} {\bibfnamefont {C.-C.}\ \bibnamefont {Zhang}},
  \bibinfo {author} {\bibfnamefont {F.-C.}\ \bibnamefont {Zhang}}, \ and\
  \bibinfo {author} {\bibfnamefont {Q.-H.}\ \bibnamefont {Wang}},\ }\href
  {\doibase 10.1103/PhysRevLett.122.027002} {\bibfield  {journal} {\bibinfo
  {journal} {Phys. Rev. Lett.}\ }\textbf {\bibinfo {volume} {122}},\ \bibinfo
  {pages} {027002} (\bibinfo {year} {2019})}\BibitemShut {NoStop}%
\bibitem [{\citenamefont {Wang}\ \emph {et~al.}(2020)\citenamefont {Wang},
  \citenamefont {Wang},\ and\ \citenamefont {Kallin}}]{Wang2020}%
  \BibitemOpen
  \bibfield  {author} {\bibinfo {author} {\bibfnamefont {Z.}~\bibnamefont
  {Wang}}, \bibinfo {author} {\bibfnamefont {X.}~\bibnamefont {Wang}}, \ and\
  \bibinfo {author} {\bibfnamefont {C.}~\bibnamefont {Kallin}},\ }\href
  {\doibase 10.1103/PhysRevB.101.064507} {\bibfield  {journal} {\bibinfo
  {journal} {Phys. Rev. B}\ }\textbf {\bibinfo {volume} {101}},\ \bibinfo
  {pages} {064507} (\bibinfo {year} {2020})}\BibitemShut {NoStop}%
\bibitem [{\citenamefont {R\o{}mer}\ \emph {et~al.}(2021)\citenamefont
  {R\o{}mer}, \citenamefont {Hirschfeld},\ and\ \citenamefont
  {Andersen}}]{Romer2021}%
  \BibitemOpen
  \bibfield  {author} {\bibinfo {author} {\bibfnamefont {A.~T.}\ \bibnamefont
  {R\o{}mer}}, \bibinfo {author} {\bibfnamefont {P.~J.}\ \bibnamefont
  {Hirschfeld}}, \ and\ \bibinfo {author} {\bibfnamefont {B.~M.}\ \bibnamefont
  {Andersen}},\ }\href {\doibase 10.1103/PhysRevB.104.064507} {\bibfield
  {journal} {\bibinfo  {journal} {Phys. Rev. B}\ }\textbf {\bibinfo {volume}
  {104}},\ \bibinfo {pages} {064507} (\bibinfo {year} {2021})}\BibitemShut
  {NoStop}%
\bibitem [{\citenamefont {Raghu}\ \emph {et~al.}(2012)\citenamefont {Raghu},
  \citenamefont {Berg}, \citenamefont {Chubukov},\ and\ \citenamefont
  {Kivelson}}]{Raghu2012}%
  \BibitemOpen
  \bibfield  {author} {\bibinfo {author} {\bibfnamefont {S.}~\bibnamefont
  {Raghu}}, \bibinfo {author} {\bibfnamefont {E.}~\bibnamefont {Berg}},
  \bibinfo {author} {\bibfnamefont {A.~V.}\ \bibnamefont {Chubukov}}, \ and\
  \bibinfo {author} {\bibfnamefont {S.~A.}\ \bibnamefont {Kivelson}},\ }\href
  {\doibase 10.1103/PhysRevB.85.024516} {\bibfield  {journal} {\bibinfo
  {journal} {Phys. Rev. B}\ }\textbf {\bibinfo {volume} {85}},\ \bibinfo
  {pages} {024516} (\bibinfo {year} {2012})}\BibitemShut {NoStop}%
\bibitem [{\citenamefont {Wolf}\ \emph {et~al.}(2018)\citenamefont {Wolf},
  \citenamefont {Schmidt},\ and\ \citenamefont {Rachel}}]{Wolf2018}%
  \BibitemOpen
  \bibfield  {author} {\bibinfo {author} {\bibfnamefont {S.}~\bibnamefont
  {Wolf}}, \bibinfo {author} {\bibfnamefont {T.~L.}\ \bibnamefont {Schmidt}}, \
  and\ \bibinfo {author} {\bibfnamefont {S.}~\bibnamefont {Rachel}},\ }\href
  {\doibase 10.1103/PhysRevB.98.174515} {\bibfield  {journal} {\bibinfo
  {journal} {Phys. Rev. B}\ }\textbf {\bibinfo {volume} {98}},\ \bibinfo
  {pages} {174515} (\bibinfo {year} {2018})}\BibitemShut {NoStop}%
\bibitem [{\citenamefont {Roig}\ \emph {et~al.}(2022)\citenamefont {Roig},
  \citenamefont {R\o{}mer}, \citenamefont {Kreisel}, \citenamefont
  {Hirschfeld},\ and\ \citenamefont {Andersen}}]{Roig2022}%
  \BibitemOpen
  \bibfield  {author} {\bibinfo {author} {\bibfnamefont {M.}~\bibnamefont
  {Roig}}, \bibinfo {author} {\bibfnamefont {A.~T.}\ \bibnamefont {R\o{}mer}},
  \bibinfo {author} {\bibfnamefont {A.}~\bibnamefont {Kreisel}}, \bibinfo
  {author} {\bibfnamefont {P.~J.}\ \bibnamefont {Hirschfeld}}, \ and\ \bibinfo
  {author} {\bibfnamefont {B.~M.}\ \bibnamefont {Andersen}},\ }\href {\doibase
  10.1103/PhysRevB.106.L100501} {\bibfield  {journal} {\bibinfo  {journal}
  {Phys. Rev. B}\ }\textbf {\bibinfo {volume} {106}},\ \bibinfo {pages}
  {L100501} (\bibinfo {year} {2022})}\BibitemShut {NoStop}%
\bibitem [{\citenamefont {R\o{}mer}\ \emph {et~al.}(2020)\citenamefont
  {R\o{}mer}, \citenamefont {Maier}, \citenamefont {Kreisel}, \citenamefont
  {Eremin}, \citenamefont {Hirschfeld},\ and\ \citenamefont
  {Andersen}}]{Romer2020b}%
  \BibitemOpen
  \bibfield  {author} {\bibinfo {author} {\bibfnamefont {A.~T.}\ \bibnamefont
  {R\o{}mer}}, \bibinfo {author} {\bibfnamefont {T.~A.}\ \bibnamefont {Maier}},
  \bibinfo {author} {\bibfnamefont {A.}~\bibnamefont {Kreisel}}, \bibinfo
  {author} {\bibfnamefont {I.}~\bibnamefont {Eremin}}, \bibinfo {author}
  {\bibfnamefont {P.~J.}\ \bibnamefont {Hirschfeld}}, \ and\ \bibinfo {author}
  {\bibfnamefont {B.~M.}\ \bibnamefont {Andersen}},\ }\href {\doibase
  10.1103/PhysRevResearch.2.013108} {\bibfield  {journal} {\bibinfo  {journal}
  {Phys. Rev. Research}\ }\textbf {\bibinfo {volume} {2}},\ \bibinfo {pages}
  {013108} (\bibinfo {year} {2020})}\BibitemShut {NoStop}%
\bibitem [{\citenamefont {Huang}\ and\ \citenamefont {Wang}(2021)}]{Huang2021}%
  \BibitemOpen
  \bibfield  {author} {\bibinfo {author} {\bibfnamefont {W.}~\bibnamefont
  {Huang}}\ and\ \bibinfo {author} {\bibfnamefont {Z.}~\bibnamefont {Wang}},\
  }\href {\doibase 10.1103/PhysRevResearch.3.L042002} {\bibfield  {journal}
  {\bibinfo  {journal} {Phys. Rev. Research}\ }\textbf {\bibinfo {volume}
  {3}},\ \bibinfo {pages} {L042002} (\bibinfo {year} {2021})}\BibitemShut
  {NoStop}%
\bibitem [{\citenamefont {K\"aser}\ \emph {et~al.}(2022)\citenamefont
  {K\"aser}, \citenamefont {Strand}, \citenamefont {Wentzell}, \citenamefont
  {Georges}, \citenamefont {Parcollet},\ and\ \citenamefont
  {Hansmann}}]{Kaeser2022}%
  \BibitemOpen
  \bibfield  {author} {\bibinfo {author} {\bibfnamefont {S.}~\bibnamefont
  {K\"aser}}, \bibinfo {author} {\bibfnamefont {H.~U.~R.}\ \bibnamefont
  {Strand}}, \bibinfo {author} {\bibfnamefont {N.}~\bibnamefont {Wentzell}},
  \bibinfo {author} {\bibfnamefont {A.}~\bibnamefont {Georges}}, \bibinfo
  {author} {\bibfnamefont {O.}~\bibnamefont {Parcollet}}, \ and\ \bibinfo
  {author} {\bibfnamefont {P.}~\bibnamefont {Hansmann}},\ }\href {\doibase
  10.1103/PhysRevB.105.155101} {\bibfield  {journal} {\bibinfo  {journal}
  {Phys. Rev. B}\ }\textbf {\bibinfo {volume} {105}},\ \bibinfo {pages}
  {155101} (\bibinfo {year} {2022})}\BibitemShut {NoStop}%
\bibitem [{\citenamefont {Georges}\ \emph {et~al.}(2013)\citenamefont
  {Georges}, \citenamefont {Medici},\ and\ \citenamefont
  {Mravlje}}]{Georges2013}%
  \BibitemOpen
  \bibfield  {author} {\bibinfo {author} {\bibfnamefont {A.}~\bibnamefont
  {Georges}}, \bibinfo {author} {\bibfnamefont {L.~d.}\ \bibnamefont {Medici}},
  \ and\ \bibinfo {author} {\bibfnamefont {J.}~\bibnamefont {Mravlje}},\ }\href
  {\doibase 10.1146/annurev-conmatphys-020911-125045} {\bibfield  {journal}
  {\bibinfo  {journal} {Annual Review of Condensed Matter Physics}\ }\textbf
  {\bibinfo {volume} {4}},\ \bibinfo {pages} {137} (\bibinfo {year}
  {2013})}\BibitemShut {NoStop}%
\bibitem [{\citenamefont {Zhang}\ \emph {et~al.}(2016)\citenamefont {Zhang},
  \citenamefont {Gorelov}, \citenamefont {Sarvestani},\ and\ \citenamefont
  {Pavarini}}]{Zhang2016}%
  \BibitemOpen
  \bibfield  {author} {\bibinfo {author} {\bibfnamefont {G.}~\bibnamefont
  {Zhang}}, \bibinfo {author} {\bibfnamefont {E.}~\bibnamefont {Gorelov}},
  \bibinfo {author} {\bibfnamefont {E.}~\bibnamefont {Sarvestani}}, \ and\
  \bibinfo {author} {\bibfnamefont {E.}~\bibnamefont {Pavarini}},\ }\href
  {\doibase 10.1103/PhysRevLett.116.106402} {\bibfield  {journal} {\bibinfo
  {journal} {Phys. Rev. Lett.}\ }\textbf {\bibinfo {volume} {116}},\ \bibinfo
  {pages} {106402} (\bibinfo {year} {2016})}\BibitemShut {NoStop}%
\bibitem [{\citenamefont {Mravlje}\ \emph {et~al.}(2011)\citenamefont
  {Mravlje}, \citenamefont {Aichhorn}, \citenamefont {Miyake}, \citenamefont
  {Haule}, \citenamefont {Kotliar},\ and\ \citenamefont
  {Georges}}]{Mravlje2011}%
  \BibitemOpen
  \bibfield  {author} {\bibinfo {author} {\bibfnamefont {J.}~\bibnamefont
  {Mravlje}}, \bibinfo {author} {\bibfnamefont {M.}~\bibnamefont {Aichhorn}},
  \bibinfo {author} {\bibfnamefont {T.}~\bibnamefont {Miyake}}, \bibinfo
  {author} {\bibfnamefont {K.}~\bibnamefont {Haule}}, \bibinfo {author}
  {\bibfnamefont {G.}~\bibnamefont {Kotliar}}, \ and\ \bibinfo {author}
  {\bibfnamefont {A.}~\bibnamefont {Georges}},\ }\href {\doibase
  10.1103/PhysRevLett.106.096401} {\bibfield  {journal} {\bibinfo  {journal}
  {Phys. Rev. Lett.}\ }\textbf {\bibinfo {volume} {106}},\ \bibinfo {pages}
  {096401} (\bibinfo {year} {2011})}\BibitemShut {NoStop}%
\bibitem [{\citenamefont {Raghu}\ \emph {et~al.}(2010)\citenamefont {Raghu},
  \citenamefont {Kivelson},\ and\ \citenamefont {Scalapino}}]{Raghu2010}%
  \BibitemOpen
  \bibfield  {author} {\bibinfo {author} {\bibfnamefont {S.}~\bibnamefont
  {Raghu}}, \bibinfo {author} {\bibfnamefont {S.~A.}\ \bibnamefont {Kivelson}},
  \ and\ \bibinfo {author} {\bibfnamefont {D.~J.}\ \bibnamefont {Scalapino}},\
  }\href {\doibase 10.1103/PhysRevB.81.224505} {\bibfield  {journal} {\bibinfo
  {journal} {Phys. Rev. B}\ }\textbf {\bibinfo {volume} {81}},\ \bibinfo
  {pages} {224505} (\bibinfo {year} {2010})}\BibitemShut {NoStop}%
\bibitem [{\citenamefont {Cho}\ \emph {et~al.}(2013)\citenamefont {Cho},
  \citenamefont {Thomale}, \citenamefont {Raghu},\ and\ \citenamefont
  {Kivelson}}]{Cho2013}%
  \BibitemOpen
  \bibfield  {author} {\bibinfo {author} {\bibfnamefont {W.}~\bibnamefont
  {Cho}}, \bibinfo {author} {\bibfnamefont {R.}~\bibnamefont {Thomale}},
  \bibinfo {author} {\bibfnamefont {S.}~\bibnamefont {Raghu}}, \ and\ \bibinfo
  {author} {\bibfnamefont {S.~A.}\ \bibnamefont {Kivelson}},\ }\href {\doibase
  10.1103/PhysRevB.88.064505} {\bibfield  {journal} {\bibinfo  {journal} {Phys.
  Rev. B}\ }\textbf {\bibinfo {volume} {88}},\ \bibinfo {pages} {064505}
  (\bibinfo {year} {2013})}\BibitemShut {NoStop}%
\bibitem [{\citenamefont {Pchelkina}\ \emph {et~al.}(2007)\citenamefont
  {Pchelkina}, \citenamefont {Nekrasov}, \citenamefont {Pruschke},
  \citenamefont {Sekiyama}, \citenamefont {Suga}, \citenamefont {Anisimov},\
  and\ \citenamefont {Vollhardt}}]{Pchelkina2007}%
  \BibitemOpen
  \bibfield  {author} {\bibinfo {author} {\bibfnamefont {Z.~V.}\ \bibnamefont
  {Pchelkina}}, \bibinfo {author} {\bibfnamefont {I.~A.}\ \bibnamefont
  {Nekrasov}}, \bibinfo {author} {\bibfnamefont {T.}~\bibnamefont {Pruschke}},
  \bibinfo {author} {\bibfnamefont {A.}~\bibnamefont {Sekiyama}}, \bibinfo
  {author} {\bibfnamefont {S.}~\bibnamefont {Suga}}, \bibinfo {author}
  {\bibfnamefont {V.~I.}\ \bibnamefont {Anisimov}}, \ and\ \bibinfo {author}
  {\bibfnamefont {D.}~\bibnamefont {Vollhardt}},\ }\href {\doibase
  10.1103/PhysRevB.75.035122} {\bibfield  {journal} {\bibinfo  {journal} {Phys.
  Rev. B}\ }\textbf {\bibinfo {volume} {75}},\ \bibinfo {pages} {035122}
  (\bibinfo {year} {2007})}\BibitemShut {NoStop}%
\bibitem [{\citenamefont {Hirayama}\ \emph {et~al.}(2018)\citenamefont
  {Hirayama}, \citenamefont {Yamaji}, \citenamefont {Misawa},\ and\
  \citenamefont {Imada}}]{Hirayama2018}%
  \BibitemOpen
  \bibfield  {author} {\bibinfo {author} {\bibfnamefont {M.}~\bibnamefont
  {Hirayama}}, \bibinfo {author} {\bibfnamefont {Y.}~\bibnamefont {Yamaji}},
  \bibinfo {author} {\bibfnamefont {T.}~\bibnamefont {Misawa}}, \ and\ \bibinfo
  {author} {\bibfnamefont {M.}~\bibnamefont {Imada}},\ }\href {\doibase
  10.1103/PhysRevB.98.134501} {\bibfield  {journal} {\bibinfo  {journal} {Phys.
  Rev. B}\ }\textbf {\bibinfo {volume} {98}},\ \bibinfo {pages} {134501}
  (\bibinfo {year} {2018})}\BibitemShut {NoStop}%
\bibitem [{\citenamefont {Hoggan}\ \emph {et~al.}(2011)\citenamefont {Hoggan},
  \citenamefont {Ruiz},\ and\ \citenamefont {Özdoǧan}}]{Hoggan2011}%
  \BibitemOpen
  \bibfield  {author} {\bibinfo {author} {\bibfnamefont {P.}~\bibnamefont
  {Hoggan}}, \bibinfo {author} {\bibfnamefont {M.~B.}\ \bibnamefont
  {Ruiz}}, \ and\ \bibinfo {author} {\bibfnamefont {T.}~\bibnamefont
  {Özdoǧan}},\ }\href@noop {} {\bibfield  {journal} {\bibinfo  {journal}
  {Quantum Frontiers of Atoms and Molecules}:\ \bibinfo {pages} {64-90}}
  (\bibinfo {year} {2011})}\BibitemShut {NoStop}%
\bibitem [{\citenamefont {R\o{}mer}\ \emph {et~al.}(2015)\citenamefont
  {R\o{}mer}, \citenamefont {Kreisel}, \citenamefont {Eremin}, \citenamefont
  {Malakhov}, \citenamefont {Maier}, \citenamefont {Hirschfeld},\ and\
  \citenamefont {Andersen}}]{Romer2015}%
  \BibitemOpen
  \bibfield  {author} {\bibinfo {author} {\bibfnamefont {A.~T.}\ \bibnamefont
  {R\o{}mer}}, \bibinfo {author} {\bibfnamefont {A.}~\bibnamefont {Kreisel}},
  \bibinfo {author} {\bibfnamefont {I.}~\bibnamefont {Eremin}}, \bibinfo
  {author} {\bibfnamefont {M.~A.}\ \bibnamefont {Malakhov}}, \bibinfo {author}
  {\bibfnamefont {T.~A.}\ \bibnamefont {Maier}}, \bibinfo {author}
  {\bibfnamefont {P.~J.}\ \bibnamefont {Hirschfeld}}, \ and\ \bibinfo {author}
  {\bibfnamefont {B.~M.}\ \bibnamefont {Andersen}},\ }\href {\doibase
  10.1103/PhysRevB.92.104505} {\bibfield  {journal} {\bibinfo  {journal} {Phys.
  Rev. B}\ }\textbf {\bibinfo {volume} {92}},\ \bibinfo {pages} {104505}
  (\bibinfo {year} {2015})}\BibitemShut {NoStop}%
\bibitem [{\citenamefont {Tamai}\ \emph {et~al.}(2019)\citenamefont {Tamai},
  \citenamefont {Zingl}, \citenamefont {Rozbicki}, \citenamefont {Cappelli},
  \citenamefont {Ricc\`o}, \citenamefont {dela Torre}, \citenamefont
  {McKeownWalker}, \citenamefont {Bruno}, \citenamefont {King}, \citenamefont
  {Meevasana}, \citenamefont {Shi}, \citenamefont
  {Radovi\ifmmode~\acute{c}\else \'{c}\fi{}}, \citenamefont {Plumb},
  \citenamefont {Gibbs}, \citenamefont {Mackenzie}, \citenamefont {Berthod},
  \citenamefont {Strand}, \citenamefont {Kim}, \citenamefont {Georges},\ and\
  \citenamefont {Baumberger}}]{Tamai2019}%
  \BibitemOpen
  \bibfield  {author} {\bibinfo {author} {\bibfnamefont {A.}~\bibnamefont
  {Tamai}}, \bibinfo {author} {\bibfnamefont {M.}~\bibnamefont {Zingl}},
  \bibinfo {author} {\bibfnamefont {E.}~\bibnamefont {Rozbicki}}, \bibinfo
  {author} {\bibfnamefont {E.}~\bibnamefont {Cappelli}}, \bibinfo {author}
  {\bibfnamefont {S.}~\bibnamefont {Ricc\`o}}, \bibinfo {author} {\bibfnamefont
  {A.}~\bibnamefont {dela Torre}}, \bibinfo {author} {\bibfnamefont
  {S.}~\bibnamefont {McKeownWalker}}, \bibinfo {author} {\bibfnamefont
  {F.~Y.}\ \bibnamefont {Bruno}}, \bibinfo {author} {\bibfnamefont {P.~D.~C.}\
  \bibnamefont {King}}, \bibinfo {author} {\bibfnamefont {W.}~\bibnamefont
  {Meevasana}}, \bibinfo {author} {\bibfnamefont {M.}~\bibnamefont {Shi}},
  \bibinfo {author} {\bibfnamefont {M.}~\bibnamefont
  {Radovi\ifmmode~\acute{c}\else \'{c}\fi{}}}, \bibinfo {author} {\bibfnamefont
  {N.~C.}\ \bibnamefont {Plumb}}, \bibinfo {author} {\bibfnamefont {A.~S.}\
  \bibnamefont {Gibbs}}, \bibinfo {author} {\bibfnamefont {A.~P.}\ \bibnamefont
  {Mackenzie}}, \bibinfo {author} {\bibfnamefont {C.}~\bibnamefont {Berthod}},
  \bibinfo {author} {\bibfnamefont {H.~U.~R.}\ \bibnamefont {Strand}}, \bibinfo
  {author} {\bibfnamefont {M.}~\bibnamefont {Kim}}, \bibinfo {author}
  {\bibfnamefont {A.}~\bibnamefont {Georges}}, \ and\ \bibinfo {author}
  {\bibfnamefont {F.}~\bibnamefont {Baumberger}},\ }\href {\doibase
  10.1103/PhysRevX.9.021048} {\bibfield  {journal} {\bibinfo  {journal} {Phys.
  Rev. X}\ }\textbf {\bibinfo {volume} {9}},\ \bibinfo {pages} {021048}
  (\bibinfo {year} {2019})}\BibitemShut {NoStop}%
\bibitem [{\citenamefont {Liu}\ \emph {et~al.}(2008)\citenamefont {Liu},
  \citenamefont {Antonov}, \citenamefont {Jepsen},\ and\ \citenamefont
  {Andersen.}}]{Liu2008}%
  \BibitemOpen
  \bibfield  {author} {\bibinfo {author} {\bibfnamefont {G.-Q.}\ \bibnamefont
  {Liu}}, \bibinfo {author} {\bibfnamefont {V.~N.}\ \bibnamefont {Antonov}},
  \bibinfo {author} {\bibfnamefont {O.}~\bibnamefont {Jepsen}}, \ and\ \bibinfo
  {author} {\bibfnamefont {O.~K.}\ \bibnamefont {Andersen}},\ }\href {\doibase
  10.1103/PhysRevLett.101.026408} {\bibfield  {journal} {\bibinfo  {journal}
  {Phys. Rev. Lett.}\ }\textbf {\bibinfo {volume} {101}},\ \bibinfo {pages}
  {026408} (\bibinfo {year} {2008})}\BibitemShut {NoStop}%
\bibitem [{\citenamefont {Isobe}\ and\ \citenamefont
  {Nagaosa}(2014)}]{Isobe2014}%
  \BibitemOpen
  \bibfield  {author} {\bibinfo {author} {\bibfnamefont {H.}~\bibnamefont
  {Isobe}}\ and\ \bibinfo {author} {\bibfnamefont {N.}~\bibnamefont
  {Nagaosa}},\ }\href {\doibase 10.1103/PhysRevB.90.115122} {\bibfield
  {journal} {\bibinfo  {journal} {Phys. Rev. B}\ }\textbf {\bibinfo {volume}
  {90}},\ \bibinfo {pages} {115122} (\bibinfo {year} {2014})}\BibitemShut
  {NoStop}%
\bibitem [{\citenamefont {Kim}\ \emph {et~al.}(2018)\citenamefont {Kim},
  \citenamefont {Mravlje}, \citenamefont {Ferrero}, \citenamefont {Parcollet},\
  and\ \citenamefont {Georges}}]{Kim2018}%
  \BibitemOpen
  \bibfield  {author} {\bibinfo {author} {\bibfnamefont {M.}~\bibnamefont
  {Kim}}, \bibinfo {author} {\bibfnamefont {J.}~\bibnamefont {Mravlje}},
  \bibinfo {author} {\bibfnamefont {M.}~\bibnamefont {Ferrero}}, \bibinfo
  {author} {\bibfnamefont {O.}~\bibnamefont {Parcollet}}, \ and\ \bibinfo
  {author} {\bibfnamefont {A.}~\bibnamefont {Georges}},\ }\href {\doibase
  10.1103/PhysRevLett.120.126401} {\bibfield  {journal} {\bibinfo  {journal}
  {Phys. Rev. Lett.}\ }\textbf {\bibinfo {volume} {120}},\ \bibinfo {pages}
  {126401} (\bibinfo {year} {2018})}\BibitemShut {NoStop}%
\bibitem [{\citenamefont {Hirayama}\ \emph {et~al.}(2019)\citenamefont
  {Hirayama}, \citenamefont {Misawa}, \citenamefont {Ohgoe}, \citenamefont
  {Yamaji},\ and\ \citenamefont {Imada}}]{Hirayama2019}%
  \BibitemOpen
  \bibfield  {author} {\bibinfo {author} {\bibfnamefont {M.}~\bibnamefont
  {Hirayama}}, \bibinfo {author} {\bibfnamefont {T.}~\bibnamefont {Misawa}},
  \bibinfo {author} {\bibfnamefont {T.}~\bibnamefont {Ohgoe}}, \bibinfo
  {author} {\bibfnamefont {Y.}~\bibnamefont {Yamaji}}, \ and\ \bibinfo {author}
  {\bibfnamefont {M.}~\bibnamefont {Imada}},\ }\href {\doibase
  10.1103/PhysRevB.99.245155} {\bibfield  {journal} {\bibinfo  {journal} {Phys.
  Rev. B}\ }\textbf {\bibinfo {volume} {99}},\ \bibinfo {pages} {245155}
  (\bibinfo {year} {2019})}\BibitemShut {NoStop}%
\bibitem [{Note1()}]{Note1}%
  \BibitemOpen
  \bibinfo {note} {$|\Delta |_{\protect \mathrm {max (min)}}$ is the maximum
  (minimum) gap magnitude over all three bands.}\BibitemShut {Stop}%
\bibitem [{\citenamefont {Lupien}\ \emph {et~al.}(2001)\citenamefont {Lupien},
  \citenamefont {MacFarlane}, \citenamefont {Proust}, \citenamefont
  {Taillefer}, \citenamefont {Mao},\ and\ \citenamefont {Maeno}}]{Lupien2001}%
  \BibitemOpen
  \bibfield  {author} {\bibinfo {author} {\bibfnamefont {C.}~\bibnamefont
  {Lupien}}, \bibinfo {author} {\bibfnamefont {W.~A.}\ \bibnamefont
  {MacFarlane}}, \bibinfo {author} {\bibfnamefont {C.}~\bibnamefont {Proust}},
  \bibinfo {author} {\bibfnamefont {L.}~\bibnamefont {Taillefer}}, \bibinfo
  {author} {\bibfnamefont {Z.~Q.}\ \bibnamefont {Mao}}, \ and\ \bibinfo
  {author} {\bibfnamefont {Y.}~\bibnamefont {Maeno}},\ }\href {\doibase
  10.1103/PhysRevLett.86.5986} {\bibfield  {journal} {\bibinfo  {journal}
  {Phys. Rev. Lett.}\ }\textbf {\bibinfo {volume} {86}},\ \bibinfo {pages}
  {5986} (\bibinfo {year} {2001})}\BibitemShut {NoStop}%
\bibitem [{\citenamefont {Leggett}(1965)}]{Leggett1965}%
  \BibitemOpen
  \bibfield  {author} {\bibinfo {author} {\bibfnamefont {A.~J.}\ \bibnamefont
  {Leggett}},\ }\href {\doibase 10.1103/PhysRevLett.14.536} {\bibfield
  {journal} {\bibinfo  {journal} {Phys. Rev. Lett.}\ }\textbf {\bibinfo
  {volume} {14}},\ \bibinfo {pages} {536} (\bibinfo {year} {1965})}\BibitemShut
  {NoStop}%
\bibitem [{\citenamefont {Chronister}\ \emph {et~al.}(2021)\citenamefont
  {Chronister}, \citenamefont {Pustogow}, \citenamefont {Kikugawa},
  \citenamefont {Sokolov}, \citenamefont {Jerzembeck}, \citenamefont {Hicks},
  \citenamefont {Mackenzie}, \citenamefont {Bauer},\ and\ \citenamefont
  {Brown}}]{Chronister2021}%
  \BibitemOpen
  \bibfield  {author} {\bibinfo {author} {\bibfnamefont {A.}~\bibnamefont
  {Chronister}}, \bibinfo {author} {\bibfnamefont {A.}~\bibnamefont
  {Pustogow}}, \bibinfo {author} {\bibfnamefont {N.}~\bibnamefont {Kikugawa}},
  \bibinfo {author} {\bibfnamefont {D.~A.}\ \bibnamefont {Sokolov}}, \bibinfo
  {author} {\bibfnamefont {F.}~\bibnamefont {Jerzembeck}}, \bibinfo {author}
  {\bibfnamefont {C.~W.}\ \bibnamefont {Hicks}}, \bibinfo {author}
  {\bibfnamefont {A.~P.}\ \bibnamefont {Mackenzie}}, \bibinfo {author}
  {\bibfnamefont {E.~D.}\ \bibnamefont {Bauer}}, \ and\ \bibinfo {author}
  {\bibfnamefont {S.~E.}\ \bibnamefont {Brown}},\ }\href {\doibase
  10.1073/pnas.2025313118} {\bibfield  {journal} {\bibinfo  {journal}
  {Proceedings of the National Academy of Sciences}\ }\textbf {\bibinfo
  {volume} {118}},\ \bibinfo {pages} {e2025313118} (\bibinfo {year}
  {2021})}\BibitemShut {NoStop}%
\bibitem [{\citenamefont {Scaffidi}\ and\ \citenamefont
  {Simon}(2015)}]{Scaffidi2015}%
  \BibitemOpen
  \bibfield  {author} {\bibinfo {author} {\bibfnamefont {T.}~\bibnamefont
  {Scaffidi}}\ and\ \bibinfo {author} {\bibfnamefont {S.~H.}\ \bibnamefont
  {Simon}},\ }\href {\doibase 10.1103/PhysRevLett.115.087003} {\bibfield
  {journal} {\bibinfo  {journal} {Phys. Rev. Lett.}\ }\textbf {\bibinfo
  {volume} {115}},\ \bibinfo {pages} {087003} (\bibinfo {year}
  {2015})}\BibitemShut {NoStop}%
\bibitem [{\citenamefont {Kirtley}\ \emph {et~al.}(2007)\citenamefont
  {Kirtley}, \citenamefont {Kallin}, \citenamefont {Hicks}, \citenamefont
  {Kim}, \citenamefont {Liu}, \citenamefont {Moler}, \citenamefont {Maeno},\
  and\ \citenamefont {Nelson}}]{Kirtley2007}%
  \BibitemOpen
  \bibfield  {author} {\bibinfo {author} {\bibfnamefont {J.~R.}\ \bibnamefont
  {Kirtley}}, \bibinfo {author} {\bibfnamefont {C.}~\bibnamefont {Kallin}},
  \bibinfo {author} {\bibfnamefont {C.~W.}\ \bibnamefont {Hicks}}, \bibinfo
  {author} {\bibfnamefont {E.-A.}\ \bibnamefont {Kim}}, \bibinfo {author}
  {\bibfnamefont {Y.}~\bibnamefont {Liu}}, \bibinfo {author} {\bibfnamefont
  {K.~A.}\ \bibnamefont {Moler}}, \bibinfo {author} {\bibfnamefont
  {Y.}~\bibnamefont {Maeno}}, \ and\ \bibinfo {author} {\bibfnamefont {K.~D.}\
  \bibnamefont {Nelson}},\ }\href {\doibase 10.1103/PhysRevB.76.014526}
  {\bibfield  {journal} {\bibinfo  {journal} {Phys. Rev. B}\ }\textbf {\bibinfo
  {volume} {76}},\ \bibinfo {pages} {014526} (\bibinfo {year}
  {2007})}\BibitemShut {NoStop}%
\bibitem [{\citenamefont {Hicks}\ \emph {et~al.}(2010)\citenamefont {Hicks},
  \citenamefont {Kirtley}, \citenamefont {Lippman}, \citenamefont {Koshnick},
  \citenamefont {Huber}, \citenamefont {Maeno}, \citenamefont {Yuhasz},
  \citenamefont {Maple},\ and\ \citenamefont {Moler}}]{Hicks2010}%
  \BibitemOpen
  \bibfield  {author} {\bibinfo {author} {\bibfnamefont {C.~W.}\ \bibnamefont
  {Hicks}}, \bibinfo {author} {\bibfnamefont {J.~R.}\ \bibnamefont {Kirtley}},
  \bibinfo {author} {\bibfnamefont {T.~M.}\ \bibnamefont {Lippman}}, \bibinfo
  {author} {\bibfnamefont {N.~C.}\ \bibnamefont {Koshnick}}, \bibinfo {author}
  {\bibfnamefont {M.~E.}\ \bibnamefont {Huber}}, \bibinfo {author}
  {\bibfnamefont {Y.}~\bibnamefont {Maeno}}, \bibinfo {author} {\bibfnamefont
  {W.~M.}\ \bibnamefont {Yuhasz}}, \bibinfo {author} {\bibfnamefont {M.~B.}\
  \bibnamefont {Maple}}, \ and\ \bibinfo {author} {\bibfnamefont {K.~A.}\
  \bibnamefont {Moler}},\ }\href {\doibase 10.1103/PhysRevB.81.214501}
  {\bibfield  {journal} {\bibinfo  {journal} {Phys. Rev. B}\ }\textbf {\bibinfo
  {volume} {81}},\ \bibinfo {pages} {214501} (\bibinfo {year}
  {2010})}\BibitemShut {NoStop}%
\bibitem [{\citenamefont {Curran}\ \emph {et~al.}(2014)\citenamefont {Curran},
  \citenamefont {Bending}, \citenamefont {Desoky}, \citenamefont {Gibbs},
  \citenamefont {Lee},\ and\ \citenamefont {Mackenzie}}]{Curran2014}%
  \BibitemOpen
  \bibfield  {author} {\bibinfo {author} {\bibfnamefont {P.~J.}\ \bibnamefont
  {Curran}}, \bibinfo {author} {\bibfnamefont {S.~J.}\ \bibnamefont {Bending}},
  \bibinfo {author} {\bibfnamefont {W.~M.}\ \bibnamefont {Desoky}}, \bibinfo
  {author} {\bibfnamefont {A.~S.}\ \bibnamefont {Gibbs}}, \bibinfo {author}
  {\bibfnamefont {S.~L.}\ \bibnamefont {Lee}}, \ and\ \bibinfo {author}
  {\bibfnamefont {A.~P.}\ \bibnamefont {Mackenzie}},\ }\href {\doibase
  10.1103/PhysRevB.89.144504} {\bibfield  {journal} {\bibinfo  {journal} {Phys.
  Rev. B}\ }\textbf {\bibinfo {volume} {89}},\ \bibinfo {pages} {144504}
  (\bibinfo {year} {2014})}\BibitemShut {NoStop}%
\bibitem [{\citenamefont {Huang}\ \emph {et~al.}(2015)\citenamefont {Huang},
  \citenamefont {Lederer}, \citenamefont {Taylor},\ and\ \citenamefont
  {Kallin}}]{Huang2015}%
  \BibitemOpen
  \bibfield  {author} {\bibinfo {author} {\bibfnamefont {W.}~\bibnamefont
  {Huang}}, \bibinfo {author} {\bibfnamefont {S.}~\bibnamefont {Lederer}},
  \bibinfo {author} {\bibfnamefont {E.}~\bibnamefont {Taylor}}, \ and\ \bibinfo
  {author} {\bibfnamefont {C.}~\bibnamefont {Kallin}},\ }\href {\doibase
  10.1103/PhysRevB.91.094507} {\bibfield  {journal} {\bibinfo  {journal} {Phys.
  Rev. B}\ }\textbf {\bibinfo {volume} {91}},\ \bibinfo {pages} {094507}
  (\bibinfo {year} {2015})}\BibitemShut {NoStop}%
\bibitem [{\citenamefont {Stone}\ and\ \citenamefont {Roy}(2004)}]{Stone2004}%
  \BibitemOpen
  \bibfield  {author} {\bibinfo {author} {\bibfnamefont {M.}~\bibnamefont
  {Stone}}\ and\ \bibinfo {author} {\bibfnamefont {R.}~\bibnamefont {Roy}},\
  }\href {\doibase 10.1103/PhysRevB.69.184511} {\bibfield  {journal} {\bibinfo
  {journal} {Phys. Rev. B}\ }\textbf {\bibinfo {volume} {69}},\ \bibinfo
  {pages} {184511} (\bibinfo {year} {2004})}\BibitemShut {NoStop}%
\bibitem [{\citenamefont {Li}\ \emph {et~al.}(2021)\citenamefont {Li},
  \citenamefont {Kikugawa}, \citenamefont {Sokolov}, \citenamefont
  {Jerzembeck}, \citenamefont {Gibbs}, \citenamefont {Maeno}, \citenamefont
  {Hicks}, \citenamefont {Schmalian}, \citenamefont {Nicklas},\ and\
  \citenamefont {Mackenzie}}]{Li2021}%
  \BibitemOpen
  \bibfield  {author} {\bibinfo {author} {\bibfnamefont {Y.-S.}\ \bibnamefont
  {Li}}, \bibinfo {author} {\bibfnamefont {N.}~\bibnamefont {Kikugawa}},
  \bibinfo {author} {\bibfnamefont {D.~A.}\ \bibnamefont {Sokolov}}, \bibinfo
  {author} {\bibfnamefont {F.}~\bibnamefont {Jerzembeck}}, \bibinfo {author}
  {\bibfnamefont {A.~S.}\ \bibnamefont {Gibbs}}, \bibinfo {author}
  {\bibfnamefont {Y.}~\bibnamefont {Maeno}}, \bibinfo {author} {\bibfnamefont
  {C.~W.}\ \bibnamefont {Hicks}}, \bibinfo {author} {\bibfnamefont
  {J.}~\bibnamefont {Schmalian}}, \bibinfo {author} {\bibfnamefont
  {M.}~\bibnamefont {Nicklas}}, \ and\ \bibinfo {author} {\bibfnamefont
  {A.~P.}\ \bibnamefont {Mackenzie}},\ }\href {\doibase
  10.1073/pnas.2020492118} {\bibfield  {journal} {\bibinfo  {journal}
  {Proceedings of the National Academy of Sciences}\ }\textbf {\bibinfo
  {volume} {118}},\ \bibinfo {pages} {e2020492118} (\bibinfo {year}
  {2021})}\BibitemShut {NoStop}%
\bibitem [{\citenamefont {Vaugier}\ \emph {et~al.}(2012)\citenamefont
  {Vaugier}, \citenamefont {Jiang},\ and\ \citenamefont
  {Biermann}}]{Vaugier2012}%
  \BibitemOpen
  \bibfield  {author} {\bibinfo {author} {\bibfnamefont {L.}~\bibnamefont
  {Vaugier}}, \bibinfo {author} {\bibfnamefont {H.}~\bibnamefont {Jiang}}, \
  and\ \bibinfo {author} {\bibfnamefont {S.}~\bibnamefont {Biermann}},\ }\href
  {\doibase 10.1103/PhysRevB.86.165105} {\bibfield  {journal} {\bibinfo
  {journal} {Phys. Rev. B}\ }\textbf {\bibinfo {volume} {86}},\ \bibinfo
  {pages} {165105} (\bibinfo {year} {2012})}\BibitemShut {NoStop}%
\bibitem [{\citenamefont {Behrmann}\ \emph {et~al.}(2012)\citenamefont
  {Behrmann}, \citenamefont {Piefke},\ and\ \citenamefont
  {Lechermann}}]{Behrmann2012}%
  \BibitemOpen
  \bibfield  {author} {\bibinfo {author} {\bibfnamefont {M.}~\bibnamefont
  {Behrmann}}, \bibinfo {author} {\bibfnamefont {C.}~\bibnamefont {Piefke}}, \
  and\ \bibinfo {author} {\bibfnamefont {F.}~\bibnamefont {Lechermann}},\
  }\href {\doibase 10.1103/PhysRevB.86.045130} {\bibfield  {journal} {\bibinfo
  {journal} {Phys. Rev. B}\ }\textbf {\bibinfo {volume} {86}},\ \bibinfo
  {pages} {045130} (\bibinfo {year} {2012})}\BibitemShut {NoStop}%
\bibitem [{\citenamefont {Tsuchiizu}\ \emph {et~al.}(2015)\citenamefont
  {Tsuchiizu}, \citenamefont {Yamakawa}, \citenamefont {Onari}, \citenamefont
  {Ohno},\ and\ \citenamefont {Kontani}}]{Tsuchiizu2015}%
  \BibitemOpen
  \bibfield  {author} {\bibinfo {author} {\bibfnamefont {M.}~\bibnamefont
  {Tsuchiizu}}, \bibinfo {author} {\bibfnamefont {Y.}~\bibnamefont {Yamakawa}},
  \bibinfo {author} {\bibfnamefont {S.}~\bibnamefont {Onari}}, \bibinfo
  {author} {\bibfnamefont {Y.}~\bibnamefont {Ohno}}, \ and\ \bibinfo {author}
  {\bibfnamefont {H.}~\bibnamefont {Kontani}},\ }\href {\doibase
  10.1103/PhysRevB.91.155103} {\bibfield  {journal} {\bibinfo  {journal} {Phys.
  Rev. B}\ }\textbf {\bibinfo {volume} {91}},\ \bibinfo {pages} {155103}
  (\bibinfo {year} {2015})}\BibitemShut {NoStop}%
\bibitem [{\citenamefont {Kaba}\ and\ \citenamefont
  {S\'en\'echal}(2019)}]{Kaba2019}%
  \BibitemOpen
  \bibfield  {author} {\bibinfo {author} {\bibfnamefont {S.-O.}\ \bibnamefont
  {Kaba}}\ and\ \bibinfo {author} {\bibfnamefont {D.}~\bibnamefont
  {S\'en\'echal}},\ }\href {\doibase 10.1103/PhysRevB.100.214507} {\bibfield
  {journal} {\bibinfo  {journal} {Phys. Rev. B}\ }\textbf {\bibinfo {volume}
  {100}},\ \bibinfo {pages} {214507} (\bibinfo {year} {2019})}\BibitemShut
  {NoStop}%
\bibitem [{\citenamefont {Ramires}\ and\ \citenamefont
  {Sigrist}(2019)}]{Ramires2019}%
  \BibitemOpen
  \bibfield  {author} {\bibinfo {author} {\bibfnamefont {A.}~\bibnamefont
  {Ramires}}\ and\ \bibinfo {author} {\bibfnamefont {M.}~\bibnamefont
  {Sigrist}},\ }\href {\doibase 10.1103/PhysRevB.100.104501} {\bibfield
  {journal} {\bibinfo  {journal} {Phys. Rev. B}\ }\textbf {\bibinfo {volume}
  {100}},\ \bibinfo {pages} {104501} (\bibinfo {year} {2019})}\BibitemShut
  {NoStop}%
\bibitem [{\citenamefont {Huang}\ \emph {et~al.}(2019)\citenamefont {Huang},
  \citenamefont {Zhou},\ and\ \citenamefont {Yao}}]{Huang2019}%
  \BibitemOpen
  \bibfield  {author} {\bibinfo {author} {\bibfnamefont {W.}~\bibnamefont
  {Huang}}, \bibinfo {author} {\bibfnamefont {Y.}~\bibnamefont {Zhou}}, \ and\
  \bibinfo {author} {\bibfnamefont {H.}~\bibnamefont {Yao}},\ }\href {\doibase
  10.1103/PhysRevB.100.134506} {\bibfield  {journal} {\bibinfo  {journal}
  {Phys. Rev. B}\ }\textbf {\bibinfo {volume} {100}},\ \bibinfo {pages}
  {134506} (\bibinfo {year} {2019})}\BibitemShut {NoStop}%
\bibitem [{\citenamefont {Altmeyer}\ \emph {et~al.}(2016)\citenamefont
  {Altmeyer}, \citenamefont {Guterding}, \citenamefont {Hirschfeld},
  \citenamefont {Maier}, \citenamefont {Valent\'{\i}},\ and\ \citenamefont
  {Scalapino}}]{Altmeyer2016}%
  \BibitemOpen
  \bibfield  {author} {\bibinfo {author} {\bibfnamefont {M.}~\bibnamefont
  {Altmeyer}}, \bibinfo {author} {\bibfnamefont {D.}~\bibnamefont {Guterding}},
  \bibinfo {author} {\bibfnamefont {P.~J.}\ \bibnamefont {Hirschfeld}},
  \bibinfo {author} {\bibfnamefont {T.~A.}\ \bibnamefont {Maier}}, \bibinfo
  {author} {\bibfnamefont {R.}~\bibnamefont {Valent\'{\i}}}, \ and\ \bibinfo
  {author} {\bibfnamefont {D.~J.}\ \bibnamefont {Scalapino}},\ }\href {\doibase
  10.1103/PhysRevB.94.214515} {\bibfield  {journal} {\bibinfo  {journal} {Phys.
  Rev. B}\ }\textbf {\bibinfo {volume} {94}},\ \bibinfo {pages} {214515}
  (\bibinfo {year} {2016})}\BibitemShut {NoStop}%
\bibitem [{\citenamefont {{R{\o}mer}}\ and\ \citenamefont
  {Andersen}(2020)}]{Romer2020a}%
  \BibitemOpen
  \bibfield  {author} {\bibinfo {author} {\bibfnamefont {A.~T.}\ \bibnamefont
  {{R{\o}mer}}}\ and\ \bibinfo {author} {\bibfnamefont {B.~M.}\ \bibnamefont
  {Andersen}},\ }\href {\doibase 10.1142/S0217984920400527} {\bibfield
  {journal} {\bibinfo  {journal} {Modern Physics Letters B}\ }\textbf {\bibinfo
  {volume} {34}},\ \bibinfo {pages} {2040052} (\bibinfo {year}
  {2020})}\BibitemShut {NoStop}%
\end{thebibliography}

%


\end{document}